\documentclass[a4paper,twocolumn,11pt,accepted=2019-12-26]{quantumarticle}
\pdfoutput=1
\usepackage[english]{babel}
\usepackage[T1]{fontenc}
\usepackage[utf8]{inputenc}
\usepackage{amsmath}
\usepackage{bbold}
\usepackage{color}
\usepackage{graphicx}
\usepackage{scalerel}
\usepackage{braket}
\usepackage{leftidx}
\usepackage[title]{appendix}
\usepackage{tikz-cd}
\usepackage{dsfont}
\usepackage{cite}
\usepackage{hyperref}
\usepackage{authblk}
\usepackage{soul}

\def\be{\begin{equation}}
\def\ee{\end{equation}}
\def\ba{\begin{eqnarray}}
\def\ea{\end{eqnarray}}

\newcommand\nn{\nonumber}
\newcommand\q{\quad}
\newcommand{\cc}{\mathcal C}

\newcommand{\cg}{\mathcal G}
\newcommand{\ch}{\mathcal H}

\newcommand{\cp}{\mathcal P}
\newcommand{\cq}{\mathcal Q}

\newcommand{\cs}{\mathcal S}
\newcommand{\ct}{\mathcal T}

\def\f{\frac}

\def\p{\partial}

\usepackage{bbm}

\newcommand{\lalg}[1]{\mathfrak{#1}}  

\renewcommand{\sl}{\lalg{sl}}

\def\q{{\quad}}

\title{A change of perspective: switching quantum reference frames via a perspective-neutral framework}

\author[1]{Augustin Vanrietvelde}
\orcid{0000-0001-9022-8655}
\author[1,2]{Philipp A.\ H\"ohn}
\thanks{corresponding author}
\orcid{0000-0001-7108-1184}
\email{hoephil@gmail.com}
\author[1,2]{Flaminia Giacomini}
\author[1,2]{Esteban Castro-Ruiz}
\affil[1]{\small Institute for Quantum Optics and Quantum Information, Austrian Academy of Sciences,\newline Boltzmanngasse 3, 1090 Vienna, Austria}
\affil[2]{\small Vienna Center for Quantum Science and Technology (VCQ), Faculty of Physics, University of Vienna, Boltzmanngasse 5, 1090 Vienna, Austria}

\date{}

\begin{document}

\maketitle

\begin{abstract}
Treating reference frames fundamentally as quantum systems is inevitable in quantum gravity and also in quantum foundations once considering laboratories as physical systems.
%
%
Both fields thereby face the question of how to describe physics relative to quantum reference systems and how the descriptions relative to different such choices are related. Here, we exploit a fruitful interplay of ideas from both fields to begin developing a unifying approach to transformations among quantum reference systems that ultimately aims at encompassing both quantum and gravitational physics. In particular, using a gravity inspired symmetry principle, which enforces physical observables to be relational and leads to an inherent redundancy in the description, we develop a perspective-neutral structure, which contains all frame perspectives at once and via which they are changed. We show that taking the perspective of a specific frame amounts to a fixing of the symmetry related redundancies in both the classical and quantum theory and that changing perspective corresponds to a symmetry transformation. We implement this using the language of constrained systems, which naturally encodes symmetries. Within a simple one-dimensional model, we recover some of the quantum frame transformations of \cite{Giacomini:2017zju}, embedding them in a perspective-neutral framework. Using them, we illustrate how entanglement and classicality of an observed system depend on the quantum frame perspective. Our operational language also inspires a new interpretation of Dirac and reduced quantized theories within our model as perspective-neutral and perspectival quantum theories, respectively, and reveals the explicit link between them. In this light, we suggest a new take on the relation between a `quantum general covariance' and the diffeomorphism symmetry in quantum gravity.
\end{abstract}

%
%

\section{Introduction}

Reference frames are essential in our description of physical phenomena. Every time we measure a physical quantity or describe a physical event, we do so with respect to a reference frame. In practice, reference frames are physical objects that are sufficiently decoupled from the system we want to describe. When the reference frame does not influence the system of interest for all practical purposes, we treat it like an external entity in our theoretical analysis. For example, in our current most successful theories, general relativity and quantum theory (incl.\ quantum field theory), such reference frames are taken as idealized classical systems that are non-dynamical and neither back-react on spacetime itself, nor on other fields contained in it. 

However, a more fundamental approach to physics should dispose of such an idealization and take seriously the fact that reference frames are always physical systems themselves and thereby subject to interactions and the laws of physics. In particular, if we accept the universality of quantum theory, we have to face the question of how to describe physics with respect to a {\it quantum} frame of reference and, subsequently, how the descriptions relative to different such quantum frames are related to one another. Classical frame transformations will not suffice to switch from the perspective of one quantum frame to that of another, as already epitomized, in its most extreme, by the Wigner's friend scenario. 
So what will take their place in a fully quantum formulation?

The answer to this question will, of course, depend on the concrete physics at hand. But our aim here and in \cite{Vanrietvelde:2018dit,Hoehn:2018aqt, Hoehn:2018whn,pqps} will be to initiate the development of a novel and systematic method for answering this question, 
that ultimately can encompass quantum reference frames in both quantum and gravitational physics. That is, this method shall be applicable in both fields to produce the sought-after transformations among quantum frame perspectives. 

Reference frames (or, more generally, reference systems) indeed provide a natural arena for an interplay of quantum and gravitational physics, appearing ubiquitously in both fields. Their recognition as quantum systems themselves dates back to at least 1967 when, in a historical coincidence, two seminal papers, by Aharonov and Susskind \cite{Aharonov:1967zza} and DeWitt \cite{DeWitt:1967yk}, separately brought this recognition to center stage in the foundations of quantum theory and quantum gravity, respectively. The ensuing study and usage of quantum reference systems took, however, rather different directions in the two fields and our goal will be to unify some of these developments.

In (quantum) gravity, there is even a necessity to employ physical systems as references with respect to which to describe the remaining physics. On account of the diffeomorphism symmetry of general relativity, which is a consequence of general covariance, 
physical systems are not localized and oriented relative to some absolute spatiotemporal structure, but with respect to one another \cite{Rovelli:2004tv}. This is spacetime relationalism and, for later purpose, we note that this is closely related to Mach's principle, which roughly states that what is an inertial frame is not determined with respect to an absolute space (as in Newtonian physics), but by the other dynamical content of the universe \cite{Barbour295,mercati2018shape,Rovelli:2004tv}. Physics is purely relational: intuitively, if one moves around the entire dynamical content of the universe while keeping the relations among its constituents intact, it will not change the physics. Mach's principle thereby implies a symmetry principle and a corresponding redundancy in the description (in general relativity this is the diffeomorphism symmetry); we shall see a toy version of this below.

This entails the {\it paradigm of relational localization}:
some dynamical matter or gravitational degrees of freedom (in the full theory these will be fields) serve as temporal or spatial reference systems for others and these relations are invariant under diffeomorphisms. That is, these relations are invariant under the gauge symmetry of general relativity and thereby physically meaningful. Such gauge invariant relations are usually referred to as {\it relational observables}. This applies not only to classical general relativity, but also to (background independent) quantum gravity approaches where one aims at quantizing all dynamical degrees of freedom, while retaining (a quantum version of) diffeomorphism invariance. As such, quantum reference systems and relational observables appear ubiquitously in quantum gravity and, given their indispensability, they have been studied extensively \cite{DeWitt:1967yk,Rovelli:2004tv,Rovelli:1990pi,Rovelli:1990ph,Kuchar:1991qf,Isham:1992ms,Brown:1994py,Dittrich:2004cb,Dittrich:2005kc,Tambornino:2011vg,Thiemann:2007zz,Dittrich:2016hvj,Dittrich:2015vfa}. 

However, owing to the challenges in a field theory context, what has {\it not} been studied extensively is how to switch among different choices of such relational quantum reference systems in quantum gravity. For systems with finitely many phase space degrees of freedom (such as quantum cosmological models), the first systematic framework for changes of quantum reference systems was developed in \cite{Bojowald:2010xp,Bojowald:2010qw,Hohn:2011us} and applied to temporal references called relational clocks, but restricted to sufficiently semiclassical states. The crucial feature of this framework is a `perspective-neutral' quantum theory which contains all clock choices at once and each clock choice corresponds to a gauge fixing.


%
%
%
%
%



In quantum information, on the other hand, quantum reference frames have been extensively discussed mainly with the purpose of devising communication tasks with physical systems serving as detectors. In the seminal papers \cite{Aharonov:1967zza, PhysRev.158.1237} it was shown that it is possible to overcome superselection rules (such as the charge superselection rule in Ref.~\cite{Aharonov:1967zza}) via the introduction of a quantum reference frame. In Ref.\ \cite{PhysRevD.30.368} it was shown that quantum mechanics can be consistently formulated without appealing to abstract reference frames of infinite mass. The subsequent literature \cite{Bartlett:2007zz,bartlett2009quantum, gour2008resource, Palmer:2013zza,bartlett2006degradation,smith2016quantum, poulin2007dynamics,PhysRevLett.111.020504} has then focused on different aspects of the introduction of quantum systems as reference frames, and mainly on a) the lack of a shared reference frame, b) bounded-size reference frames, and c) the possibility of overcoming general superselction rules by employing quantum reference frames (e.g., see \cite{Bartlett:2007zz} for a review). 
These approaches resort to an encoding of quantum information into relational degrees of freedom. The latter are invariant under an averaging over the external symmetry group, defining decoherence free subsystems. A relational approach to quantum reference frames has been considered also in \cite{loveridge2017relativity, pienaar2016relational, angelo2011physics}. 
The transformation between two quantum reference frames is in general not considered in this applied quantum information context, with the important exception of \cite{Palmer:2013zza}. More foundationaly, quantum reference frames have also been used to derive the Lorentz group from operational conditions on quantum communication without presupposing a specific spacetime structure \cite{Hoehn:2014vua}; this exemplifies how quantum information protocols can constrain the spacetime structures in which they are feasible.

A suitable starting point for establishing a connection between these efforts in quantum information and in quantum gravity is the approach to quantum reference frames developed in \cite{Giacomini:2017zju}, which we shall further exploit in the present paper. The main idea of Ref.~\cite{Giacomini:2017zju} is to formulate changes between quantum reference frames in an operational and fully relational way, without referring to any external or absolute entity. Within this formulation, Ref.~\cite{Giacomini:2017zju} investigates the extension of the covariance of physical laws under such reference frame transformations, paving the way for a formulation of a notion of ``quantum general covariance''. Such developments give concrete meaning to the idea of describing physics from the point of view of a quantum frame of reference. This approach might be particularly relevant in the context of quantum gravity, where a fixed notion of spacetime (spacetime metric) is not available. As an indication of the concrete possibility of formulating physics on a non-fixed spacetime metric, Ref.~\cite{Giacomini:2017zju} develops an extension of Galileo's weak equivalence principle for quantum reference frames, which holds when the reference frame is a system falling in a superposition of accelerations. This approach is in resonance with works aiming at formulating physics on indefinite causal structures from an observer-dependent perspective \cite{guerin2018observer, oreshkov2012quantum,Hardy:2018kbp}. In particular, it is closely related to Hardy's proposal for a quantum equivalence principle \cite{Hardy:2018kbp}, stating that it is always possible to find a quantum reference frame having a definite causal structure in the local vicinity of any point.



Our main ambition will be to synthesize these developments in quantum gravity and quantum information
 into a unifying method for switching perspectives in the quantum theory that includes both spatial and temporal quantum reference systems and applies in both fields. 
 Indeed, in the course of this work, here and in \cite{Vanrietvelde:2018dit,Hoehn:2018aqt, Hoehn:2018whn,pqps}, we shall show how (some of) the quantum reference frame transformations of \cite{Giacomini:2017zju} and the method of relational clock changes \cite{Bojowald:2010xp,Bojowald:2010qw,Hohn:2011us} can be accommodated and reproduced within one framework. This will be achieved by adopting key ingredients from both 
sides.

In particular, we shall adopt a gravity inspired symmetry principle to develop, as proposed in \cite{Hoehn:2017gst}, 
a perspective-neutral super structure that encodes, so to speak, all perspectives at once and requires additional choices to `jump' into the perspective of a specific frame. Technically, we will achieve this by availing ourselves of the tools and concepts of constrained Hamiltonian systems \cite{Dirac,Henneaux:1992ig} that also play a key role in the canonical formulation of general relativity and quantum gravity \cite{Rovelli:2004tv,Thiemann:2007zz} and that were also used for the relational clock changes in \cite{Bojowald:2010xp,Bojowald:2010qw,Hohn:2011us}. 

The symmetry principle will, as mentioned above, entail two related key features: (i) an inevitable redundancy in the description of the physics (gauge freedom {and constraints}), and (ii) that the physically meaningful (i.e.\ gauge invariant) information is purely relational. The inherent redundancy will permit us to treat all possible reference frames as part of a larger physical system at once and on an equal footing; a priori no choice of frame is preferred and no frame is described externally. However, in order to make operational sense out of physical phenomena, we must make additional choices to fix these redundancies. We will show that {\it choosing a system from the perspective-neutral picture to serve as our reference frame is equivalent to fixing these redundancies} and that classically this is a choice of gauge. Accordingly, (at least classically) switching from the internal perspective of one frame to another will amount to a symmetry transformation as in \cite{Bojowald:2010xp,Bojowald:2010qw,Hohn:2011us}. 
Our approach thereby connects with, but also extends the discussion in \cite{Rovelli:2013fga}, where it is argued that the redundancy in gauge theories is not just a mathematical artifact, but expresses the fact that physics is relational and provides the `handles' through which systems can couple (and relate to one another) in different ways.\footnote{A complementary extension of these ideas, which does not rely on gauge fixings to define frames, has also recently been put forward for the field theory context in \cite{Gomes:2018dxs}.}



Conversely, the operational language of quantum foundations and, specifically, the approach to quantum reference frames in \cite{Giacomini:2017zju} will 
supply our operational interpretation of the formalism. In particular, it will inspire compelling new insights into the quantization of constrained systems. These insights will thereby be of  relevance for quantum gravity. Indeed, there exist two main strategies in the literature for canonically quantizing constrained systems: 
\begin{description}
\item[Reduced quantization:] One solves the constraints first at the classical level and only quantizes non-redundant gauge invariant degrees of freedom.
\item[Dirac quantization:] One quantizes first all (incl.\ redundant and gauge) degrees of freedom and solves the constraints in the quantum theory.
\end{description}
There has been an ample discussion in the literature as to how these two quantization strategies are related -- with the general conclusion that `constraint imposition and quantization do not commute' -- and about when one or the other should be applied \cite{guillemin1982geometric, tian1998analytic,hochs2008guillemin,gotay1986constraints,Ashtekar:1982wv,kucha1986covariant,Ashtekar:1991hf,Schleich:1990gd,Kunstatter:1991ds,Hajicek:1990eu,Romano:1989zb,Dittrich:2016hvj,Dittrich:2015vfa,Loll:1990rx,plyushchay1996dirac}. Adopting the operational language of \cite{Giacomini:2017zju}, we will shed {some} new light on this discussion, both technically and conceptually. 

{In this article, we will begin with a technically rather simple model, subject to a {\it linear} constraint, on a finite dimensional phase space.} {Within this model both quantization} methods are necessary for a complete relational interpretation. As we shall see, Dirac quantization will yield a perspective-neutral quantum theory, containing all quantum reference frame perspectives at once, while reduced quantization produces the quantum physics as seen by a specific frame. We will also provide the transformations that take us from one to the other and will exploit this to establish switches between different quantum reference frames. 
{In particular, the transformation linking Dirac with a given reduced quantization constitutes a \emph{quantum symmetry reduction procedure}, i.e.\ the quantum analog of phase space reduction. This reduction is always formulated relative to a choice of quantum reference frame. In this simple model, all these transformations will be valid globally on phase and Hilbert space, so that no technical subtleties cloud our main arguments and interpretation.}

{However, in generic systems, it will not always be true that the quantum theory obtained by applying the quantum symmetry reduction to Dirac quantization coincides with a specific reduced phase space quantization, in line with the observations in \cite{Ashtekar:1982wv,kucha1986covariant,Ashtekar:1991hf,Schleich:1990gd,Kunstatter:1991ds,Hajicek:1990eu,Romano:1989zb,Dittrich:2016hvj,Dittrich:2015vfa,Loll:1990rx,plyushchay1996dirac}. This will not be a problem for our approach, as we explain in more detail later: in general, we will interpret the result of applying the quantum symmetry reduction to Dirac quantization, which removes redundancy in that description, as the perspective of the associated frame.} 

{Furthermore, in generic systems, globally valid perspectives of quantum reference frames will} become impossible (this is analogous to the Gribov problem in gauge theories) and, accordingly, {the} transformations between them will likewise not be of global validity. This is illustrated in the companion paper \cite{Vanrietvelde:2018dit}, where we extend our discussion to the three-dimensional $N$-body problem, and in \cite{Hoehn:2018aqt, Hoehn:2018whn}, where it will be shown how to change relational quantum clocks, using our new method. In particular, in contrast to \cite{Bojowald:2010xp,Bojowald:2010qw,Hohn:2011us}, these clock changes will also be valid beyond a semiclassical regime.

In connection with discussions of relative and global states in the literature, 
we emphasize that our perspective-neutral structure itself will not admit an immediate operational interpretation, only the description relative to a given perspective. Concretely, this means that there will be global quantum states for the entire physics at once, namely those of the perspective-neutral Dirac quantized theory. However, there will be no global operational states and only states relative to a frame (which will not include its own degrees of freedom) will admit an operational interpretation. This will be exploited in \cite{Hoehn:2018whn} to develop a novel take on the `wave function of the universe' in quantum cosmology, as proposed in \cite{Hoehn:2014uua,Hoehn:2017gst}, and suggests a new conception of the relative states of {\it relational quantum mechanics} \cite{Rovelli:1995fv,rovelli2018space} and their interrelations.

Quantum foundations and (quantum) gravity are usually considered independently. However, our results are a clear testimony to how a fruitful interplay of their tools and perspectives can lead to new conceptual and technical insights in both fields. 

The rest of this article is organized as follows. In sec.\ \ref{sec_meta}, we explain the interplay of perspective-neutral structures and internal perspectives in physics more carefully; a quick reader can skip it on a first reading. Subsequently, in sec.\ \ref{classical}, we introduce a toy model of $N$ particles in one-dimensional Newtonian space in which we impose a symmetry principle, namely global translation invariance, which will serve as a toy version of Mach's principle. Here we show how frame perspectives are related to gauge choices. In sec.~\ref{quantum} we quantize the classical model and explicitly reveal the conceptual and technical relation between the Dirac and reduced quantization {of our toy model}, which here give equivalent expectation values. Finally, in sec.~\ref{operational} we analyze some of the operational consequences of describing physics from the point of view of a quantum system. In particular, a concrete example will illustrate the quantum frame dependence of the degree of entanglement of an observed system. Finally, we conclude in sec.~\ref{sec_conc} with an outlook on further applications of our approach. Details have been moved into appendices.

\section{A meta-perspective on perspectives}\label{sec_meta}

{\it The quick reader can skip this section and proceed directly to sec.~\ref{classical}.} 

The purpose of this section is to motivate and specify more clearly what we mean by a perspective-neutral theory, as proposed in \cite{Hoehn:2017gst}. To this end, we shall adapt the abstract language introduced in \cite{Hoehn:2014vua} to explain from a very general standpoint how different perspectives can fit into one framework and how one can switch between them. We shall thus revisit some fairly basic questions, illustrating along the way how perspective-neutral structures already appear ubiquitously in all of physics. This discussion will also highlight some peculiarities, such as an absence of global perspectives in most systems of interest, and explain within a broad context the structure of the sought-after perspective changes which we will encounter in the course of our work. 

The aim of physics is to describe the physical world, or at least a subset thereof. Usually, this is done by choosing a perspective from which to describe the physical situation at hand. Abstractly, choosing a perspective is thus tantamount to choosing a map from the physics of interest to some suitable mathematical description 
thereof. More precisely, 
denote by $\cs_{\rm phys}$ the set of possible physical situations one wishes to describe (and could, in principle, measure) and by $\cs_{\rm des}$ the set of mathematical objects used for the description of these situations. Then choosing a perspective defines a map
\ba
\varphi:\cs_{\rm phys}\rightarrow\cs_{\rm des}\nn
\ea
from the actual physics to its description. 

The important point to notice is that {\it the actual physics, encoded in $\cs_{\rm phys}$, is, in fact, perspective-neutral}. For instance, suppose the physical situation is that a billiard ball flies through space so that $\cs_{\rm phys}$ denotes the set of all its possible spatial velocities. The statement of such a physical situation per se does not require the perspective of some reference frame, but it {\it can} be described from many different perspectives. Indeed, suppose there is an observer Alice who measures the (components of the) velocity of the ball in three spatial directions and reads it off the scales of her measurement device. Then Alice would usually take $\cs_{\rm des}$ to be $\mathbb{R}^3$ and $\varphi_A$ associates to each physical velocity a three-dimensional vector, corresponding to the three real numbers she reads off her measurement device, thereby specifying the velocity relative to her frame of reference. This is a second point to notice: {\it the choice of a map, i.e.\ a perspective, $\varphi_A$ is (usually) associated with a choice of reference frame}, which is why we have now attached a frame label to it. Note that only a concrete perspective has an immediate operational interpretation.

Of course, this structure is completely general. For example, $\cs_{\rm phys}$ could also represent the quantum states associated with (possibly an ensemble of) a physical system that one can try to estimate, using tomography, in a laboratory. A physically distinguished choice for $\cs_{\rm des}$ would be the appropriate set of density matrices. Clearly, a concrete description $\varphi_A$ of the quantum states depends on a choice of reference frame as it involves the choice of a Hilbert space basis that Alice associates with certain measurement outcomes, say, of spin in her $z$-direction on which another observer Bob may not agree.

In the previous two examples, while the actual physics is perspective-neutral, the theories describing it are arguably {\it not}. For example, if one wrote down a standard Lagrangian for the billiard ball, it would fail to be invariant under general time-dependent changes of coordinates in configuration space; it does not abide by a full symmetry principle and thereby presupposes a special class of (e.g., inertial) frames with respect to which it is formulated. Similarly, at least the standard textbook formulation of quantum mechanics implicitly assumes the frame of the observer and her measurement and preparation devices at the outset.

By contrast, a prime example of a {\it perspective-neutral theory} is general relativity. The Einstein-Hilbert action is completely independent of coordinates and choices of reference frame (it is diffeomorphism invariant) and so the theory does not dictate the choice of perspective from which to interpret and describe the physics {\it in} spacetime; it contains all frame perspectives at once and on equal footing and it is up to the physicist to pick one. {For example, when considering the dynamics {\it in} a given spacetime} in general relativity, $\cs_{\rm phys}$ may represent {the possible physical situations happening in that} spacetime.{\footnote{{By contrast, when considering the dynamics  {\it of} spacetime in general relativity, $\cs_{\rm phys}$ may represent the space of solutions or, in the canonical formulation, the constraint surface (see also comments below).}}}  
Given a reference frame associated to some observer Alice (usually an orthonormal tetrad), $\cs_{\rm des}$ is then normally taken to be $\mathbb{R}^4$. Her perspective $\varphi_A$ defines a (usually only locally valid) coordinate description {of the physics in that given spacetime}, e.g., encoding the tangent vector corresponding to the motion of a massive object in a four-vector whose components describe the velocity relative to Alice.

In the course of our work, 
we shall show how to embed the discussion of quantum reference frames into such a meta-framework. In particular, in analogy to general relativity, we will use a symmetry principle, in the form of invariant Lagrangians and (first class) constraints, to formulate perspective-neutral theories of reference frames. We shall use these theories to argue for a novel, more general interpretation of key structures of constrained systems, incl.\ canonical gravity. 

Indeed, for a system with first class constraints, we shall propose to interpret the classical constraint surface and the gauge invariant physical Hilbert space $\ch_{\rm phys}$ of its Dirac quantization as the perspective-neutral physics $\cs_{\rm phys}$ of the classical and quantum theory, respectively. Correspondingly, we shall argue that the reduced (gauge fixed) phase spaces and their reduced quantizations, {the reduced Hilbert spaces} $\ch_{\rm red}$, assume the role of the descriptions $\cs_{\rm des}$ of the classical and quantized physics, respectively. 
Hence, a perspective $\varphi_A$ will define a mapping from the constraint surface/$\ch_{\rm phys}$ to the reduced phase space/$\ch_{\rm red}$ and we shall only assign an operational interpretation to the latter reduced structures; these are the physics described with respect to a given classical or quantum reference frame.\footnote{There is an interesting analogy to the relation between shape dynamics and general relativity \cite{Gomes:2010fh,mercati2018shape}. The two theories are related via a `linking theory' \cite{Gomes:2011zi} that can be regarded as the perspective-neutral theory. When restricted to solutions admitting constant-mean-curvature slicings, shape dynamics and general relativity (as reductions of the linking theory) can be regarded as two different descriptions of the same physics.}

For {\it any} of the above examples and theories, it is now also clear how to switch from the perspective of, say Alice's frame, to another, say Bob's, namely through the following transformation:
\ba
T_{A\rightarrow B}=\varphi_B\circ\varphi_A^{-1}\,.\label{gentrafo}
\ea
Note that, while $T_{A\rightarrow B}:\cs_{\rm des}\rightarrow\cs_{\rm des}$ is a map from description to description,\footnote{Similarly, one can construct transformations $T_{A\rightarrow B}^{\rm phys}:=\varphi_B^{-1}\circ\varphi_A:\cs_{\rm phys}\rightarrow\cs_{\rm phys}$ that are actual operations on the physics \cite{Hoehn:2014vua}. Since we are only interested in perspectives and their relations, such operations are not relevant here.}  it {\it always} proceeds via the perspective-neutral structure $\cs_{\rm phys}$ in-between, thanks to its compositional form. This is the general form of our sought-after transformations and we shall encounter it repeatedly throughout our work, i.e.\ below and in \cite{Vanrietvelde:2018dit,Hoehn:2018aqt, Hoehn:2018whn,pqps}. Hence, we will always {\it switch perspectives via the perspective-neutral meta-structure} in both the classical and the quantum theory. Notice also the structural resemblance to coordinate changes on a manifold. However, here it is more than just a coordinate transformation: it is a change of perspective. 

The transformation (\ref{gentrafo}) clearly assumes the perspective map $\varphi_A$ to be invertible {\it somewhere}. The example of general relativity above makes it clear, however, that this will in general not be possible globally; $\varphi_A$ need not be defined everywhere on the perspective-neutral physics $\cs_{\rm phys}$. In other words, in general we will find that {\it global perspectives on the physics (with operational interpretation) do not exist in most interesting systems}. 
This will also be illustrated 
in the companion articles \cite{Vanrietvelde:2018dit,Hoehn:2018aqt, Hoehn:2018whn}.

In consequence, the perspective changes (\ref{gentrafo}) will generally constitute non-global transformations and it will become a non-trivial question whether (and where) Bob's perspective $\varphi_B$ can be concatenated with the inverse $\varphi_A^{-1}$ of Alice's perspective. Hence, in general it will be a non-trivial question too whether perspective changes (\ref{gentrafo}) can be concatenated and constitute a group or more general structures such as a groupoid. Such questions are crucial as a lot of information about the physics resides in perspectives and 
their relations. For instance, the information about a spacetime's geometry is encoded in the relations among its reference frames and this is also where symmetries reside.

In the following, we shall now transition from a perspective-neutral structure to internal perspectives and study operational consequences of the ensuing transformations (\ref{gentrafo}). By contrast, the constructions in \cite{Hoehn:2014vua,Hoehn:2018rfe} can be regarded as pursuing in the opposite direction: they start with operational conditions on relations among internal perspectives and attempt to reconstruct a perspective-neutral structure.


\section{Classical reference frame perspectives as gauge-fixings} \label{classical}

We now construct a simple model, which incorporates {a toy version of} Mach's principle for $N$ interacting particles in one-dimensional Newtonian space through a global translational invariance. We will argue that it constitutes a perspective-neutral theory in which no reference frame has been chosen yet, and in which physical quantities are relational. We will then show that going to the perspective of a particular reference frame amounts to a gauge fixing, and, correspondingly, that switches from one frame perspective to another are gauge transformations.

For intuition: `jumping' into the perspective of a given reference frame defines, e.g., what it means to be `at the origin' in position space, fixing the translational symmetry. Conversely, starting from the assumption that one can always `jump' into a frame that is `at the origin', one is led to a symmetry, because our ability to `fix' any system at the origin means its `absolute position' is not physical. 

The technical simplicity of the model will permit us to illustrate in sec.\ \ref{quantum} the general method for changing perspectives via a perspective-neutral structure in the quantum theory and, in particular, to derive the quantum reference frame transformations constructed in \cite{Giacomini:2017zju} for the one-dimensional case from first principles. In this manuscript we will thus not need to worry about technical subtleties that cloud the main arguments and which will be studied in more complicated models in \cite{Vanrietvelde:2018dit,Hoehn:2018aqt, Hoehn:2018whn,pqps}.

\subsection{A toy model for Mach's principle in 1D space}\label{sec_lag1d}

For simplicity, we shall take the $N$ particles to be of unit mass\footnote{It is straightforward to generalize the model to arbitrary individual particle masses and we come back to this.} and the configuration space as $\cq=\mathbb{R}^N$ so that the phase space is simply $\mathbb{R}^{2N}$. We use canonical pairs $(q_i,p_i)_{i=1}^N$ as coordinates. It turns out (see Appendix \ref{app_lag1d}) that a Lagrangian with global translation invariance necessarily leads to a (primary) constraint, namely that the center of mass momentum vanishes
\begin{equation}\label{constraint}
P = \sum_{i=1}^N p_i \approx 0\,,
\end{equation}
so that the momenta of the individual particles are not all independent. Note that this equation defines a $(2N-1)$-dimensional {\it constraint surface} in phase space. The symbol $\approx$ denotes a {\it weak equality}, i.e.\ an equality that only holds on this constraint surface. (See \cite{Dirac,Henneaux:1992ig} for an introduction to constrained Hamiltonian systems.)


On the constraint surface defined by (\ref{constraint}), the Hamiltonian (following from the Lagrangian of Appendix \ref{app_lag1d}) will be of the form
\ba
H=\frac{1}{2}\sum_i p_i^2 + V(\{q_i - q_j\}_{i,j}).
\ea
Clearly, the constraint is preserved by the dynamics 
$\{P,H\} = 0$ (where $\{.,.\}$ denotes the Poisson bracket)
and so, in the terminology of Dirac, no secondary constraints arise to enforce the conservation of $P$ and it is automatically first-class. It is therefore a generator of gauge transformations \cite{Dirac,Henneaux:1992ig}. Indeed, in line with the symmetry of the Lagrangian of Appendix \ref{app_lag1d}, it generates global translations, infinitesimally given by
\begin{equation}\label{gaugeTransfo}
\begin{cases}
q_i \to q_i + \{q_i, P\}\, \varepsilon = q_i + \varepsilon \\
p_i \to p_i + \{p_i, P\}\, \varepsilon = p_i
\end{cases}\,.
\end{equation}
The physical interpretation is here (see Appendix \ref{app_lag1d} for a discussion): the localizations $q_i(t)$ and motions $\dot{q}_i(t)$ of the $N$ particles with respect to the Newtonian background space have no physical meaning, but are gauge dependent. Only the relative localization and motion of the particles are physically relevant, thereby providing a toy model for Mach's principle. Thanks to the symmetry, physics is here relational.\footnote{In this simple model, only the spatial physics is relational, while we have kept the absolute Newtonian time as physical. One can also make the temporal physics of such $N$ particle models relational, see e.g.\ \cite{mercati2018shape,Barbour295,Barbour:2014bga,Barbour:2015sba}.}

Given the gauge symmetry, we need to find physical quantities that are gauge invariant and thus do not depend on the localization and motion relative to the Newtonian background space. Technically, these are phase space functions $O$, which Poisson-commute with the gauge generator on the constraint surface $\{O,P\}\approx0$ (i.e., are invariant under the gauge flow generated by $P$) and are called {\it Dirac observables}. In this simple model, there are obvious examples: For instance, all $N$ momenta $p_i$ and all $\binom{N}{2}$ relative distances $q_i-q_j$, $i,j=1,\ldots,N$ are Dirac observables. Clearly, $H$ is also a Dirac observable and the {\it total Hamiltonian} (sum of a gauge invariant Hamiltonian plus a linear combination of gauge generators \cite{Dirac,Henneaux:1992ig}) thereby reads
\begin{equation}\label{H}
H_{\rm tot} = \frac{1}{2}\sum_i p_i^2 + V(\{q_i - q_j\}) +  \lambda P\,,
\end{equation}
where $\lambda$ is a Lagrange multiplier, namely an arbitrary function of time which encodes the gauge freedom (eq.\ (\ref{gauge1d}) in the Lagrangian formulation in Appendix \ref{app_lag1d}) in the canonical equations of motion. Intuitively, it is clear that an arbitrary function of time will have to appear in the evolution generator, for otherwise the evolution of {\it all} variables would be unambiguously determined, given initial data, leaving no room for gauge freedom. Note that $\lambda$ will get fixed below when fixing the gauge. It is evident that the equations of motion of Dirac observables (generated by $H_{\rm tot}$) will not depend on $\lambda$ on the constraint surface; their dynamics thus features no arbitrariness, given suitable initial data.

However, there is redundancy among the Dirac observables mentioned. Thanks to (\ref{constraint}), only $N-1$ of the $p_i$ are independent on the constraint surface. Similarly, only $N-1$ of the relative distances are independent, as $q_i-q_k$ is just the sum of $q_i-q_j$ and $q_j-q_k$. Altogether, we therefore have $2(N-1)$ independent gauge invariant phase space functions. Indeed, given that $P$ generates one-dimensional gauge orbits in its $(2N-1)$-dimensional constraint surface, the reduced (i.e.\ gauge invariant) phase space \cite{Henneaux:1992ig} is $2(N-1)$-dimensional for this model.

We propose to interpret what we have described thus far as a perspective-neutral super theory. Using this structure, we derived the gauge invariant degrees of freedom, but we have not described them from the perspective of, e.g., any of the $N$ particles, each of which could serve as a physical reference system. 
That is to say, we have not chosen any reference frame from which to describe the physics. The perspective-neutral super structure contains, so to speak, all perspectives at once and thereby does not by itself admit an immediate operational interpretation. Instead, we shall now argue that choosing the internal perspective of a reference system on the physics is tantamount to choosing a particular gauge fixing. In particular, the perspective-neutral structure tells us there is a redundancy among the basic Dirac observables, but it does not by itself choose which of the Dirac observables to consider as the redundant ones. Gauge fixing will take care of this.



%


\subsection{Choosing an internal perspective $=$ choosing a gauge}\label{choosingRF}

We shall now reduce the phase space, getting rid of gauge freedom altogether and working only with the physical quantities written in a particular gauge. 
Suppose we want to describe the physics from the internal perspective of particle $A$. We are free to define $A$ as the origin from which to measure distances in coordinates, imposing (emphasized through the symbol !)
\begin{equation}\label{constraintChi}
\chi = q_A \overset{!}{=} 0\,,
\end{equation}
which is a global gauge fixing, as $\{\chi,P\} = 1$\footnote{This implies that $\chi=0$ intersects every $P$-generated gauge orbit once and only once.} and technically this implies that the constraints become second class.

%
%
%
%
%


This gauge choice indeed corresponds to `taking the point of view of $A$', since now all relative distances between $A$ and the other $N-1$ particles (these are a complete set of independent configuration Dirac observables) simply become
\ba
q_i-q_A\q\mapsto \q q_i\,,\q\q\q\q i\neq A\,.
\ea
Accordingly, we can consistently interpret the $q_i$ as position measurements of the remaining particles relative to particle $A$. (The relative distances among the remaining particles are clearly redundant information.)

It is clear that we should also solve the redundancy among the basic momenta for $p_A$,
\ba
p_A\approx-\sum_{i\neq A}\,p_i\,,
\ea
so that all the $N-1$ $p_{i\neq A}$ become the independent momentum variables. Note that $p_A$ is not proportional to $\dot{q}_A$ alone {(see (\ref{leg}) in Appendix \ref{app_lag1d})}  so that the fact that generally now $p_A\neq0$ does not mean the motion of $A$ is not fixed.

In fact, we have to ensure that defining $A$ as the origin is consistent at all times. This fixes the Lagrange multiplier $\lambda$. Indeed, the equations of motion are
\begin{subequations}\label{EOM}
\begin{equation}\label{EOMq}
\dot{q}_i = \frac{\partial H_{\rm tot}}{\partial p_i} \approx p_i + \lambda\,,
\end{equation}
\begin{equation}\label{EOMp}
\dot{p}_i = - \frac{\partial H_{\rm tot}}{\partial q_i} \approx - \frac{\partial V}{\partial q_i}\,,
\end{equation}
\end{subequations}
so that the conservation of (\ref{constraintChi}), namely $\dot{q}_A\overset{!}{=}0$, imposes
\begin{equation}\label{LagrangeMultiplier}
\lambda = - p_A\,,
\end{equation}
thereby fixing any arbitrariness in the equations of motion.
Inserting (\ref{LagrangeMultiplier}) in (\ref{EOM}) gives us the 
dynamics of all particles in the chosen gauge, i.e.\ 
`as seen by $A$'.

%

%


We noted above that the reduced phase space is $2(N-1)$-dimensional and it is clear that it is coordinatized by the $(q_i,p_i)$ where $i\neq A$. 
However, being a new gauge-fixed phase space, we have to specify the bracket structure on it that is inherited from the original phase space $\mathbb{R}^{2N}$. 
For constrained systems, this amounts to replacing the Poisson bracket with the Dirac bracket \cite{Dirac,Henneaux:1992ig}. In the present model it simply reads:
\begin{equation}\label{}
 \{F,G\}_D = \{F,G\} - \{F,P\}  \{\chi,G\} + \{F,\chi\}  \{P,G\}\,,
\end{equation}
for any phase space functions $F,G$, where $\{.,.\}$ denotes the usual Poisson bracket. The Dirac brackets of our basic phase-space variables are then:
\begin{equation}\label{dirac1d}
\{q_A,p_A\}_D = 0\,,\q\,\,
 \{q_i,p_j\}_D = \delta_{ij}\,,\q\,\, \forall\, i,j \neq A\,.\end{equation}

Hence, this reduction simply discards particle $A$'s position and momentum from among the physical degrees of freedom and we pick the remaining ones as coordinates of the reduced phase space. We 
thus end up with a theory for $N-1$ particles -- as seen by $A$. The corresponding reduced Hamiltonian can be computed from (\ref{H}) and (\ref{constraint}, \ref{constraintChi}):

\begin{equation}\label{HReduced}
H_A^{\rm red} =  \sum_{i\neq A} p_i^2 + \sum_{\substack{i \neq j \\i,j\neq A}} p_i p_j + V(\{q_i\}_{i\neq A})\,.
\end{equation}

This Hamiltonian is of a somewhat non-standard form: the usual 1/2 factor in the kinetic energy is not present and there are couplings between the $p_i$'s. However, it encodes the relational physics correctly. 
Indeed, restricting ourselves to the $N=3$ case for clarity, the equations of motion give the accelerations (writing $\p_i:=\p/\p q_i$):

\begin{subequations}\label{acc3}
\begin{equation}\label{accB}
\ddot{q}_B = - 2 \partial_B V - \partial_C V\,,
\end{equation}
\begin{equation}\label{}
\ddot{q}_C = - 2 \partial_C V - \partial_B V\,.
\end{equation}
\end{subequations}
Recall that the variables $q_B$ and $q_C$ encode the \textit{relative} positions of $B$ and $C$ with respect to $A$ in the reduced phase space. Thus, if we take for example (\ref{accB}), the factor 2 in $\partial_B V $ stems from the fact that the effect of, for instance, an attractive force between $A$ and $B$ has to be counted twice, as it both pulls $B$ towards $A$ and $A$ towards $B$. As for the presence of a $\partial_C V$, it is due to the fact that, even in the absence of an interaction between $A$ and $B$, an interaction between $A$ and $C$ will affect the position of $A$, and thus the position of $B$ relative to it. These considerations generalize to arbitrary $N$.


As an aside, it is interesting to also look at what Hamiltonian (\ref{HReduced}) becomes had we permitted the particles to have differing masses $m_i$ in (\ref{Lag1d}):

\begin{align}\label{HReducedMasses}
H^{\rm red}_A =& \,\,\frac{1}{2} \sum_{i\neq A} \left(\frac{1}{m_i} + \frac{1}{m_A}\right) p_i^2 \nn\\
&\q\q\q+ \sum_{\substack{i \neq j\\i,j\neq A}} \frac{p_i p_j}{m_A} + V(\{q_i\}_{i\neq A})
\end{align}
In the limit $m_A \to \infty$, (\ref{HReducedMasses}) becomes the usual Hamiltonian, in agreement with the fact that a reference system with infinite mass can be used as an inertial frame. This limit also recovers the standard situation in quantum mechanics, where the description is given with respect to a classical reference frame.

In summary, we interpret the reduced phase in a particular gauge as the physics described relative to a reference frame, which corresponds to that gauge.

\subsection{Switching internal perspectives}\label{sec_clswitch}

It is clear that going from one reference frame to another amounts to a finite gauge transformation on the constraint surface (and a corresponding swap of which Dirac observables to treat as redundant, achieved simply through an exchange of $A,C$ labels). That is, in order to switch perspective, we have to go back to the perspective-neutral structure of the original phase space, into which the reduced phase space embeds. We shall only be schematic here as the situation is geometrically evident {(see fig.\ \ref{fig_geometry})}; the details of the following discussion can be found in Appendix \ref{app_clswitch}.

{\it The quick reader may skip the following paragraphs and proceed directly to sec.\ \ref{quantum}}.
\begin{figure*}
	\begin{center}
		\includegraphics[scale=0.64]{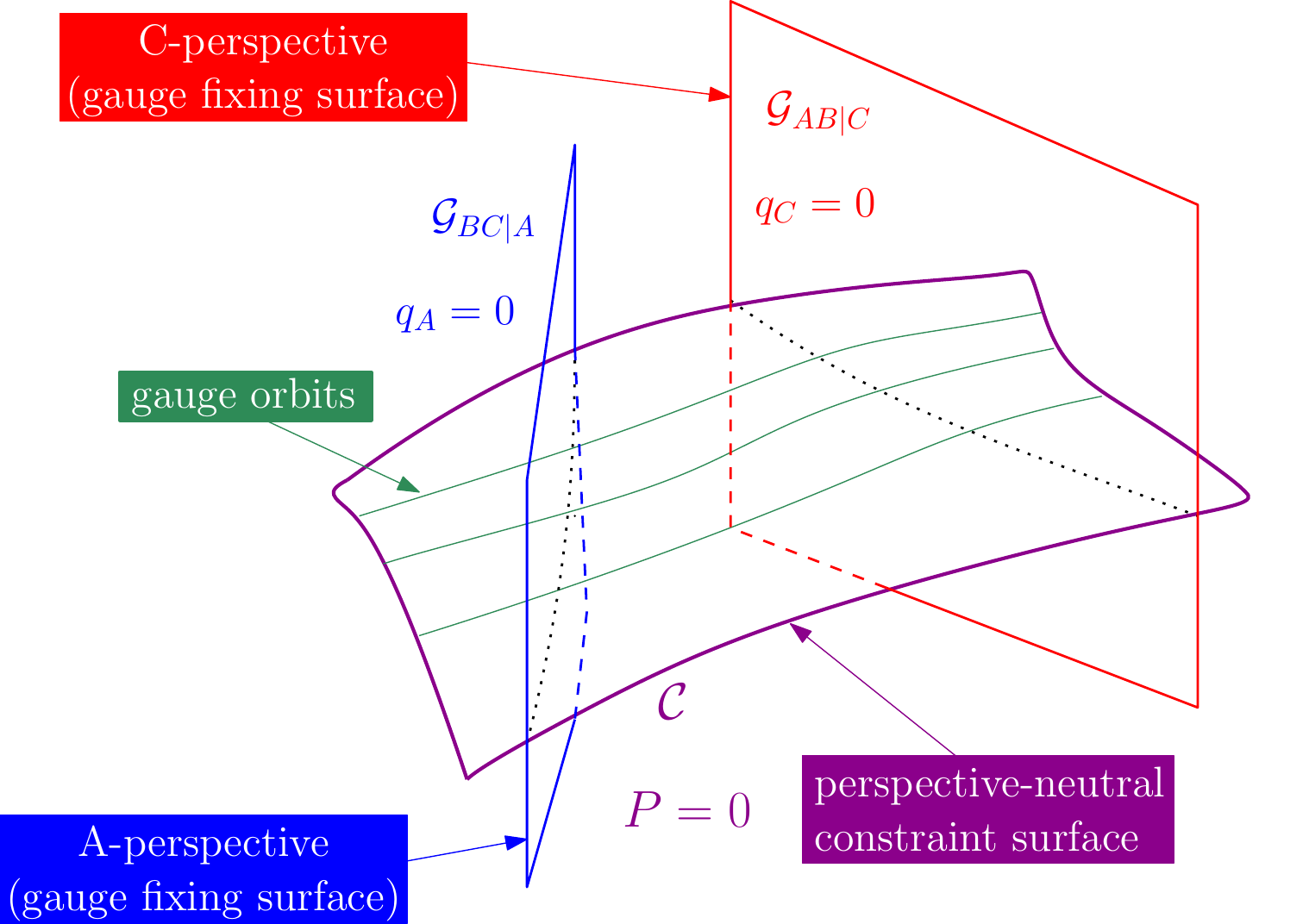}
		\caption{\label{fig_geometry}\small{ Phase space geometry of classical frame perspective switches.}}
	\end{center}
\end{figure*}

Denote the reduced phase space in $A$ perspective by $\cp_{BC|A}$. As discussed above its canonical coordinates are $(q_i,p_i)_{i\neq A}$. Next, denote the constraint surface in the original phase space $\mathbb{R}^{2N}$, defined through $P$, by $\cc$. $\cp_{BC|A}$ canonically embeds into $\cc$ as the intersection $\cc\cap\cg_{BC|A}$ where $\cg_{BC|A}$ is the gauge fixing surface defined by the gauge $\chi=0$. This defines an embedding map
\ba
\iota_{BC|A}:\cp_{BC|A}\hookrightarrow\cc\,.
\ea
whose image is $\cc\cap\cg_{BC|A}$.  {It is important here that $\cp_{BC|A}$ is equipped with the interpretation as the physics seen by $A$ to avoid ambiguities in the embedding map. Indeed, abstractly, the reduced phase space is the space of gauge orbits (i.e., every gauge orbit corresponds to one physical state) and thus simply the quotient $\cp_{\rm red}=\cc/\sim$, where $\sim$ is the equivalence relation that identifies points in the same orbit. This abstract $\cp_{\rm red}$ can be interpreted as the {\it perspective-neutral phase space}: it is gauge invariant and coordinatized by Dirac observables (which really are functions on the set of orbits). It is also isomorphic to every (globally) gauge fixed reduced phase space. Without further information, the embedding of the abstract reduced phase space $\cp_{\rm red}$ into $\cc$ would be highly ambiguous. However, here it is the physical interpretation of the gauge fixed $\cp_{BC|A}$ that singles out its embedding.}
Note that $\cc\cap\cg_{BC|A}$ indeed defines a $2(N-1)$-dimensional hypersurface in the original phase space $\mathbb{R}^{2N}$. 

Similarly, one can define a `projection'
\ba
\pi_{BC|A}:\cc\cap\cg_{BC|A}\rightarrow\cp_{BC|A}\,,
\ea 
so that $\pi_{BC|A}\circ\iota_{BC|A}=\text{Id}_{\cp_{BC|A}}$. This projection essentially drops all redundant embedding information. The same construction can, of course, also be carried out for the reduced phase space $\cp_{AB|C}$ in $C$ perspective, by simply exchanging $A$ and $C$ labels. In particular, $\cg_{AB|C}$ is now defined by $q_C=0$.

In order to switch from $A$ to $C$ perspective, we now need the gauge transformation $\alpha_{A\to C}$, generated by the constraint $P$, that takes us from one embedding $\cc\cap\cg_{BC|A}$ to the other $\cc\cap\cg_{AB|C}$. In Appendix \ref{app_clswitch}, we show that this defines a map $
\cs_{A\to C}:\cp_{BC|A}\rightarrow\cp_{AB|C}
$
that produces the expected result 
\begin{widetext}
\ba
(q_B,p_B,q_C,p_C)\mapsto\left(q'_A=-q_C,p'_A=-p_B-p_C,q'_B=q_B-q_C,p'_B=p_B\right)\,\label{AtoC}
\ea
and satisfies the following commutative diagram, {where $\zeta_C$ denotes the invertible map that associates to each orbit in $\cc$ its intersection point with $\cg_{AB|C}$ (and similarly for $\zeta_A$):}
\begin{center}
\begin{tikzcd}[row sep=huge, column sep = huge]
& \cp_{\rm red}=\mathcal{C}/\sim \arrow[rd, "\zeta_C"]& \\
\mathcal{C} \cap \mathcal{G}_{BC|A} \arrow[ru, "\zeta^{-1}_A"] \arrow[rr, "\alpha_{A\to C}"]&& \mathcal{C} \cap \mathcal{G}_{AB|C}  \arrow[d, "\pi_{AB|C}"]\\
\mathcal{P}_{BC|A} \arrow[u, "\iota_{BC|A}"] \arrow[rr, "\cs_{A \to C}"] && \mathcal{P}_{AB|C}
\end{tikzcd}
\end{center}
\end{widetext}

{Taking $\cs_{\rm phys}:=\cc/\sim$, $\varphi_A:=\pi_{BC|A}\circ\zeta_A$ and $\varphi_C:=\pi_{AB|C}\circ\zeta_C$, this perspective change is indeed of the form (\ref{gentrafo}),
$
\cs_{A\to C}:=\varphi_{C}\circ\varphi_A^{-1},
$
proceeding via the perspective-neutral phase space. Equivalently, we could also take $\cs_{\rm phys}:=\cc$, $\tilde\varphi_A:=\pi_{BC|A}$ and $\tilde\varphi_C:=\pi_{AB|C}$ to write
$
\cs_{A\to C}:=\tilde\varphi_{C}\circ\alpha_{A\to C}\circ\tilde\varphi_A^{-1},
$
exploiting the shortcut via the intermediate gauge transformation $\alpha_{A\to C}$. (Such a shortcut will be absent in the quantum theory.) This yields a similar compositional structure as in (\ref{gentrafo}) via the perspective-neutral constraint surface. 
}

%
%
%

\subsection{Remark on the preferred role of the position basis}

Note that in our present construction of frame transformations the position basis plays a special role, in contrast to \cite{Giacomini:2017zju}. This is a consequence of our symmetry principle which is formulated at the level of the Lagrangian (in Appendix \ref{app_lag1d}) in position and velocity space. Upon transition to phase space, it is clear that such symmetries are generated by (primary) constraints that necessarily involve momenta. As such, the gauge fixing {\it must} include position information and relative positions are here indispensable relational observables, as opposed to, for example, relative momenta. This is also reflected in the interpretation of the frames and their relations. By contrast, if, as in \cite{Giacomini:2017zju}, one also wanted to switch the roles of the configuration and momentum basis, one ideally would like to have a constraint $Q=\sum_i\,q_i$ as a symmetry generator in the one-dimensional $N$-body problem. However, this is a so-called {\it holonomic} constraint (involving only configuration data) and such constraints can only arise through equations of motion as secondary ones, and are usually second class (thus not symmetry generators). Hence, one would have to proceed differently than in our construction here.

\section{Quantum reference frames in 1D space}   \label{quantum}

Our task is now to translate the perspective-neutral super structure and the inside perspectives into the quantum theory. This will permit us to switch between different quantum reference frame perspectives {and suggests a new interpretation of two quantization methods for constrained systems.}

The two most commonly used strategies for canonically quantizing constrained systems are:
\begin{description}
\item[Reduced quantization:] Solve the constraints (and possibly gauge fix) first at the classical level, then quantize the reduced theory. 
\item[Dirac quantization:] Quantize the system first (incl.\ unphysical degrees of freedom), then solve the constraints in the quantum theory.
\end{description}
There is a general debate in the literature about the relation between these two methods and, in particular, about when one or the other is the correct method to be employed.

{In the context of the Guillemin-Sternberg conjecture, the two methods have been shown to be equivalent for the case of compact phase spaces and compact symmetry groups acting on them \cite{guillemin1982geometric, tian1998analytic} (see also \cite{hochs2008guillemin} for an attempt at a generalization). It follows from  \cite{gotay1986constraints,kucha1986covariant,Kunstatter:1991ds} that for the more interesting case of non-compact phase spaces, such as the cotangent bundle $\mathbb{R}^{2N}$ used in the present model, an equivalence of the two methods can also be sometimes established. This happens provided certain factor ordering choices are made, the gauge transformations take the form of point transformations, as in (\ref{gaugeTransfo}), and there are no global obstructions to either separating gauge from gauge-invariant degrees of freedom or fixing a gauge. We note that constraints which are \emph{linear} in the momenta, as in our model, always generate point transformations. However, the two methods are known to yield unitarily inequivalent results in more general settings \cite{Ashtekar:1982wv,kucha1986covariant,Ashtekar:1991hf,Schleich:1990gd,Kunstatter:1991ds,Hajicek:1990eu,Romano:1989zb,Dittrich:2016hvj,Dittrich:2015vfa,Loll:1990rx,plyushchay1996dirac}.}

Usually the Dirac method is invoked because classically solving constraints and fixing a gauge can become arbitrarily complicated. For instance, in general relativity {constructing the reduced phase space is tantamount to} also solving the dynamics of the theory, which {for the full theory} seems a hopeless endeavor. Furthermore, globally valid gauge choices are generically absent, e.g., not only in Yang-Mills theories or general relativity (the Gribov problem), but also in much simpler systems \cite{Hajicek:1986ky,Bojowald:2010xp,Bojowald:2010qw,Hohn:2011us,Dittrich:2016hvj,Dittrich:2015vfa}. There even exist extreme cases where a reduced quantization is outright impossible (without changing other structure in the model), while the Dirac method can be applied \cite{Dittrich:2016hvj,Dittrich:2015vfa}. 

Nevertheless, {in} our simple model 
{both methods are necessary for a complete relational interpretation. In particular, we will establish a new systematic quantum symmetry reduction procedure of the Dirac quantized theory which is the quantum analog of the classical phase space reduction through gauge-fixing discussed in Sec.~\ref{choosingRF}. In particular, this quantum symmetry reduction procedure removes the redundancy in the description and is always relative to a choice of quantum reference frame: it picks out the reference frame degrees of freedom as the redundant ones because we do not want to describe the reference frame relative to itself. }

{In our simple model, this procedure can, in fact, be interpreted as the `quantization' of the phase space symmetry reduction  in the sense that symmetry reduction and quantization `commute'. That is, first quantizing according to the Dirac method, then performing the new quantum symmetry reduction procedure yields a quantum theory, which coincides with the quantization of the classical symmetry reduced phase space. In our model, this quantum symmetry reduction procedure will thus map the Dirac quantized theory bijectively into the reduced quantized theory. This permits us to interpret the reduced quantum theory (in a specific gauge) as the description of the quantum physics relative to a (quantum) reference frame, while the Dirac quantum theory will assume the role of the perspective-neutral quantum theory via which internal perspectives will be changed. The new quantum symmetry reduction maps will thereby be the key to the quantum reference frame switches. 
Given that the Dirac quantized theory constitutes the perspective-neutral super structure, it does not by itself admit an immediate operational interpretation (recall the discussion in sec.\ \ref{sec_meta}). }

Of course, our model {is essentially as simple a constrained system} as it gets: it features a single linear constraint, global gauge fixing conditions and accordingly, we have globally valid inside perspectives in both the classical and quantum theory. {In generic models, not only do global challenges (such as the Gribov problem) arise, which may inhibit the existence of globally valid inside perspectives \cite{Vanrietvelde:2018dit}, nor will it always be true that symmetry reduction and quantization  commute. The quantum theory obtained by applying the new quantum symmetry reduction procedure to the Dirac quantized theory will not always coincide with the quantization of a classical reduced phase space, due to the well-known inequivalence of Dirac and reduced quantization in more general situations \cite{Ashtekar:1982wv,kucha1986covariant,Ashtekar:1991hf,Schleich:1990gd,Kunstatter:1991ds,Hajicek:1990eu,Romano:1989zb,Dittrich:2016hvj,Dittrich:2015vfa,Loll:1990rx,plyushchay1996dirac}. However, this will not be a problem for our interpretation and we shall comment on this further in the conclusions, Sec.~\ref{sec_conc}.}

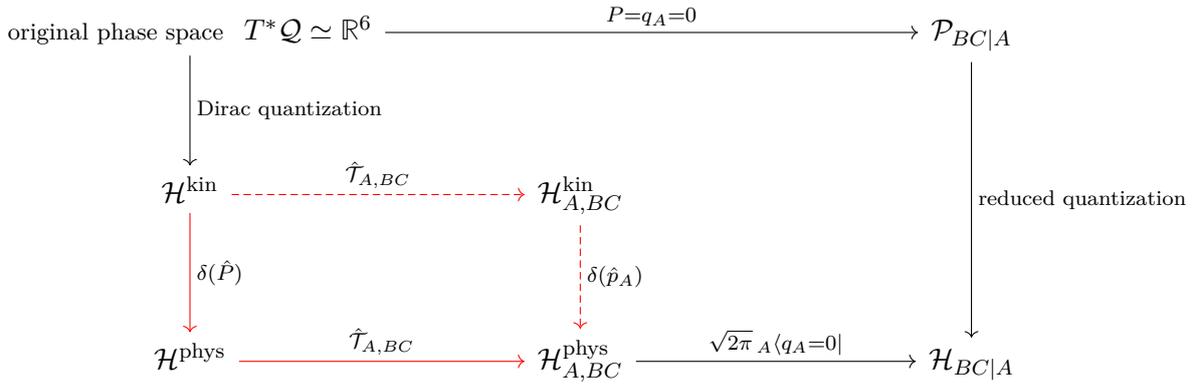
\begin{figure*}
\centering
    \begin{tikzcd}[row sep=huge, column sep = huge]
  \text{\footnotesize original phase space \hspace*{.05cm} }  T^*\cq\simeq\mathbb{R}^6 \arrow[rrr, "P=q_A = 0"] \arrow[d, "\textrm{Dirac quantization}"] & & & \cp_{BC|A} \arrow[dd, "\textrm{reduced quantization}"]\\
\mathcal{H}^{\rm kin} \arrow[red]{d}[black]{\delta(\hat{P})} \arrow[red,dashrightarrow]{r}[black]{\Hat{\mathcal{T}}_{A,BC}}
& \mathcal{H}^{\rm kin}_{A,BC} \arrow[red, dashrightarrow]{d}[black]{\delta(\hat{p}_A)}
\\
\mathcal{H}^{\rm phys} \arrow[red]{r}[black]{\Hat{\mathcal{T}}_{A,BC}}
& \mathcal{H}^{\rm phys}_{A,BC} \arrow[rr, "\sqrt{2\pi}\,{}_A\bra{q_A=0}"]
& 
&\mathcal{H}_{BC|A} 
\end{tikzcd}
\caption{\small Diagram of the two quantization methods and their relation for three particles. 
The horizontal arrows between Hilbert spaces are all isometries. The red diagram is commutative.\protect\footnotemark\,  
 The quantum symmetry reduction procedure from the perspective-neutral physical Hilbert space $\ch^{\rm phys}$ of the Dirac quantization to the reduced Hilbert space, say, in A-perspective $\mathcal{H}_{BC|A} $ involves two steps: 1.\ a constraint trivialization $\ct_{A,BC}$ which transforms the constraint in such a way that it {\it only} acts on the reference frame variables; 2.\ the reference frame variables, having become redundant, are discarded by projecting onto the classical gauge fixing conditions.} \label{commutativeDiagram}
\end{figure*}

Since we will be dealing with a number of distinct Hilbert spaces through the two methods, we have organized the various steps of the construction and their relation in fig.\ \ref{commutativeDiagram} for visualization. 

\footnotetext{{The entire diagram may also be considered commutative in the sense that one obtains the same quantum theory in the bottom right corner, although clearly individual elements of $T^*\cq$ cannot be associated with individual elements of $\ch_{BC|A}$. By contrast, in the red  diagram every element of $\ch^{\rm kin}$ will be mapped to a unique element of $\ch_{A,BC}^{\rm phys}$.}}

\subsection{Reduced quantization -- quantum theory from an internal perspective}\label{reducedQuantization}

We begin with the reduced quantization of the model from the previous section, as we aim to recover it subsequently from the Dirac quantization. That is, we return to the gauge corresponding to, say, $A$'s internal perspective and simply quantize the reduced phase space of sec.\ \ref{choosingRF}. This is standard and amounts to promoting the $q_i,p_i$, $i\neq A$, to operators and the Dirac brackets (\ref{dirac1d}) to commutators\footnote{Henceforth, we shall work in units where $\hbar=1$.} 
\begin{equation}\label{}
[\hat{q}_i,\hat{p}_j] = i \,\delta_{ij},\q\,\, [\hat{q}_i,\hat{q}_j]=[\hat{p}_i,\hat{p}_j]=0\,,\q\, i,j\neq\,A,
\end{equation}
on an $L^2(\mathbb{R}^{N-1})$ Hilbert space. 

In order to simplify the equations, we restrict to $N=3$ in the sequel. 
For instance, the quantized Hamiltonian (\ref{HReduced}) for 3 particles reads:
\begin{equation}\label{HTQReduced3}
\hat{H}_{BC|A}:=\hat{H}^{\rm red}_A = \hat{p}_B^2 + \hat{p}_C^2 + \hat{p}_B \hat{p}_C + V(\hat{q}_B,\hat{q}_C)\,,
\end{equation}
and arbitrary reduced states can be represented as follows:
\ba
|\psi\rangle_{BC|A}=\int\,dp_B\,dp_C\,\psi_{BC|A}(p_B,p_C)\,|p_B,p_C\rangle\,.\nn\\\label{redAstate}
\ea
The corresponding reduced Hilbert space will be denoted by $\ch_{BC|A}$, see fig.\ \ref{commutativeDiagram}. It is clear that generalizing to arbitrary $N$ (or differing masses) poses no efforts. 


In line with the classical case, we interpret this reduced quantum theory as the description of the quantum dynamics of the remaining particles
as seen from the quantum reference frame of particle $A$. Yet, 
as we moved to the reference frame of $A$ (that is, fixed the gauge accordingly) \textit{before} quantizing, we have washed out the perspective-neutral information and the reduced quantum description alone lacks structure to tell us how to switch from this frame to the perspective of another one (for instance, $B$), while staying in the quantum theory. To switch perspectives within quantum theory, we need to relate the reduced descriptions to the Dirac method, which quantizes the classical perspective-neutral structure.



\subsection{Dirac quantization -- the perspective-neutral quantum theory}

We now quantize first, then solve the constraints. This requires two distinct Hilbert spaces, see fig.\ \ref{commutativeDiagram}. 

First we promote {\it all} canonical pairs $(q_i,p_i)_{i=1}^N$ (i.e., incl.\ physically redundant and gauge degrees of freedom), coordinatizing the original phase space $T^*\cq\simeq\mathbb{R}^{2N}$ of sec.\ \ref{sec_lag1d}, to operators and the Poisson brackets to commutators on a {\it kinematical} (or auxiliary) Hilbert space  
$\mathcal{H}^{\rm kin}=L^2(\mathbb{R}^N)$. Next, we employ this Hilbert space to quantize the total momentum constraint (\ref{constraint}) and solve the latter in the quantum theory by requiring that {\it physical} states $\ket{\psi}^{\rm phys}$ of our system are annihilated by it. Returning in the sequel to the $N=3$ case for simplicity, we thus impose
\begin{equation}\label{hatp}
\hat{P} \ket{\psi}^{\rm phys} = (\hat{p}_A + \hat{p}_B + \hat{p}_C) \ket{\psi}^{\rm phys} \overset{!}{=} 0\,.
\end{equation}
Physical states are zero-eigenstates of the constraint.

One subtlety arises: $\hat{P}$ has a continuous spectrum around zero and so physical states are {\it not} normalized with respect to the standard inner product on $\mathcal{H}^{\rm kin}$; they are thus not actually contained in the kinematical Hilbert space. Instead, we have to construct a new inner product for physical states, to turn the space of solutions to (\ref{hatp}) into a proper physical Hilbert space $\ch^{\rm phys}$. (But see also \cite{Kempf:2000qz} for an alternative method using a modification of the Hilbert space topology.)

To this end, we employ {\it group averaging} (or refined algebraic quantization) \cite{Marolf:1995cn,Marolf:2000iq,Thiemann:2007zz} and define an (improper) projector onto solutions of (\ref{hatp}):
\begin{equation}\label{projp}
\begin{split}
\delta(\hat{P}):  & \,\,\mathcal{H}^{\rm kin}  \to  \mathcal{H}^{\rm phys} \\
&\!\!\!\!\ket{\phi}^{\rm kin}  \mapsto  \ket{\phi}^{\rm phys}:=\Big(\f{1}{2\pi} \int_{-\infty}^{+\infty} \mathrm{d}s \ e^{i s \hat{P}}  \Big) \ket{\phi}^{\rm kin}. 
\end{split}
\end{equation}
Projecting an arbitrary state of $\ch^{\rm kin}$ in momentum representation,
\begin{widetext}
\ba
\ket{\phi}^{\rm kin}=\int\,\mathrm{d}p_A\ \mathrm{d}p_B\ \mathrm{d}p_C\,\phi^{\rm kin}(p_A,p_B,p_C)\,\ket{p_A}\ket{p_B}\ket{p_C}\,,\nn
\ea
a general solution becomes, depending on which particle's momentum is solved for,
\ba
\ket{\phi}^{\rm phys}&=& \int \mathrm{d} p_B \ \mathrm{d}p_C \ \phi_{BC|A}(p_B,p_C) \ket{-p_B-p_C}_A \ket{p_B}_B \ket{p_C}_C\nn\\
&=& \int \mathrm{d} p_A \ \mathrm{d}p_C \ \phi_{AC|B}(p_A,p_C) \ket{p_A}_A \ket{-p_A-p_C}_B \ket{p_C}_C\label{physstate1d}\\
&=& \int \mathrm{d} p_A \ \mathrm{d}p_B \ \phi_{AB|C}(p_A,p_B) \ket{p_A}_A \ket{p_B}_B \ket{-p_A-p_B}_C\,.\nn
\ea
\end{widetext}
where for later use we have defined
\ba
\phi_{BC|A}(p_B,p_C)&:=&\phi^{\rm kin}(-p_B-p_C,p_B,p_C)\,,\nn\\
\phi_{AC|B}(p_A,p_C)&:=&\phi^{\rm kin}(p_A,-p_A-p_C,p_C)\,,\q\q\label{redphysstate}\\
\phi_{AB|C}(p_A,p_B)&:=&\phi^{\rm kin}(p_A,p_B,-p_A-p_B)\,.\nn
\ea
All three lines in (\ref{physstate1d}) give different descriptions of the same physical state $\ket{\phi}^{\rm phys}$ and we shall exploit this below.
Note that $\delta(\hat{P})$ is an improper projector since $\delta(\hat{P})^2$ is clearly singular. 

It turns out (see Appendix \ref{app_PIP}) that the sought-after inner product between physical states is
\begin{equation}
(\psi^{\rm phys},\phi^{\rm phys})_{\rm phys} := {}^{\rm kin}\bra{\psi} \delta(\hat{P})\ket{\phi}^{\rm kin}\,,
\end{equation}
where $\langle\cdot|\cdot\rangle$ is the original inner product of $\ch^{\rm kin}$. Through Cauchy completion (and other technical subtleties which we shall here ignore), the space of solutions to (\ref{hatp}) can thereby be turned into a proper Hilbert space $\ch^{\rm phys}$.

Clearly, in analogy to the classical case, observables $\hat{O}$ on $\ch^{\rm phys}$ must satisfy $[\hat{O},\hat{P}]=0$, for otherwise they would map out of the space of solutions. Any such $\hat{O}$ is thus gauge invariant and a quantum Dirac observable. For instance, in this simple model, the quantization of the elementary classical Dirac observables, relative distances and momenta, are obviously quantum Dirac observables,
\ba
\hat{q}_B-\hat{q}_A\,,\,\,\hat{q}_C-\hat{q}_A\,,\,\,\hat{q}_B-\hat{q}_C\,,\,\, \hat{p}_A\,,\,\,\hat{p}_B\,,\,\,\hat{p}_C\,,\nn\\\label{qDirac1d}
\ea
as is the total Hamiltonian, which on $\ch^{\rm phys}$ reads
\ba\label{HDirac}
\hat{H}_{\rm tot} &=& \frac{1}{2} (\hat{p}_A^2 + \hat{p}_B^2 + \hat{p}_C^2) \\
&&\q\q+ V(\hat{q}_B - \hat{q}_A,\hat{q}_C - \hat{q}_A,\hat{q}_B-\hat{q}_C)\,.\nn
\ea

Just as in the classical case, the observables (\ref{qDirac1d}) are redundant and only define four independent Dirac observables on $\ch^{\rm phys}$. {Related to this, (\ref{physstate1d}) shows that we also have a redundancy in the description of a fixed physical state.} Dirac quantization by itself does not tell us which of the Dirac observables to treat as the redundant ones. We thus interpret the gauge invariant physics in $\ch^{\rm phys}$ as the perspective-neutral quantum theory. Here, we have not chosen a quantum reference frame from which to describe the non-redundant physics and precisely the redundancy (originating in gauge symmetry) permits us to choose from among a multitude of perspectives. 

%
%
%


\subsection{From Dirac to reduced quantum theory: recovering relative states}\label{relativeState}

Classically, solving constraints means restricting to the constraint surface in phase space and this by itself does not lead to gauge invariance because first class constraints still generate gauge flows on the constraint surface. We have exploited this in our classical construction: choosing an internal perspective corresponded to imposing an additional gauge fixing condition to break the flow of the constraint (see sec.\ \ref{choosingRF}). 

In Dirac quantization, the situation is different: solving the constraint in the quantum theory already implies gauge invariance. Indeed, $\ch^{\rm phys}$ (`the quantum constraint surface') is invariant under the flow of the constraint since, owing to (\ref{hatp}), $\exp(i\,s\,\hat{P})\,\ket{\phi}^{\rm phys} = \ket{\phi}^{\rm phys}$. Intuitively, this difference to the classical case can be understood through the Heisenberg uncertainty relations: gauge dependent quantities do {\it not} commute with the constraint. For example, the center of mass position $\hat{q}_{\rm cm}=1/3\,(\hat{q}_A+\hat{q}_B+\hat{q}_C)$ is conjugate to the constraint, $[\hat{q}_{\rm cm},\hat{P}]=i$. Hence, physical states as zero-eigenstates of $\hat{P}$ must be maximally spread out over $q_{\rm cm}$. But this is gauge invariance: to smear/average over the gauge orbit. Indeed, this is precisely what the improper projector (\ref{projp}) does.

We thus have to proceed differently in the quantum theory, in order to map from the perspective-neutral structure to the perspective of a specific reference frame, i.e.\ to map from the Dirac to a reduced quantum theory. In particular, we can {\it not} fix a gauge.\footnote{{While gauge-fixing is thus not feasible in the operatorial Dirac quantization after imposing the constraints,  in the path integral formulation the gauge-fixing happens inside the path integral, for example using the Faddeev-Popov trick for canonical gauges \cite{Faddeev:1967fc}, or the Batalin-Fradkin-Vilkovisky framework for arbitrary relativistic gauge systems \cite{Fradkin:1975cq,Batalin:1977pb,Fradkin:1977xi}. In light of our work, some of the different gauge choices inside the path integral may thus encode different quantum reference frame perspectives. In particular, expressing the gauge-invariant path integral in a suitably gauge-fixed fashion would then correspond to describing the dynamics relative to a choice of quantum reference frame.}}  Instead, {quantum symmetry reduction proceeds as follows:
\begin{enumerate}
\item Pick a reference system.
\item Transform the quantum constraint(s) in such a way that the result only acts on the reference system variables, which thereby become fixed. This step is called {\it constraint trivialization}.
\item Discard the now redundant reference system degrees of freedom through a projection onto the classical gauge fixing conditions.
\end{enumerate}}

{This quantum symmetry reduction procedure is the quantum analog of phase space reduction through gauge fixing. In particular, the entire procedure will define a map from the gauge-invariant states and observables on the physical Hilbert space $\ch^{\rm phys}$ to the corresponding states and observables on the appropriate reduced Hilbert space. This map can be interpreted as the `quantum coordinate map' taking us from the perspective-neutral description to the perspective of a reference frame. It maps from the gauge-invariant description on $\ch^{\rm phys}$ to what can be understood as the quantum analog of a `gauge-fixed' description of this gauge-invariant formulation. In particular, this map will preserve inner products, algebraic properties of observables and their expectation values, despite transforming observables and states. Accordingly, when reducing later relative to different reference frame choices, we will still always describe the \emph{same} physical situation, however, from different frame perspectives.}

{We illustrate the procedure by moving} to the perspective of particle $A$ and recovering the corresponding reduced quantum theory of sec.\ \ref{reducedQuantization} (see fig.\ \ref{commutativeDiagram} for illustration of the following steps). Hence, the degrees of freedom corresponding to $A$ are the redundant ones and we need to remove them. To this end, use (\ref{redphysstate}) and write an arbitrary physical state (\ref{physstate1d}) as
\begin{widetext}
\begin{equation}\label{Psi}
\ket{\psi}^{\rm phys} = \int \mathrm{d} p_B \ \mathrm{d}p_C \ \psi_{BC|A}(p_B,p_C) \ket{-p_B-p_C}_A \ket{p_B}_B \ket{p_C}_C\,.
\end{equation}
\end{widetext}
Next, on $\ch^{\rm kin}$ we define the unitary transformation:
\begin{equation}\label{TA}
\hat{\mathcal{T}}_{A,BC} = \exp{ \Big( i\, \Hat{q}_A (\Hat{p}_B + \Hat{p}_C) \Big) }.
\end{equation}
Understanding physical states as distributions on $\ch^{\rm kin}$, we can apply this transformation also to physical states. However, given that $\Hat{\mathcal{T}}_{A,BC}$ does not commute with $\hat{P}$, this transformation will actually map out of $\ch^{\rm phys}$. Yet, it will define an isometry to a transformed set of distributions on $\ch^{\rm kin}$ without losing physically relevant information. That is, the end product can be considered just a new representation $\ch^{\rm phys}_{A,BC}:=\Hat{\mathcal{T}}_{A,BC}(\ch^{\rm phys})$ of the physical Hilbert space. Indeed, we obtain
\begin{widetext}
\begin{equation}\label{}
\begin{split}
    \ket{\psi}_{A,BC} &:=  \Hat{\mathcal{T}}_{A,BC} \ket{\psi}^{\rm phys} = \ket{p=0}_A \otimes \Big( \int \mathrm{d} p_B \ \mathrm{d}p_C \  \  \psi_{BC|A}(p_B,p_C) \ket{p_B}_B \ket{p_C}_C \Big)\,,
\end{split}
\end{equation}
\end{widetext}
so that we can write:
\begin{equation}\label{PsiA,BC}
    \ket{\psi}_{A,BC}  = \ket{p=0}_A \otimes \ket{\psi}_{BC|A}\,.
\end{equation}
It is important to note that this step does {\it not} correspond to `gauge fixing' to $p_A=0$ (there is no gauge symmetry left and $p_A$ is in any case a Dirac observable). Instead, this is really a rewriting -- a trivialization\footnote{Classically, it is also often useful to implement canonical transformations that trivialize constraints in the sense that they become new momentum variables. If the constraints are first class then the gauge degrees of freedom can be made directly conjugate to them, while the other new canonical pairs would be directly Dirac observables. For examples of this method, see, e.g., \cite{Henneaux:1992ig,Dittrich:2004cb,Dittrich:2005kc,Dittrich:2011ke,Dittrich:2013jaa,Hoehn:2014aoa,Hoehn:2014qxa,Hoehn:2014wwa}. In the present model, this would amount to the linear canonical transformation 
\ba
(q_i,p_i)_{i=A,B,C}\mapsto (q_A,P)\,,\,\,\,(q_B-q_A,p_B)\,,\,\,\,(q_C-q_A,p_C)\,.\nn
\ea
Upon gauge fixing $q_A=0$, implementing the Dirac bracket and dropping the redundant $A$-variables, this is equivalent to what we constructed in sec.\ \ref{choosingRF}. Here, we are implementing the quantum analog of that procedure -- except that it does (and can) not employ gauge fixing.} -- of the constraint {to system $A$}, since
\ba
\hat{P}_{A,BC}:=\Hat{\mathcal{T}}_{A,BC} \ \hat{P}  \ (\Hat{\mathcal{T}}_{A,BC})^\dagger = \hat{p}_A\,,
\ea
where $\dag$ is defined with respect to $\ch^{\rm kin}$, so that 
\ba
\hat{P}_{A,BC}\,\ket{\psi}_{A,BC}=\hat{p}_{A}\,\ket{\psi}_{A,BC}=0
\ea
and $\ket{\psi}_{A,BC}$ is actually a physical state, but in a different representation. It is clear that, in contrast to the classical case, there is no sense in which we can talk about additionally gauge fixing $A$'s position.

Crucially, observe that the information in the $A$-slot of $\ket{\psi}_{A,BC}$ contains no relevant information about the original state (\ref{Psi}). We may thus consider $A$ as redundant and, consequently, interpret the remainder of the state $\ket{\psi}_{BC|A}$ preliminarily as the \textit{quantum state of $B$ and $C$ relative to $A$}, corresponding to the perspective-neutral state $\ket{\psi}^{\rm phys}$. This is subject to further justification, but notice already that $\ket{\psi}_{BC|A}$ is now precisely of the form of the reduced states (\ref{redAstate}). 

Given the redundancy of $\ket{p=0}_A$, 
it is natural to discard it altogether and consider only $\ket{\psi}_{BC|A}$, which 
contains all the physical (that is, relational) information about $\ket{\psi}^{\rm phys}$. We can achieve this -- in some analogy to the Page-Wootters construction \cite{Page:1983uc} -- by projecting the factor of the reference system {onto the classical gauge fixing condition (\ref{constraintChi})}:
\ba\label{relativePsi}
        \ket{\psi}_{BC|A} &=&
       {\sqrt{2\pi}\,{}_A\braket{q=0|\psi}_{A,BC}}\nn\\
       &=&\!\int\,\mathrm{d}p'\,{}_A \braket{p'| \psi}_{A,BC}\\
        &=&\!\! \int \mathrm{d}p_B \, \mathrm{d}p_C\,  \psi_{BC|A}(p_B,p_C) \ket{p_B}_B \ket{p_C}_C.\nn
\ea
As shown in Appendix \ref{app_PIP}, $\hat{\mathcal{T}}_{A,BC}$, followed by the projection (\ref{relativePsi}), defines an isometry from  $\ch^{\rm phys}$ to the reduced Hilbert space $\ch_{BC|A}$ of sec.\ \ref{reducedQuantization}. {That is, the procedure preserves the inner product.}

Before claiming that (\ref{TA}) defines a correct transformation from the perspective-neutral quantum theory to the one described from $A$'s perspective, we have to check that the relevant Dirac observables from (\ref{qDirac1d}) transform correctly to those of the reduced theory. Indeed, we find 
\begin{equation}\label{xConjugationB}
\begin{split}
\Hat{\mathcal{T}}_{A,BC} \ (\hat{q}_B - \hat{q}_A) \ (\Hat{\mathcal{T}}_{A,BC})^\dagger &= \hat{q}_B\,,\\
\Hat{\mathcal{T}}_{A,BC} \ \hat{p}_B \ (\Hat{\mathcal{T}}_{A,BC})^\dagger &= \hat{p}_B\,,\\
\!\!\!\!\!\!\Hat{\mathcal{T}}_{A,BC} \ (\hat{q}_C - \hat{q}_A) \ (\Hat{\mathcal{T}}_{A,BC})^\dagger &= \hat{q}_C\,,\\
\Hat{\mathcal{T}}_{A,BC} \ \hat{p}_C  \ (\Hat{\mathcal{T}}_{A,BC})^\dagger &= \hat{p}_C\,.
\end{split}
\end{equation}
Hence, the operator $\hat{q}_B - \hat{q}_A$ on $\ket{\psi}^{\rm phys}$ corresponds to the operator $\hat{q}_B$ on $\ket{\psi}_{A,BC}$, and therefore also on $\ket{\psi}_{BC|A}$. In other words, the position information stored in the $B$-slot of $\ket{\psi}_{BC|A}$ is indeed the relative position of $B$ with respect to $A$, and the same goes for $C$. 

Let us also check that the total Hamiltonian (\ref{HDirac}) transforms as desired. The Hamiltonian $\hat{H}_{A,BC}$ for $\ket{\psi}_{A,BC}$ becomes (assuming $V$ can be Taylor expanded)
\begin{equation}
\begin{split}
\Hat{H}_{A,BC} &= \Hat{\mathcal{T}}_{A,BC} \ \hat{H}_{\rm tot} \ (\Hat{\mathcal{T}}_{A,BC})^\dagger \\
&= \frac{1}{2} \hat{p}_A^2 + \hat{p}_B^2 + \hat{p}_C^2 \\
&\q+ \hat{p}_B \hat{p}_C - \hat{p}_A \hat{p}_B - \hat{p}_A \hat{p}_C + V(\hat{q}_B,\hat{q}_C)\,.\nn
\end{split}\end{equation}
Yet, $\hat{p}_A$ annihilates $\ket{\psi}_{A,BC}$; it is thus equivalent to eliminate the terms containing it from $\Hat{H}_{A,BC}$, which has then no component acting on the $A$-factor of $\ket{\psi}_{A,BC}$, and which can therefore also be considered as a Hamiltonian for the relative state $\ket{\psi}_{BC|A}$:
\begin{equation}\label{HDiracReduced}
\Hat{H}_{BC|A} =  \hat{p}_B^2 + \hat{p}_C^2 + \hat{p}_B \hat{p}_C + V(\hat{q}_B,\hat{q}_C)\,.
\end{equation}
This is precisely the Hamiltonian (\ref{HTQReduced3}) of the reduced quantum theory in $A$-perspective. Hence, the Schr\"odinger equation on $\ch^{\rm phys}$ implies the Schr\"odinger equation on $\ch_{BC|A}$:
\ba\label{seqn}
i\,\p_t\,|\psi\rangle^{\rm phys} &=& \hat{H}_{\rm tot}\,|\psi\rangle^{\rm phys}\nn\\
\Rightarrow\q\, i\,\p_t\,|\psi\rangle_{BC|A} &=& \hat{H}_{BC|A}\,|\psi\rangle_{BC|A}\,.
\ea

In conjunction, it follows (see Appendix \ref{app_PIP} for more detail) that expectation values of relevant Dirac observables on $\ch^{\rm phys}$ are identical to those of the transformed observables on the reduced Hilbert space $\ch_{BC|A}$ and we do not lose any physically relevant information through our transformation $\Hat{\mathcal{T}}_{A,BC}$, despite mapping out of $\ch^{\rm phys}$. Indeed, our procedure illustrated in fig.\ \ref{commutativeDiagram} exploits that $\Hat{\mathcal{T}}_{A,BC}(\ch^{\rm phys})$ is just a new, but equivalent representation of the physical Hilbert space. We thus conclude that the constraint trivialization map (\ref{TA}), followed by the projection (\ref{relativePsi}), indeed constitutes the desired transformation\footnote{In fact, as discussed in Appendix \ref{app_nonunique}, this constraint trivialization map is {\it mathematically} not unique. However, this non-uniqueness only affects the irrelevant information in the redundant $A$ slot and thus has no physical consequences.} from the perspective-neutral to the quantum theory `seen from $A$'s perspective'.
{Regardless of} our model's simplicity, this sheds new light on both the conceptual and technical relation between the Dirac and reduced quantization methods. {Indeed, in the companion articles \cite{Vanrietvelde:2018dit,Hoehn:2018aqt, Hoehn:2018whn}, we shall corroborate this with more complicated models}.

\subsection{Switching internal perspectives in the quantum theory}\label{sec_qswitch1d}

In the previous section, we could equally well have chosen $C$ as the reference system, starting with the respective expressions in the last lines of each of (\ref{physstate1d}, \ref{redphysstate}, \ref{PIP2}) and repeating the same steps by switching $A$ and $C$ labels. It is thus clear how to change from the internal perspective of quantum reference frame $A$ to that of $C$ {\it via} the perspective-neutral Dirac quantum theory: invert the transformations to $A$-perspective and apply the transformations to $C$-perspective.  This is the quantum analog of the classical procedure in sec.\ \ref{sec_clswitch}. Concretely, this defines a map
\ba
 \hat{\mathcal{S}}_{A\to C}:\ch_{BC|A}\rightarrow\ch_{AB|C}\,,
 \ea
of the form 
 \begin{equation}\label{eq41}
 \begin{split} \hat{\mathcal{S}}_{A\to C}&:=\int\mathrm{d}p'\,{}_C\bra{p'}\hat{\ct}_{C,AB}\,\\
 &\q\q\q\q\q\times(\hat{\ct}_{A,BC})^\dag\ket{p=0}_A\otimes[\cdot]\\
 &=\int\,\mathrm{d}p'\,{}_C\bra{p'}\,\exp{ \Big( i\, \Hat{q}_C (\Hat{p}_A + \Hat{p}_B) \Big) }\\
 &\,\times\exp{ \Big( -i\, \Hat{q}_A (\Hat{p}_B + \Hat{p}_C) \Big) }\,\ket{p=0}_A\,\otimes[\cdot],
 \end{split}
 \end{equation}
{where the reduced state $\ket{\psi}_{BC|A}$ of interested has to be inserted into the empty slot $[\cdot]$ of the tensor factor associated to particles $B$ and $C$.}

{A few comments are in place which bring us back to the classical discussion of Sec.~\ref{sec_clswitch}. There we emphasized that it is the physical interpretation of the reduced description as the perspective of reference frame $A$ that singles out the otherwise highly ambiguous embedding of the gauge-fixed reduced phase space into the perspective-neutral constraint surface. This interpretation is, of course, added information compared to the reduced description alone, but it is crucial. Note that this is also qualitatively analogous to what happens in coordinate changes on a manifold. Suppose one picks the coordinate map associated to some observer in general relativity to map a spacetime neighbourhood into some coordinate description thereof. If one only kept the coordinate description of the neighbourhood but discarded all information about the coordinate map itself (and thus the interpretation of this coordinate description), it would be impossible to map back into the spacetime manifold and thus to consistently change from one frame perspective to another. One has to keep track of which coordinates are associated to which spacetime point in order to compare descriptions of it.}

{From the previous section it follows that the `quantum coordinate map' from the perspective-neutral physical Hilbert space $\ch^{\rm phys}$ to the perspective of $A$ is $\varphi_A:=\sqrt{2\pi}{}_A\bra{q=0}\,\hat\ct_{A,BC}$. Just like in the case of classical coordinate changes, we cannot discard the information about $\varphi_A$ (and thereby the interpretation of the reduced theory) and must invert it to return to $\ch^{\rm phys}$. Appending, e.g., the $A$ tensor factor $\ket{p=0}_A$ (rather than another state of $A$) to the reduced state in (\ref{eq41}) in order to map back into the perspective-neutral $\ch^{\rm phys}$ has to be understood in this sense.\footnote{{Appendix~\ref{app_nonunique} entails that the `quantum coordinate map' $\varphi_A$ is mathematically non-unique up to the number to which we fix $A$'s momentum. The same thus holds true for the inverse map $\varphi_A^{-1}$. However, this non-uniqueness is physically irrelevant as discussed in appendix~\ref{app_nonunique} and as long as one sticks to a convention one will always relate the same reduced state $\ket{\psi}_{BC|A}$ with the same original physical state $\ket{\psi}^{\rm phys}$.}}}

In Appendix \ref{transformation}, we show that (i) indeed $\hat{\mathcal{S}}_{A\to C} \ket{\psi}_{BC|A} = \ket{\psi}_{AB|C}$, where $\ket{\psi}_{BC|A},\ket{\psi}_{AB|C}$ correspond via (\ref{physstate1d}, \ref{redphysstate}) to the same state $\ket{\psi}^{\rm phys}$, and that (ii) this transformation is equivalent to 
\begin{equation}\label{SAC}
 \hat{\mathcal{S}}_{A\to C} = \hat{\mathcal{P}}_{CA} e^{i\, \Hat{q}_C \Hat{p}_B}\,,
\end{equation}
where $\hat{\cp}_{CA}$ is the parity-swap operator defined in \cite{Giacomini:2017zju}, which, acting on momentum eigenstates of $A$ yields:
\begin{equation}\label{}
\begin{split}
\hat{\mathcal{P}}_{CA} \ket{p}_C = \ket{-p}_A\,.
\end{split}
\end{equation}
Crucially, (\ref{SAC}) is precisely the transformation between quantum reference frame perspectives constructed in a different approach in \cite{Giacomini:2017zju} for particle systems in one-dimensional Newtonian space. In Ref. \cite{Giacomini:2017zju}, this transformation arises as a specific instance of a more general class of quantum reference frames transformations, including also a generalization of extended Galilean transformations. The present construction permits us to derive the specific transformation \eqref{SAC} from first (symmetry) principles and via an associated perspective-neutral quantum structure into which all perspectives can be embedded. It also is clear that the reduced observables transform correctly from $\ch_{BC|A}$ to $\ch_{AB|C}$
\begin{equation} \label{eq:QuantAtoC}
\begin{split}
		\mathcal{S}_{A\to C} \,\hat{q}_B \,\mathcal{S}_{A\to C}^\dagger &= \hat{q}'_B - \hat{q}'_A, \\ \mathcal{S}_{A\to C}\, \hat{q}_C\, \mathcal{S}_{A\to C}^\dagger &= - \hat{q}'_A,\\
		\mathcal{S}_{A\to C} \,\hat{p}_B \,\mathcal{S}_{A\to C}^\dagger&= \hat{p}'_B, \\ \mathcal{S}_{A\to C}\, \hat{p}_C\, \mathcal{S}_{A\to C}^\dagger &= - \hat{p}'_B - \hat{p}'_A,
	\end{split}
\end{equation}
where the primed operators represent the position and momentum operators in the reference frame of C. Note that the way the operators transform coincides with the results found in Ref.~\cite{Giacomini:2017zju} and matches the classical case of Eq.\ \eqref{AtoC}.

For clarity, we summarize this internal perspective change in the following {\it commutative} diagram:
\newpage
\begin{widetext}
\begin{center}
\begin{tikzcd}[row sep=huge, column sep = huge]
& \mathcal{H}^{\rm phys} \arrow[rd, "\hat{\mathcal{T}}_{C,AB}"]& \\
\mathcal{H}^{\rm phys}_{A,BC} \arrow[ru, " \hat{\mathcal{T}}_{A,BC}^\dagger"]& & \mathcal{H}^{\rm phys}_{C,AB} \arrow[d, "{\sqrt{2\pi}\,{}_C\bra{q=0}}"]\\
\mathcal{H}_{BC|A} \arrow[u, "\ket{p=0}_A \otimes (\cdot) "] \arrow[rr, "\hat{\mathcal{S}}_{A \to C}"] & & \mathcal{H}_{AB|C}
\end{tikzcd}
\end{center}
\end{widetext}
Notice that setting $\cs_{\rm phys}:=\ch^{\rm phys}$ as the perspective-neutral structure, $\varphi_A:={\sqrt{2\pi}\,{}_A\bra{q=0}}\,\hat{\mathcal{T}}_{A,BC}$ as $A$'s perspective map and $\varphi_C:={\sqrt{2\pi}\,{}_C\bra{q=0}}\,\hat{\mathcal{T}}_{C,AB}$ as $C$'s perspective map, we find that $  \hat{\mathcal{S}}_{A\to C} = \varphi_B\circ\varphi_A^{-1}$ is indeed of the general form (\ref{gentrafo}).

\section{Some operational consequences of switching perspectives in the classical and quantum theory} \label{operational}




The operational consequences of the transformation between two quantum reference frames have been thoroughly analyzed in Ref.~\cite{Giacomini:2017zju}. There, it was shown that entanglement and superposition depend on the {quantum reference frame}, and this is operational in that it can in principle be tested experimentally. In other words, a state which appears as ``classical'' (for instance, in a coherent state) from the point of view of a certain quantum reference frame, might appear entangled, or in a superposition state from the point of view of a different {quantum reference frame}. Additionally, the notion of {quantum reference frame} can turn out to be extremely useful in concrete applications. For instance, the approach in Ref.~\cite{Giacomini:2017zju} allows one to identify the transformation to jump into the rest frame of a quantum system, intended as a system moving in a superposition of velocities. This operation would be impossible with a standard reference frame transformation. 

\begin{figure}
	\begin{center}
		\includegraphics[scale=0.5]{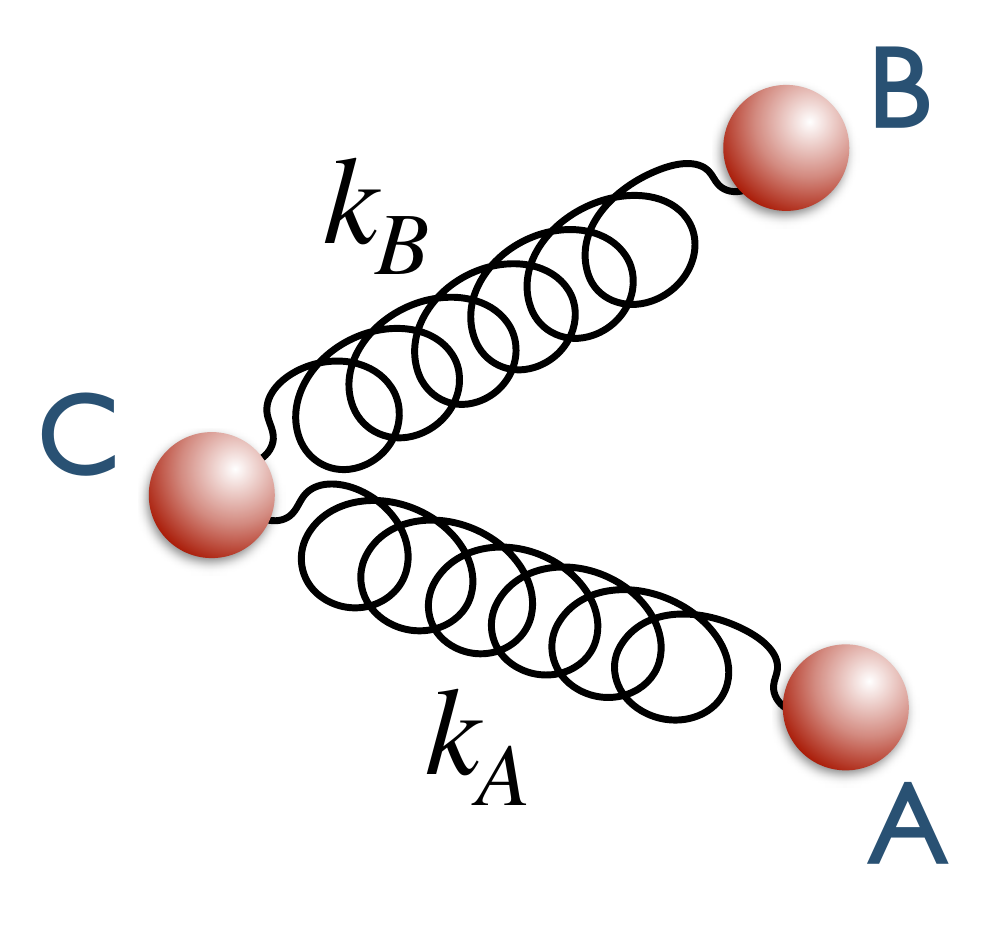}
		\caption{\label{fig:oscillators} In the {\it perspective-neutral description}, the three systems A, B, and C behave like two harmonic oscillators, with springs being attached to system C and A (with spring constanst $k_A$), and to systems C and B (with spring constant $k_B$). From this perspective, the Hamiltonian (both in the classical and quantum case) is $H= \frac{p_A^2}{2m_A} + \frac{p_B^2}{2m_B} + \frac{p_C^2}{2m_C} + \frac{1}{2}k_A (q_C - q_A)^2 + \frac{1}{2}k_B (q_C - q_B)^2$.}
	\end{center}
\end{figure}

In the following, we analyse a simple model of two harmonic oscillators (as illustrated in fig.~\ref{fig:oscillators}), both in the classical and in the quantum case. Our goal is to show how the behaviour of the different systems is described in two different reference frames. In particular, in the quantum case we will recover the dependence of entanglement on the quantum reference frame, compatibly with what has been found in \cite{Giacomini:2017zju}. Here, we additionally provide a study of the entanglement in different frames in a dynamical setting, i.e., by studying the solutions of the equations of motion. Let us consider a system of two harmonic oscillators, as seen from the perspective of C. In the reduced theory, the Hamiltonian is
\begin{equation}\begin{split}
	H_{AB|C} &= \frac{\xi_A^2}{2m_A} + \frac{\xi_B^2}{2m_B} + \frac{(\xi_A+\xi_B)^2}{2m_C} \\
	&\q\q+ \frac{1}{2}k_A x_A^2 + \frac{1}{2}k_B x_B^2,
	\end{split}
\end{equation}
where $m_A, m_B, m_C$ are respectively the mass of system A, B, and C, and $k_A, k_B$ are the spring constants of systems A and B respectively. Note that in this section we have renamed the relative coordinates in C's reference frame as $x_A$ and $x_B$ and the momenta in C's reference frame as $\xi_A$ and $\xi_B$. Under the assumption that $m_C \gg m_A, m_B$, the systems B and C behave as two decoupled oscillators, moving along the trajectories
	\begin{equation}
	\begin{split}
		x_A(t) &= A_0 \cos \left( \omega_A t + \phi_A \right),\\
		x_B(t)& = B_0 \cos \left( \omega_B t + \phi_B \right),
	\end{split}\end{equation}
where $A_0, B_0, \phi_A, \phi_B$ are fixed by the initial conditions and $\omega_i^2 \approx \frac{k_i}{m_i}$, $i=A,B$.

\begin{figure}
	\begin{center}
		\includegraphics[scale=0.5]{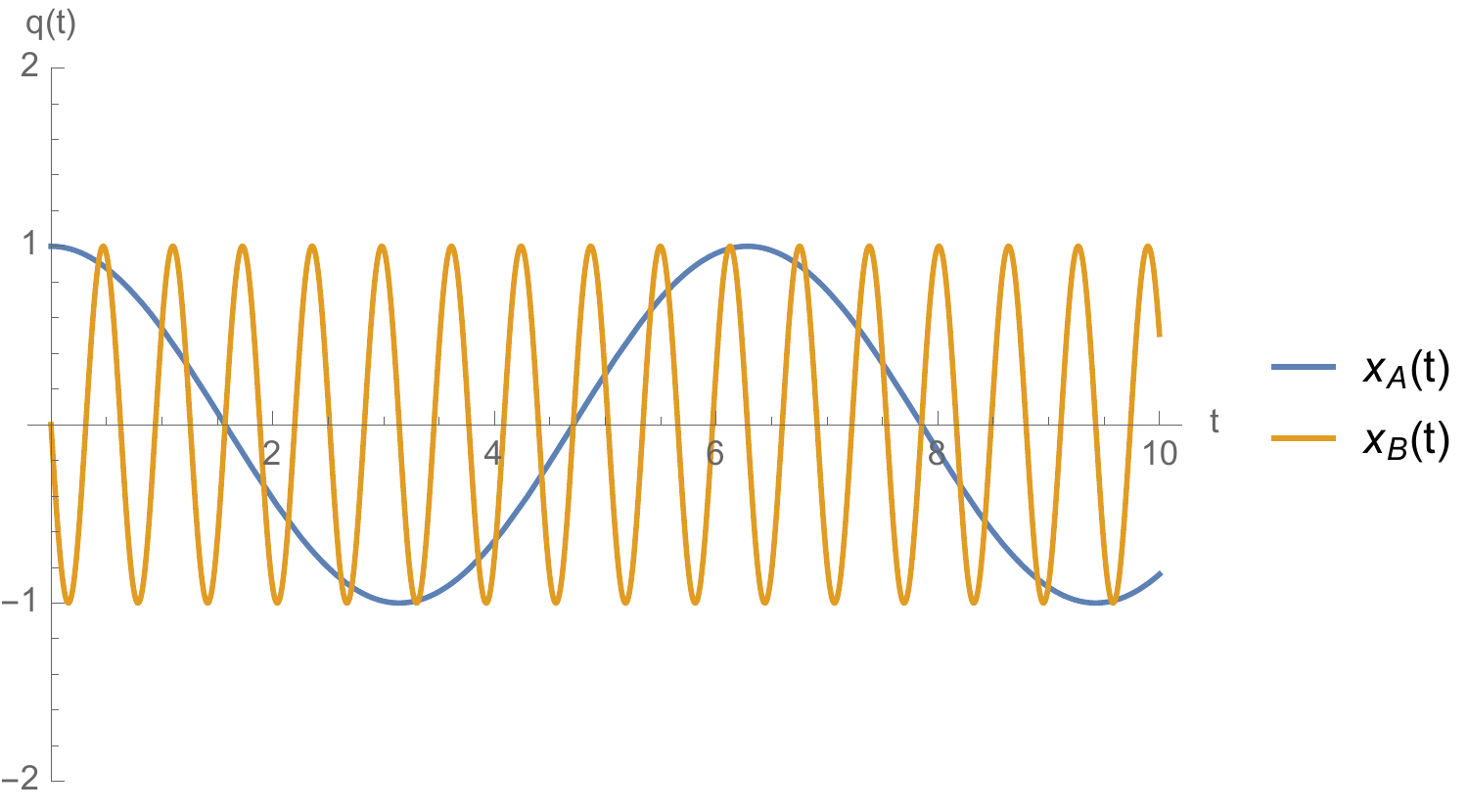}\q\q\q\q\q\q\q\\\vspace*{.5cm}		\includegraphics[scale=0.5]{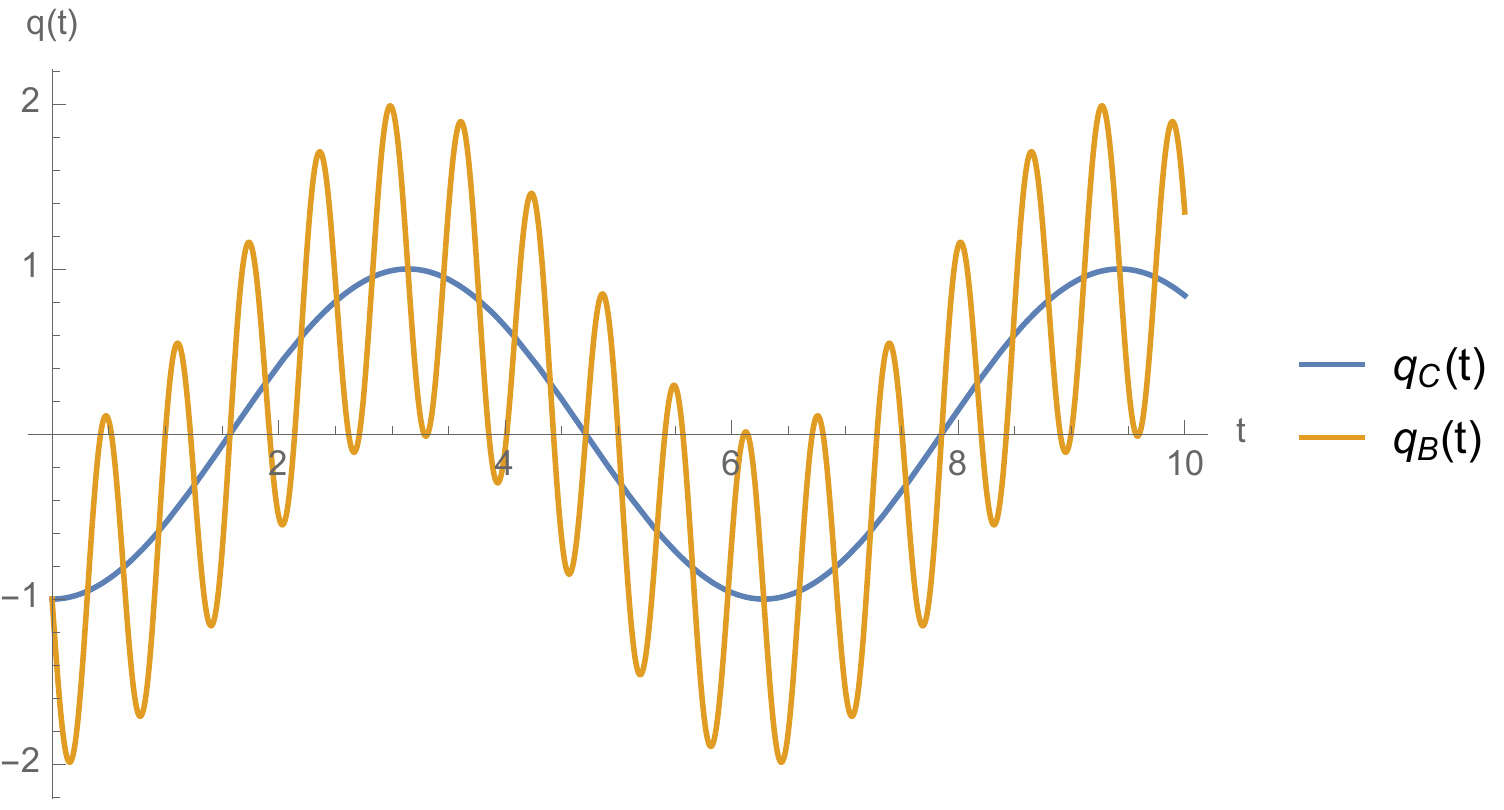}
		\caption{\label{fig:Classical1} Up: $x_A(t)$ and $x_B(t)$ when $A_0 = B_0 =1$, $\omega_A = 1$, $\omega_B=10$, $\phi_A =0$ and $\phi_B= \pi/2$. Down: the solutions of the equations of motion $q_B(t)$ and $q_C(t)$ in A's reference frame.}
	\end{center}
\end{figure}

\begin{figure}
	\begin{center}
		\includegraphics[scale=0.5]{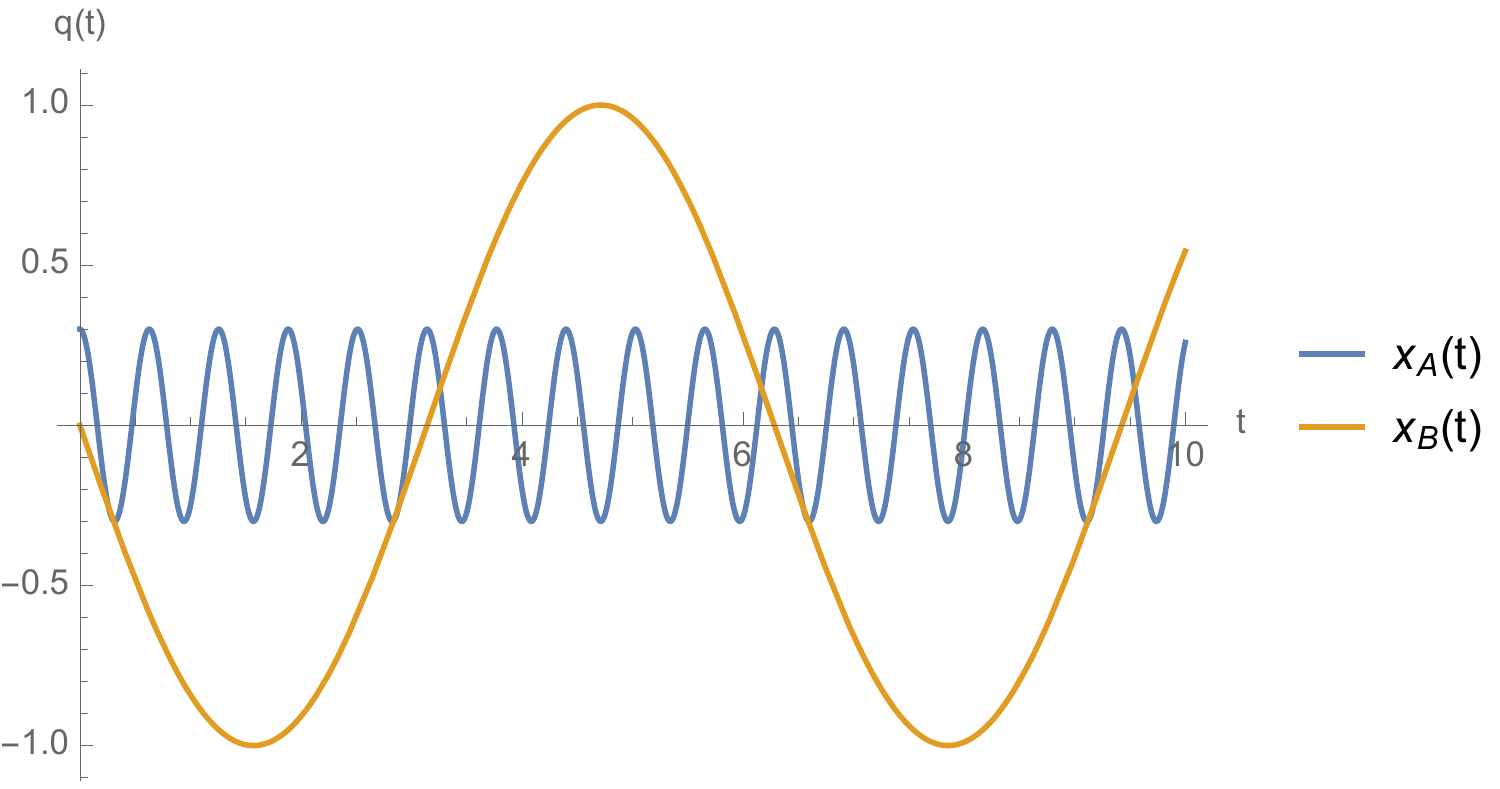}\q\q\q\q\q\q\q	\\\vspace*{.5cm}\includegraphics[scale=0.5]{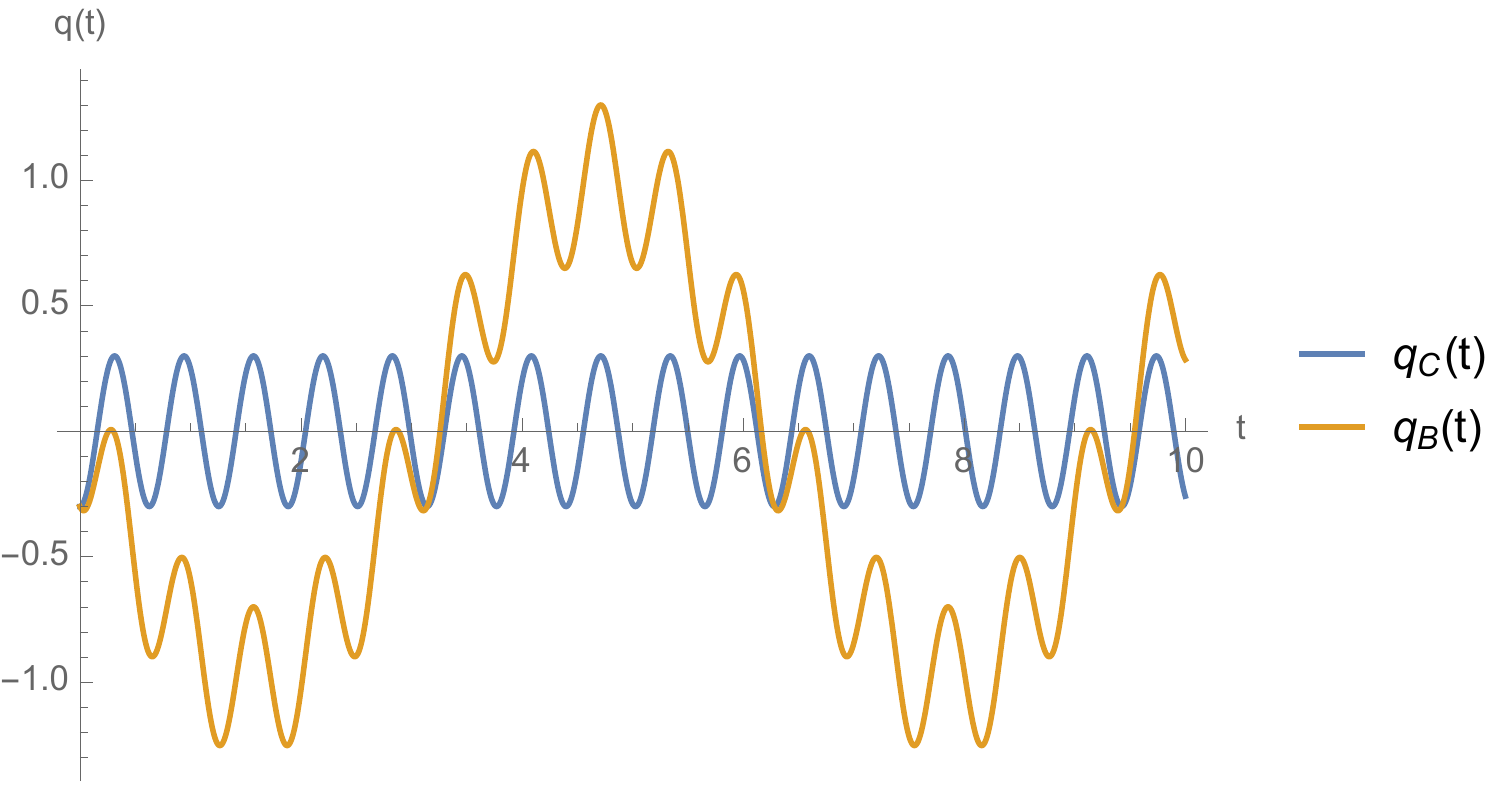}
		\caption{\label{fig:Classical2} Up: $x_A(t)$ and $x_B(t)$ when $A_0 =0.3$ $B_0 =1$, $\omega_A = 10$, $\omega_B=1$, $\phi_A =0$ and $\phi_B= \pi/2$. Down: the solutions of the equations of motion $q_B(t)$ and $q_C(t)$ in A's reference frame.}
	\end{center}
\end{figure}

If we now change to the reference frame of A  by changing to the coordinates $q_C = -x_A$ and $q_B = x_B - x_A$ using Eq.~\eqref{AtoC}, the Hamiltonian becomes
\begin{equation}\begin{split}
	H_{BC|A} &= \frac{\pi_B^2}{2m_B} + \frac{\pi_C^2}{2m_C} + \frac{(\pi_B+\pi_C)^2}{2m_A}  \\&\q\q+ \frac{1}{2}k_A q_C^2 + \frac{1}{2}k_B (q_B - q_C)^2,
\end{split}\end{equation}
where $q_B$ and $q_C$ are the new coordinates and $\pi_B$, $\pi_C$ the new momenta in A's reference frame. The solutions of the equations of motion, matched with the transformed initial conditions, read
	\begin{equation}\begin{split}
		q_B(t) &= B_0 \cos \left( \omega_B t + \phi_B \right) - A_0 \cos \left( \omega_A t + \phi_A \right),\\
		q_C(t) &= -A_0 \cos \left( \omega_A t + \phi_A \right).
	\end{split}\end{equation}
Note that these solutions coincide with $q_B (t) = x_B (t) - x_A(t)$ and $q_C(t) = - x_A (t)$. The solutions of the equations of motion in the initial and final reference frames are illustrated, for different values of the parameters, in fig.~\ref{fig:Classical1} and in fig.~\ref{fig:Classical2}. In particular, we notice that, while in the reference frame C the two solutions are independent, in the new reference frame correlations arise.

In the particular case when $\omega_A = \omega_B$ and when the oscillators are in phase, one finds the solution $q_B (t)=0$. Physically, this means that if the two oscillators are perfectly in phase and oscillate at the same amplitude and frequency, from the point of view of A the system B doesn't move.

After quantization, as we have shown in sec.~\ref{quantum}, the Hamiltonian acting on the reduced phase space is quantized as
\begin{equation}\begin{split}
	\hat{H}_{AB|C}&= \frac{\hat{\xi}_A^2}{2m_A} + \frac{\hat{\xi}_B^2}{2m_B} + \frac{(\hat{\xi}_A + \hat{\xi}_B)^2}{2m_C} \\
	&\q\q\q+ \frac{1}{2}k_A \hat{x}_A^2 + \frac{1}{2}k_B \hat{x}_B^2,
\end{split}\end{equation}
where the parameters and the operators have the same meaning as in the classical case. For simplicity, we assume that the system is initially prepared in an eigenstate of this Hamiltonian (so that the time evolution of the state only amounts to a global phase, which we can then discard. Notice that this method is general, because any other state of the Hibert space $L^2(\mathbb{R}^2)$ can be obtained by linear combinations of the eigenstates of the harmonic oscillator).

Under the assumptions that the derivatives of the total eigenstates of A and B $\Psi^n (x_A, x_B)$, $n \in \mathbb{N}$, are of the same order, and that $m_C \gg m_A, m_B$, we can consider a perturbative expansion in $\frac{m_{A(B)}}{m_C}$. To the lowest order in perturbation theory, we have two decoupled harmonic oscillators with frequency $\omega_A = \sqrt{k_A/m_A}$ and $\omega_B = \sqrt{k_B/m_B}$. The eigenstate can then be split into the two eigenstates of A and B, which are easily expressed in terms of the Hermite polynomials. For concreteness, we shall focus on the first two eigenstates
\begin{equation}
\begin{split}
	\psi^0_i (x_i) &= \left( \frac{\alpha_i}{\pi}\right)^{1/4} e^{- \frac{\alpha_i x_i^2}{2}},\\
	\psi^1_i (x_i)&= \sqrt{2} \left( \frac{\alpha_i^3}{\pi}\right)^{1/4} x_i e^{- \frac{\alpha_i x_i^2}{2}},
	\end{split}
\end{equation}
where $\alpha_i = \frac{m_i \omega_i}{\hbar}$ and $i=A, B$. Since the quantum reference frame transformation is unitary, the transformed state is also an eigenstate of the new Hamiltonian with the same eigenvalue. Therefore, if in the initial reference frame we have $\left| \Psi (t) \right\rangle_{AB|C} = e^{-\frac{i}{\hbar}(E_A^n + E_B^m)t}\left| \psi^n \right\rangle_{A|C}\left| \psi^m \right\rangle_{B|C}$, where $n=0,1$, in A's reference frame this state is transformed to $\left| \Psi (t) \right\rangle_{BC|A}= \hat{\mathcal{S}}_{C\rightarrow A}\left| \Psi (t) \right\rangle_{AB|C}$. Explicitly, 
\begin{equation} \label{eq:psiharmA}
\begin{split}
	\left| \Psi (t) \right\rangle_{BC|A} &=e^{-\frac{i}{\hbar}(E_A^n + E_B^m)t} \int dq_B \,dq_C\, \psi^n (-q_C) \,\\
	&\q\q\times\psi^m (q_B-q_C) \left| q_B \right\rangle_{B|A} \left| q_C \right\rangle_{C|A}.
	\end{split}
\end{equation}
We can see that the state of B and C from the point of view of A is an entangled state. Thus, we have mapped, via a quantum reference frame transformation, a product state into an entangled state, showing the dependence of entanglement on the quantum reference frame. The Hamiltonian from the viewpoint of A can easily be calculated as
\begin{equation}
\begin{split}
	\hat{H}_{BC|A}&= \hat{\mathcal{S}}_{C\rightarrow A} \hat{H}_{AB|C}\hat{\mathcal{S}}_{C\rightarrow A}^\dagger \\
	&= \frac{\hat{\pi}_B^2}{2m_B} + \frac{\hat{\pi}_C^2}{2m_C} + \frac{(\hat{\pi}_B + \hat{\pi_C})^2}{2m_A} \\
	&\q\q\q+ \frac{1}{2}k_A \hat{q}_C^2 + \frac{1}{2}k_B (\hat{q}_B - \hat{q}_C)^2,
\end{split}
\end{equation}
where $\hat{q}_i$ and $\hat{\pi}_i$, $i=B, C$ are the position and momentum operator in the reduced phase space from the point of view of A.

In order to analyse the dependence of quantum features on the reference frame, it is convenient, in this particular example, to look at the Wigner function of the relative states in the two reference frames. In the initial reference frame C, the Wigner function of the state of A and B is the product of the two Wigner functions  $f_{W, i|C} (x_i, \xi_i)$, with $i=A,B$. In particular, the Wigner function of the ground state $\psi^0(x_i)$ of the harmonic oscillator is
\begin{equation}
	f_{W,i|C}^0 (x_i, \xi_i) = \frac{1}{\pi \hbar} e^{- \alpha_i x_i^2} e^{- \frac{\xi_i^2}{\hbar^2 \alpha_i}},
\end{equation}
and the Wigner function of the first excited state $\psi^1(x_i)$ is
\begin{equation}
\begin{split}
	f_{W, i|C}^1 (x_i, \xi_i) &= \frac{1}{\pi \hbar} \left( 2 \alpha_i x_i^2 + \frac{2\xi_i^2}{\alpha_i \hbar^2}-1\right) \\
	&\q\q\q\q\times e^{- \alpha_i x_i^2} e^{- \frac{\xi_i^2}{\hbar^2 \alpha_i}}.
	\end{split}
\end{equation}

\begin{figure*}
	\begin{center}
		\includegraphics[scale=0.4]{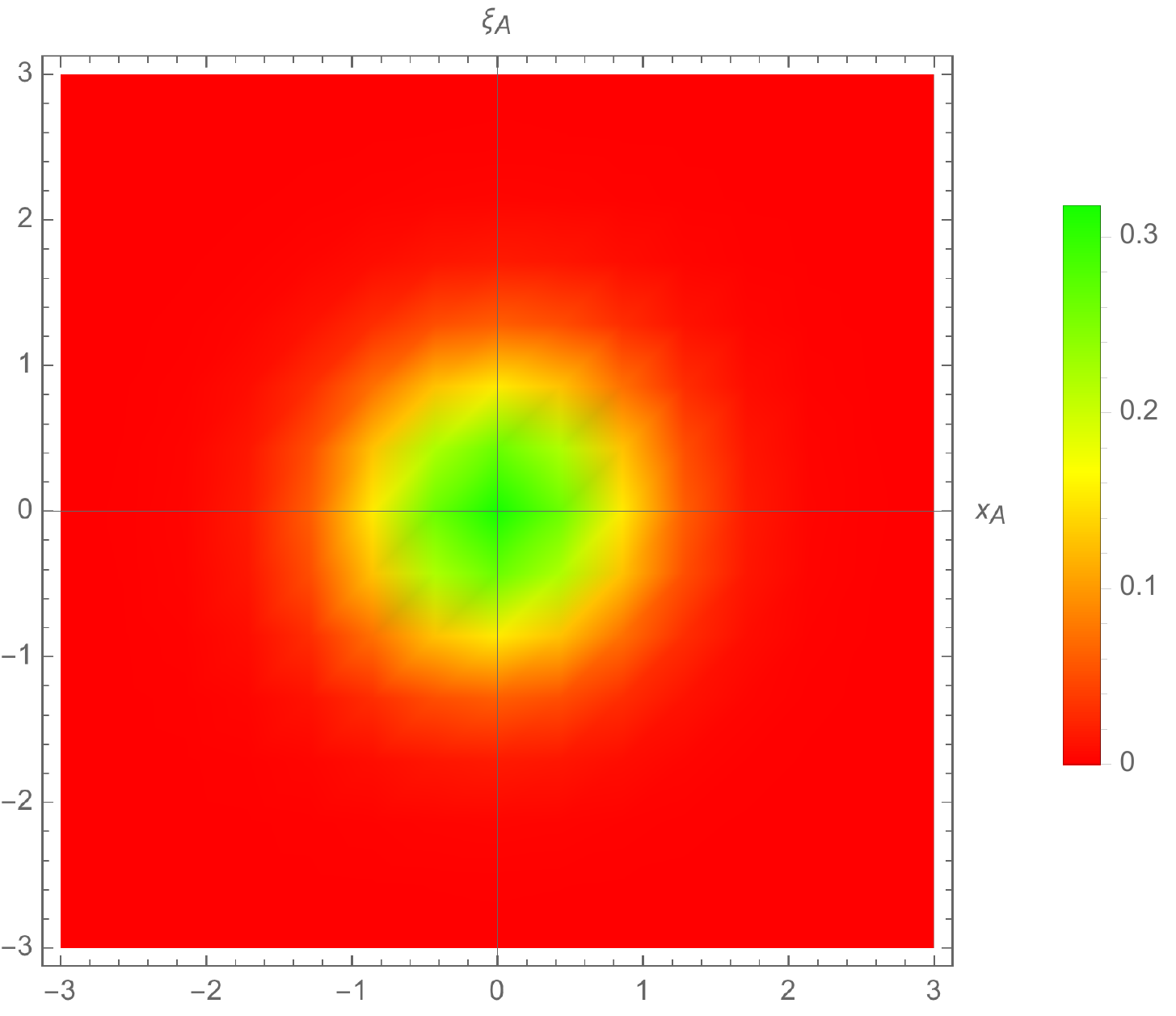}\q\q\q\q\q\q\q		\includegraphics[scale=0.4]{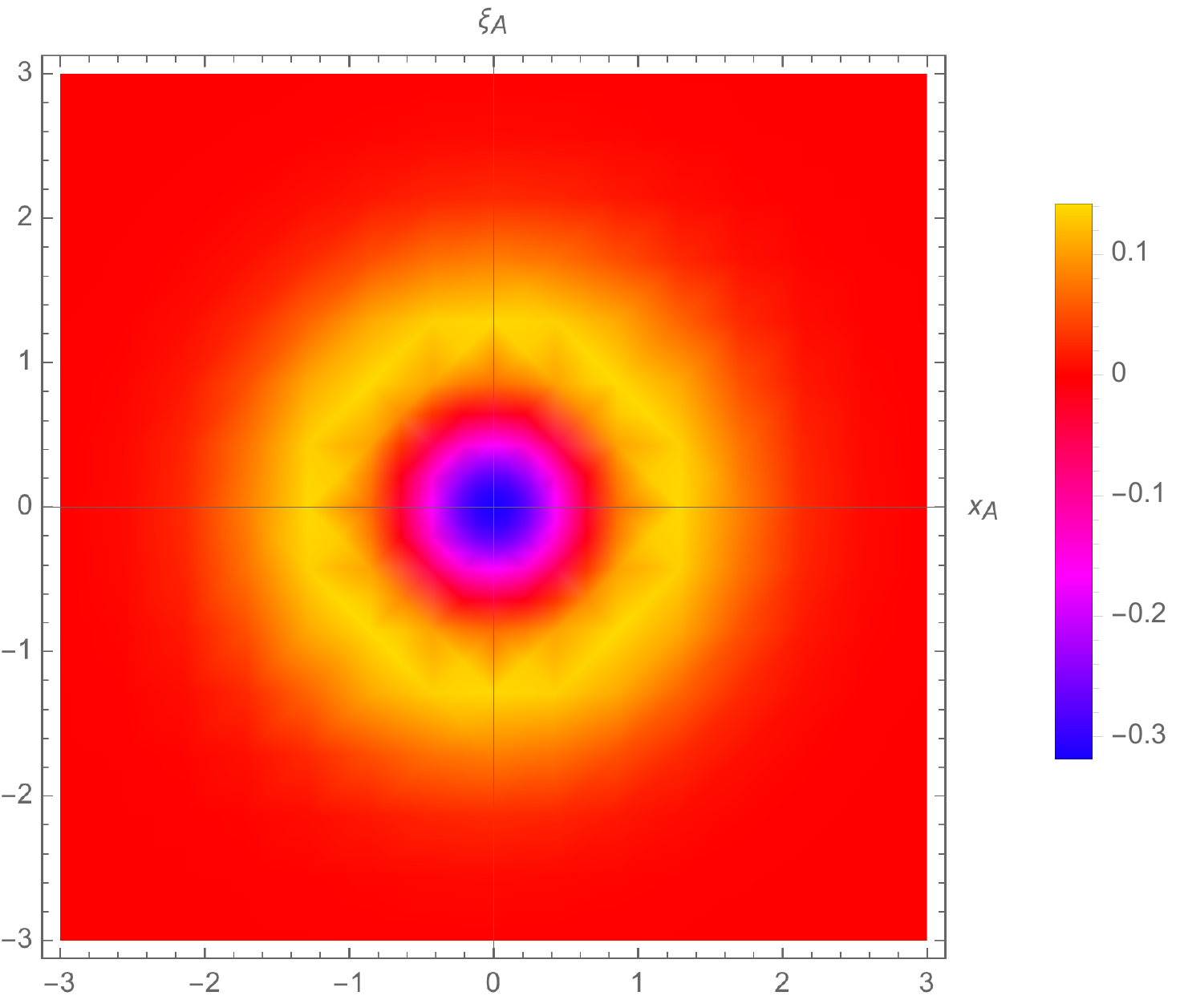}
		\caption{\label{fig:quantum1} On the left, the Wigner function of the ground state of the Hamiltonian, $f_{W, A|C}^0(x_A, \xi_A)$, with $\alpha=1$. On the right, the Wigner function of the first excited eigenstate of the Hamiltonian, $f_{W, A|C}^1(x_A, \xi_A)$, with $\alpha=1$. The Wigner functions for system B are analogous to those of system A, and are calculated by taking the marginals on either system A or B in the initial reference frame C.}
	\end{center}
\end{figure*}

\begin{figure*}
	\begin{center}
		\includegraphics[scale=0.4]{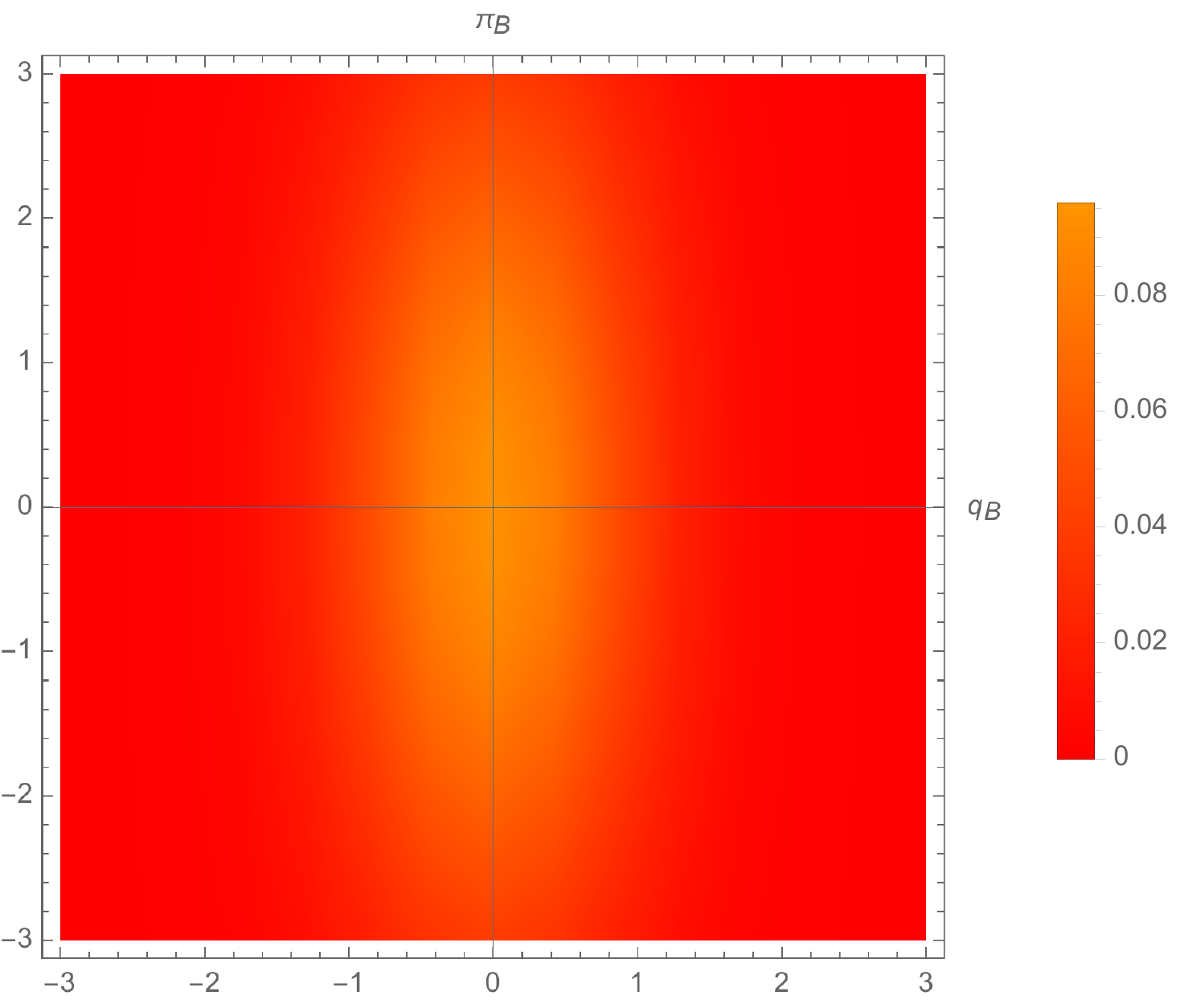}\q\q\q\q\q\q\q		\includegraphics[scale=0.4]{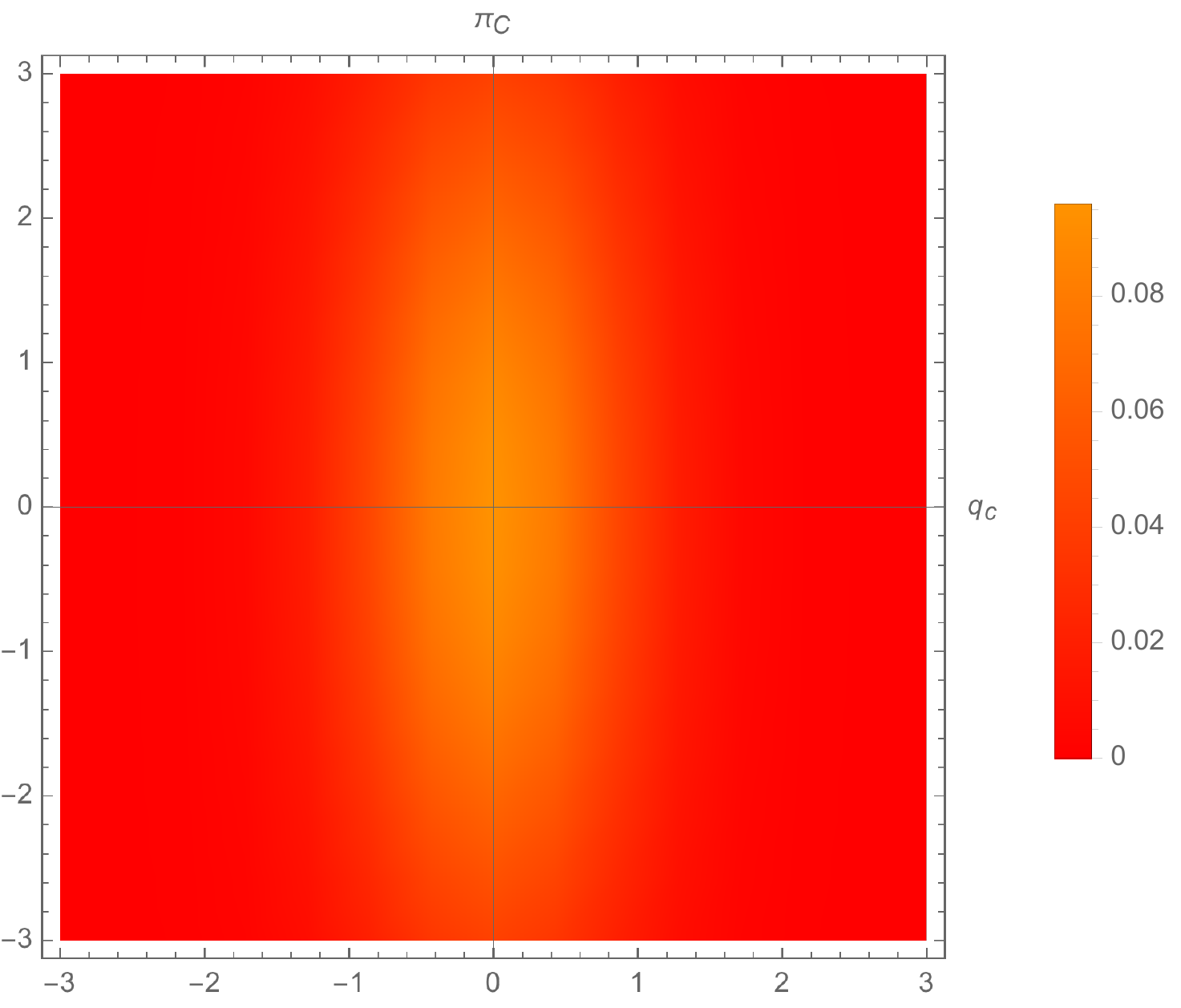}
		\caption{\label{fig:quantum2} In the final {quantum reference frame} A, the marginals of the total Wigner function representing the reduced state of system B (on the left) and C (on the right) when both A and B were initially in the ground state. In both cases, $\frac{\alpha_A}{\alpha_B}= 0.1$.}
	\end{center}
\end{figure*}

\begin{figure*}
	\begin{center}
		\includegraphics[scale=0.4]{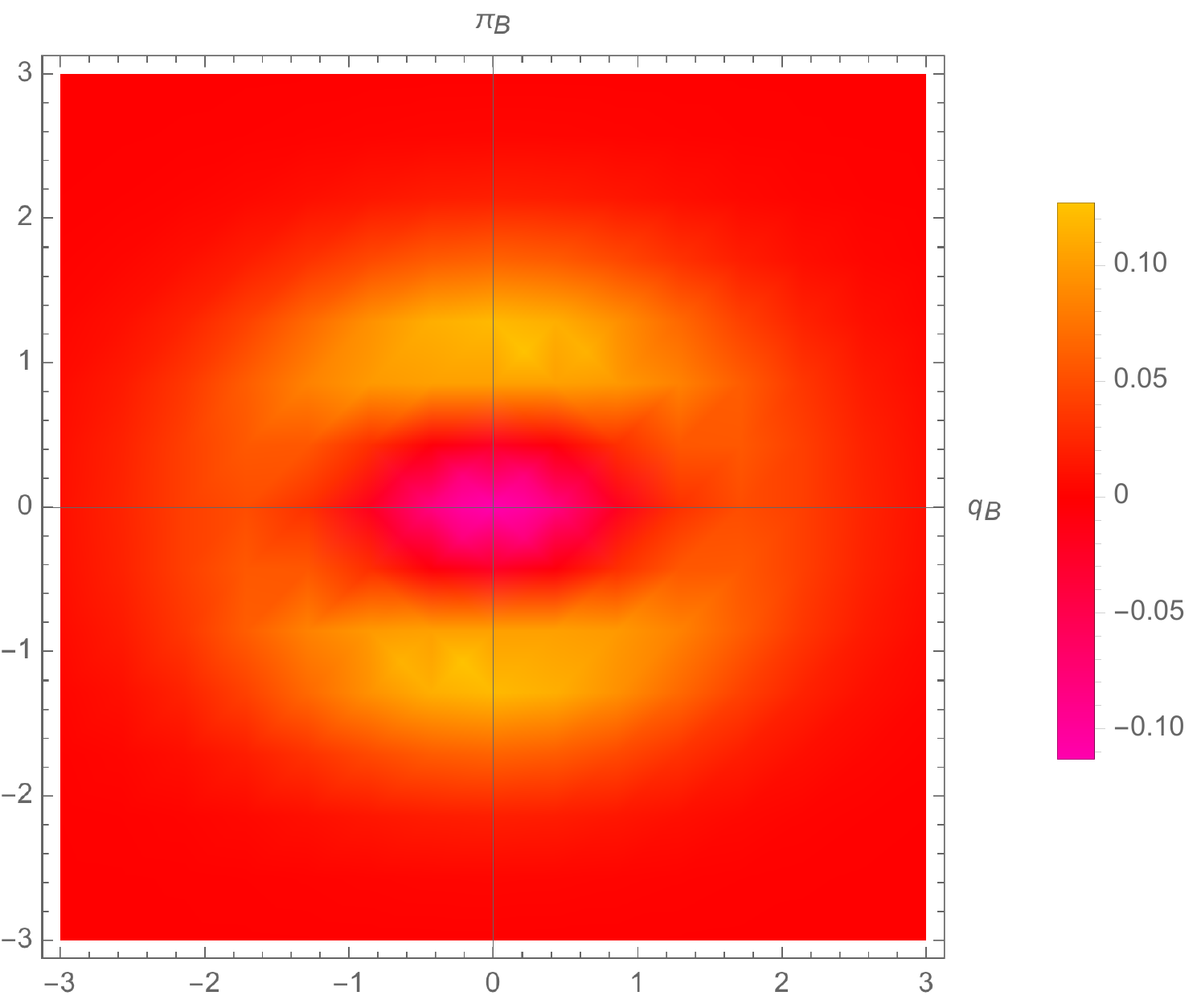}\q\q\q\q\q\q\q		\includegraphics[scale=0.4]{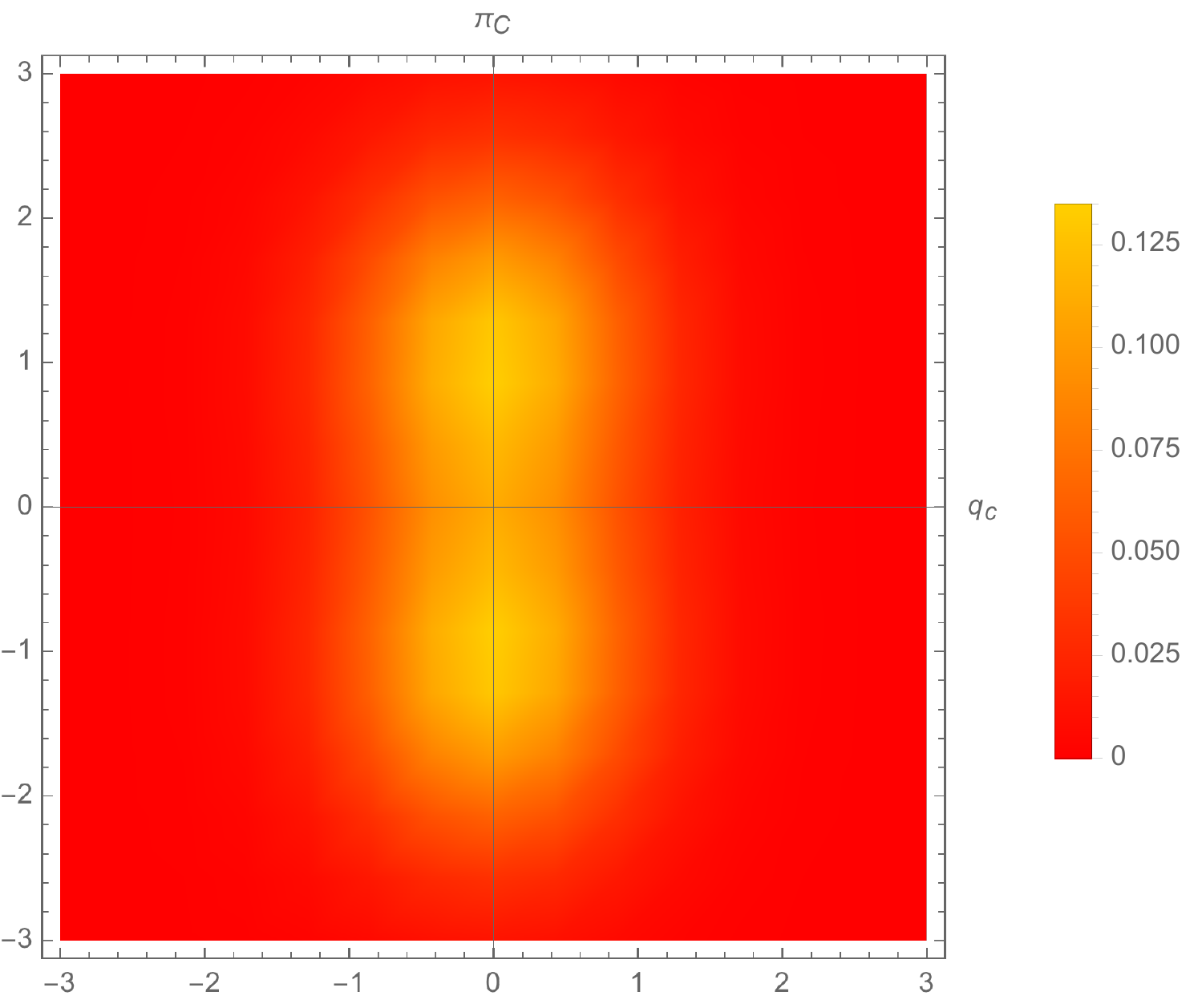}
		\caption{\label{fig:quantum3} In the final {quantum reference frame} A, the marginals of the total Wigner function representing the reduced state of system B (on the left) and C (on the right) when A was initially in the ground state and B in the first excited state. In both cases, $\frac{\alpha_A}{\alpha_B}= 1$.}
	\end{center}
\end{figure*}
\begin{figure*}
	\begin{center}
		\includegraphics[scale=0.4]{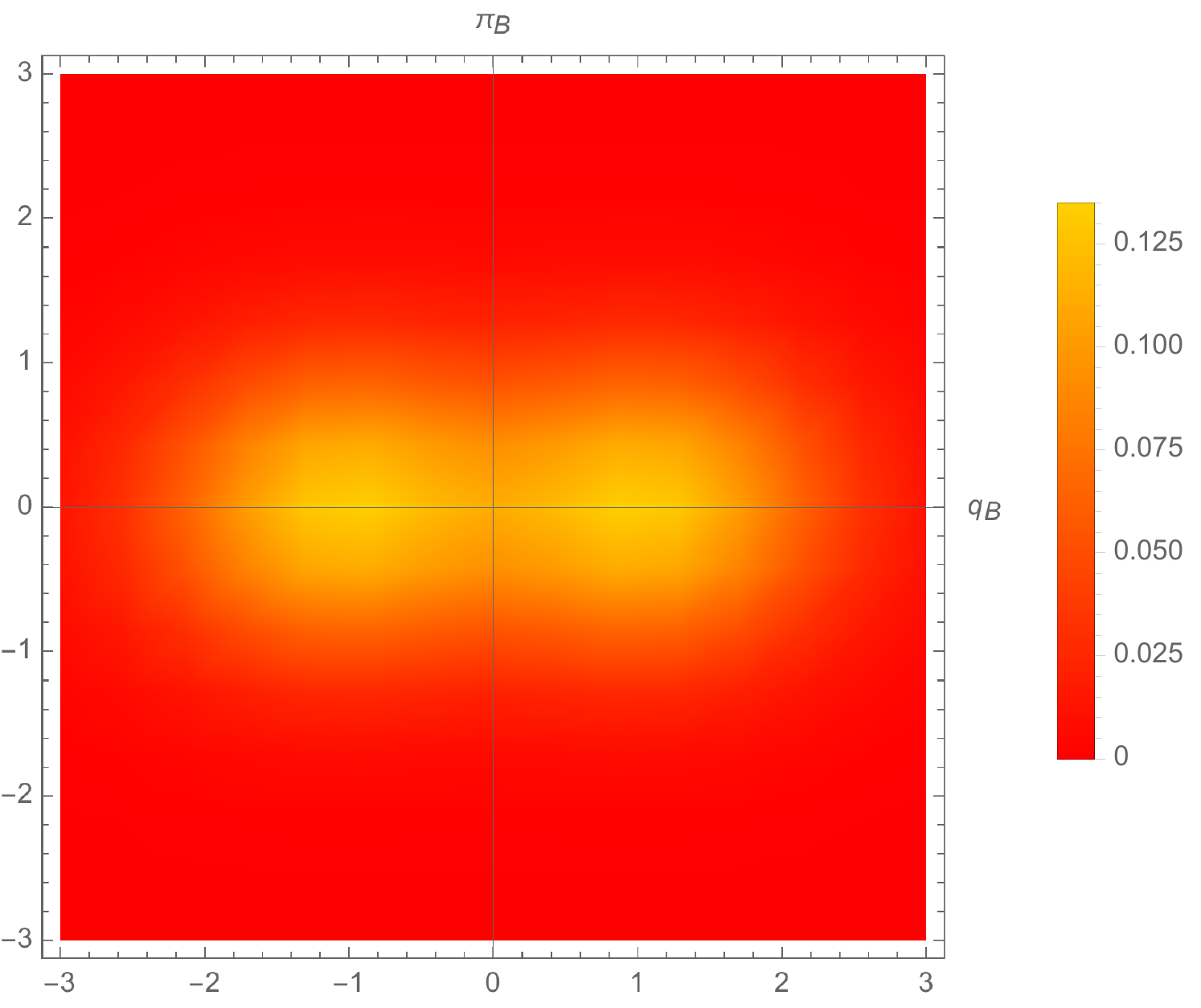}\q\q\q\q\q\q\q		\includegraphics[scale=0.4]{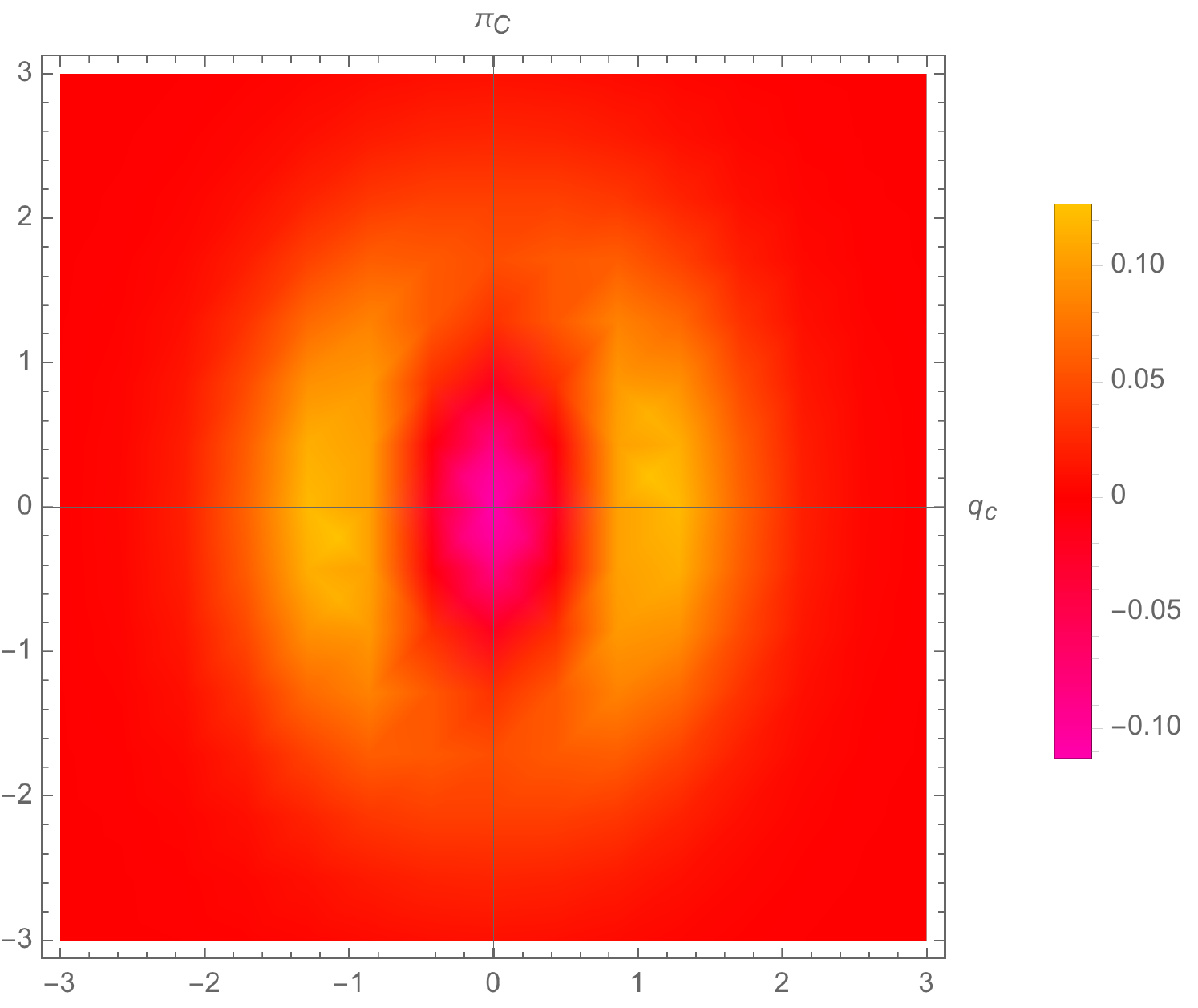}
		\caption{\label{fig:quantum4} In the final {quantum reference frame} A, the marginals of the total Wigner function representing the reduced state of system B (on the left) and C (on the right) when A was initially in the first excited state and B in the ground state. In both cases, $\frac{\alpha_A}{\alpha_B}= 1$.}
	\end{center}
\end{figure*}

\begin{figure*}
	\begin{center}
		\includegraphics[scale=0.4]{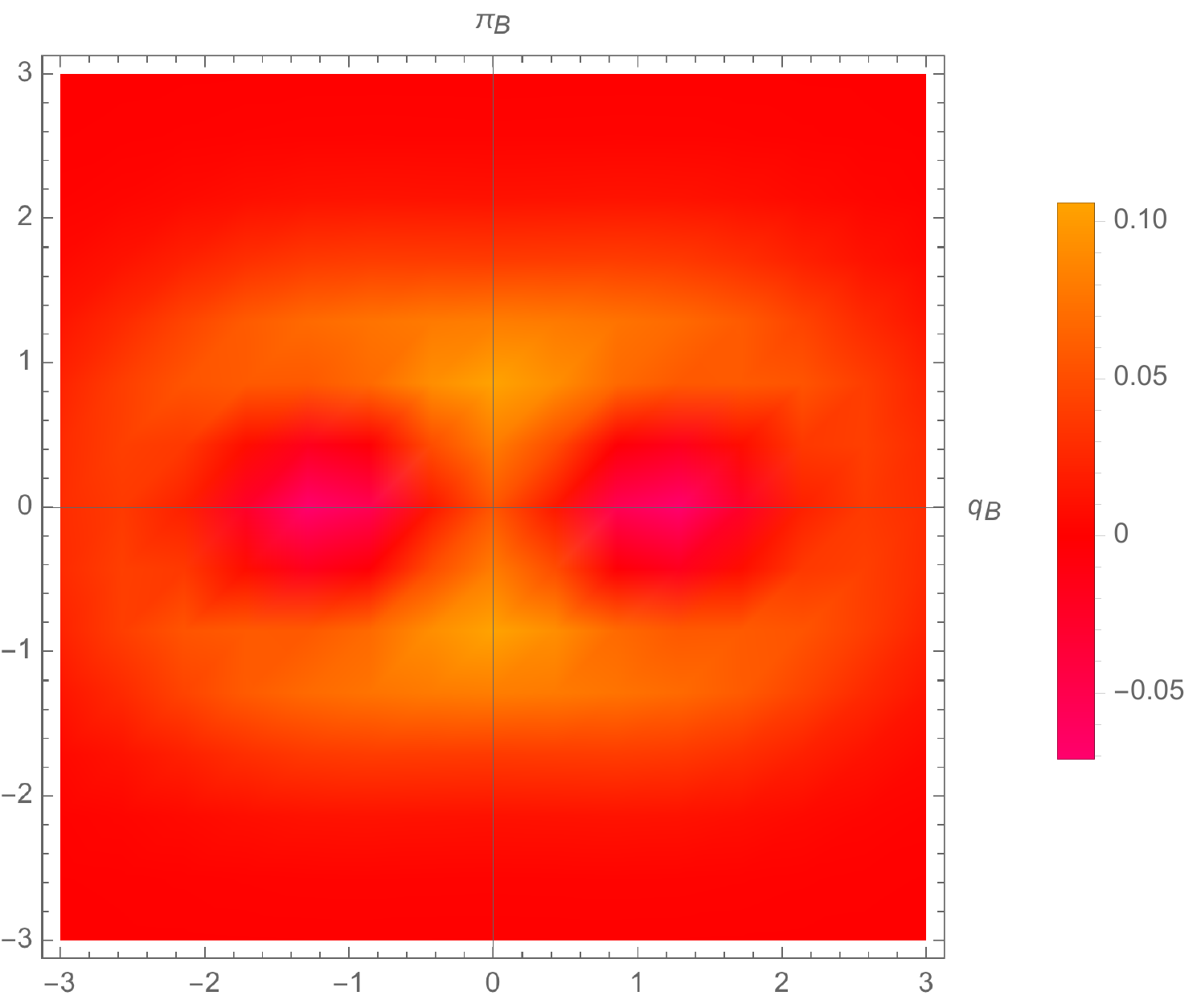}\q\q\q\q\q\q\q
		\includegraphics[scale=0.4]{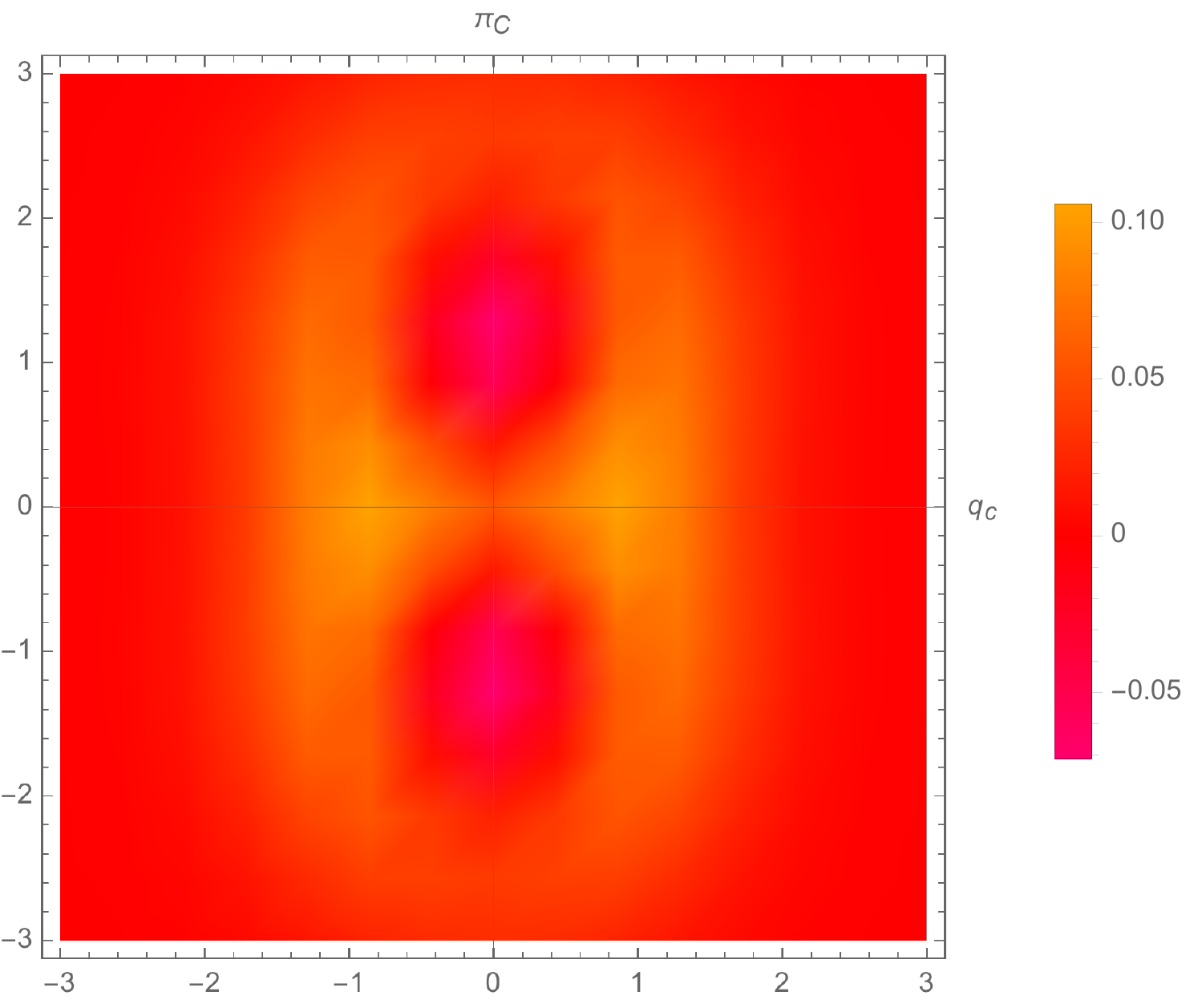}
		\caption{\label{fig:quantum5} In the final {quantum reference frame} A, the marginals of the total Wigner function representing the reduced state of system B (on the left) and C (on the right) when both A and B were initially in the first excited state. In both cases, $\frac{\alpha_A}{\alpha_B}= 1$.}
	\end{center}
\end{figure*}

When we change to the reference frame A, we get
\begin{widetext}
\begin{equation}\begin{split}
	f_{W, BC|A}(q_B, q_C, \pi_B, \pi_C) = f_{W,A|C}^j (-q_C, -\pi_B-\pi_C) 
	 f_{W,B|C}^k (q_B-q_C, \pi_B),
\end{split}\end{equation}
where $j,k=0,1$ and $A,B$ label the initial Wigner functions of systems A and B respectively. In order to find the Wigner function of B or C, it is enough to take the marginals 
\begin{equation}
	f_{W,B|A} (q_B, \pi_B) = \int dq_C\, d\pi_C\, f_{W, BC|A}(q_B, q_C, \pi_B, \pi_C),
\end{equation}
\begin{equation}
	f_{W,C|A} (q_C, \pi_C) = \int dq_B \,d\pi_B\, f_{W, BC|A}(q_B, q_C, \pi_B, \pi_C).
\end{equation}
\end{widetext}
Different combinations of these Wigner functions are plotted in the figures \ref{fig:quantum1}--\ref{fig:quantum5}. In particular, in fig.~\ref{fig:quantum1} the Wigner functions of the ground and excited state of the harmonic oscillator are illustrated. These functions can refer to both system A and B from the viewpoint of the initial reference frame C. On the right in fig.~\ref{fig:quantum1}, the negativity of the Wigner function indicates the nonclassicality of the excited state. In the figures \ref{fig:quantum2}, \ref{fig:quantum3}, \ref{fig:quantum4}, and \ref{fig:quantum5} the Wigner functions of the reduced state of B (on the left) and of C (on the right) are shown in the new reference frame A for different combinations of states. In particular, fig.~\ref{fig:quantum2} shows the Wigner functions of B and C from the point of view of A when the state of A and B from the point of view of C was the product of the ground state eigenstates in the initial reference frame. Figures~\ref{fig:quantum3} and \ref{fig:quantum4} show the Wigner functions of B and C relative to A when the state of A and B from the point of view of C was the product of the ground state and the excited state. Finally, fig.~\ref{fig:quantum5} shows the Wigner functions of B and C when in C's reference frame the total state was the product of the two excited states. Compared to the states in fig.~\ref{fig:quantum1}, the states in the reference frame A appear more spread out, and the characteristic quantumness (i.e., the negativity of the Wigner function, an indicator of quantum behaviour) is sharply reduced. This happens because in the new reference frame the total state of B and C is entangled, as can easily be seen in Eq.~\eqref{eq:psiharmA}, in such a way that the marginals describe a mixed state.

This concludes our quantum discussion of using a perspective-neutral structure in order to switch from one particle reference frame in one-dimensional space to another. 

\section{Conclusions and outlook}\label{sec_conc}

In this work, we have exploited a fruitful interplay of ideas from quantum gravity and quantum foundations to begin developing a unifying approach to transformations among quantum reference systems -- of both temporal and spatial character -- that ultimately should be applicable in both fields. Methodologically, we have combined tools and concepts from constrained systems, also inherently used in the relational clock changes of \cite{Bojowald:2010xp,Bojowald:2010qw,Hohn:2011us}, with the operational approach to quantum reference frames recently put forward in \cite{Giacomini:2017zju}. In particular, as proposed in \cite{Hoehn:2017gst}, we took recourse to a gravity inspired symmetry principle to formulate a perspective-neutral super structure that, so to speak, contains all perspectives at once and via which one can switch among the individual perspectives of the different classical or quantum reference systems. This extends the method of \cite{Bojowald:2010xp,Bojowald:2010qw,Hohn:2011us}, equips it with a novel operational interpretation thanks to \cite{Giacomini:2017zju} and embeds the approach of \cite{Giacomini:2017zju} in a perspective-neutral framework. Our construction offers a systematic method for transforming quantum reference systems, with possible applications in both quantum foundations and gravity.

Using this novel perspective-neutral framework, we have been able to recover  one of the transformations between quantum reference frames in one-dimensional space constructed, among other things, in \cite{Giacomini:2017zju} through a different approach. Finally, we have also studied some striking operational consequences of these quantum frame switches. Specifically, we have illustrated how 
entanglement and classicality of a system interacting with two quantum reference frames depends on whether the perspective of one or the other is chosen.

As we showed, {in our new approach} classically choosing the perspective of a specific frame amounts to a choice of gauge and perspective changes require a gauge transformation within the perspective-neutral constraint surface. In the quantum theory, on the other hand, it was the reduced quantum theories which assumed the role of the quantum physics as seen from a particular quantum reference frame, while the Dirac quantized theory constitutes the perspective-neutral quantum theory, without immediate operational interpretation, via which quantum reference frame perspectives have to be switched. In particular, {for our model} we have clarified the quantum symmetry reduction procedure that maps Dirac to reduced quantization.

{Our quantum symmetry reduction procedure relative to a choice of quantum reference frame can be adapted to more complicated models, as shown in the companion articles \cite{Vanrietvelde:2018dit,Hoehn:2018aqt, Hoehn:2018whn}. We suspect that the general steps of the procedure -- 1.\ choose a quantum reference frame, 2.\ suitably trivialize the constraint with respect to this choice of quantum reference frame, and 3.\ project onto the classical gauge-fixing conditions -- may be adaptable to an even much more general class of models. Of course,  due to the well-known inequivalence of Dirac and reduced quantization in more general situations \cite{Ashtekar:1982wv,kucha1986covariant,Ashtekar:1991hf,Schleich:1990gd,Kunstatter:1991ds,Hajicek:1990eu,Romano:1989zb,Dittrich:2016hvj,Dittrich:2015vfa,Loll:1990rx,plyushchay1996dirac}, the result of applying the quantum symmetry reduction procedure to the Dirac quantized theory will usually not coincide with the quantization of a reduced phase space, in contrast to the simple model discussed here. However, this is not a problem for our perspective-neutral approach.}

{We propose to always interpret the quantum symmetry reduced theory as the description of the quantum physics of the remaining degrees of freedom relative to the associated quantum reference frame. The reason is that Dirac quantization is more general than reduced quantization in the sense that it also encodes quantum fluctuations of the reference frame degrees of freedom. By contrast, there are no such quantum fluctuations of the reference frame in reduced quantization since the reference frame degrees of freedom have been removed altogether prior to quantization. Nonetheless, the classical symmetry reduction procedure, such as in Sec.~\ref{choosingRF}, leading to reduced quantization is conceptually important because it clarifies that we can think of our new quantum symmetry reduction procedure as being its proper quantum analog. This supports the interpretation of the quantum symmetry reduced theory as the `perspective' of an associated quantum reference frame, which is why we have also discussed reduced quantization in this article.}

{We emphasize that in more general systems another property will feature: unlike in our toy model, globally valid gauge-fixing conditions will be absent (globally valid means that every gauge orbit is intersected once and only once by the gauge-fixing surface). This has the consequence that internal frame perspectives, which are associated to a choice of gauge, will not be globally valid, neither classically nor in the quantum theory. More precisely, both the classical and quantum symmetry reduction procedures will not be defined on the entire perspective-neutral structure, i.e.\ classically the constraint surface and in the quantum theory the physical Hilbert space. }

{This too is not a fundamental problem, but rather has to be expected from the general discussion in Sec.~\ref{sec_meta}: globally valid internal perspectives are special. We have seen that changing from one quantum reference frame perspective to another has a compositional structure analogous to coordinate changes on a manifold. The quantum symmetry reduction maps assume the role of  `quantum coordinate maps' and in analogy to classical coordinate maps they will generically not be globally defined. We propose to nevertheless use such reduction maps to describe the physics from the internal perspective of a dynamical reference frame wherever defined. In \cite{Vanrietvelde:2018dit} it will be shown through the relational $N$-body problem in 3D that our new approach indeed remains valid also in more complicated systems where global internal perspectives will be absent. As long as one can locally (in a phase space sense) fix a gauge, one can, in principle, construct local reduced quantum descriptions (see also \cite{Bojowald:2010xp,Bojowald:2010qw,Hohn:2011us}).}

Furthermore, in \cite{Hoehn:2018aqt, Hoehn:2018whn} it will be demonstrated how our method can  be employed to switch temporal reference systems, i.e.\ relational quantum clocks (such as in quantum gravity and cosmology) where subtleties due to the quadratic nature of the constraints arise. 
Finally, in \cite{pqps}, it will be established that our new method is indeed equivalent to that developed in \cite{Bojowald:2010xp,Bojowald:2010qw,Hohn:2011us} when restricted to a semiclassical regime within which the latter was formulated. 

None of these systems include {\it internal} degrees of freedom. In forthcoming work \cite{giacomini2019relativistic}, relativistic particles with spin will be incorporated into the original quantum reference frame approach of \cite{Giacomini:2017zju} and the operational consequences of quantum frame transformations will be explored in this setting.

We conclude with an outlook on some problems where our approach may inspire new perspectives:
\begin{description}

\item[{\it Wigner's friend.}] A paradigmatic example for the challenges of fitting different perspectives in quantum theory into one picture is the Wigner friend scenario on which much has been written (e.g., see \cite{wigner1995remarks,deutsch1985quantum,Rovelli:1995fv,brukner2017quantum,frauchiger2016single}). Including a perspective-neutral meta-structure, similar to here, may open up a new approach to the problem. Of course, this would require the inclusion of measurement interactions into the perspective-neutral structure that lead to `collapses' in the respective internal perspectives.

\item[{\it Quantum general covariance and diffeomorphism}] \emph{symmetry.} Classical general covariance and diffeomorphism symmetry, while intimately related, are not the same concept \cite{Rovelli:2004tv}. Indeed, within the language of sec.\ \ref{sec_meta}, general covariance refers to the {\it operational} level of frame perspectives onto the physics and their relations (all the laws of physics are the same in every reference frame). Diffeomorphism symmetry, on the other hand, refers to the perspective-neutral structure (the diffeomorphism equivalence class of a spacetime) that contains and connects all these different individual frame perspectives. 

Our approach suggests to extend this interplay to the quantum case and we now see how the `quantum general covariance', as advocated in \cite{Giacomini:2017zju}, in principle fits, through the language of sec.\ \ref{sec_meta}, into a bigger picture together with the diffeomorphism symmetry in quantum gravity \cite{Rovelli:2004tv,Thiemann:2007zz,Ashtekar:1991hf}. The `quantum general covariance' of \cite{Giacomini:2017zju}, again, refers to the operational level of quantum reference frames and their relations, which, in our new approach, is encoded in the perspectives and their corresponding reduced quantum theories. The diffeomorphism symmetry in canonical quantum gravity \cite{Rovelli:2004tv,Thiemann:2007zz,Ashtekar:1991hf}, on the other hand, refers to the Dirac quantized theory where one attempts to implement the Hamiltonian and diffeomorphism constraints, which constitute the (first class) Dirac hypersurface deformation algebra that generates the diffeomorphism symmetry. The corresponding diffeomorphism invariant physical Hilbert space, solving these constraints, in the language of sec.\ \ref{sec_meta}, defines the perspective-neutral meta-structure. In line with our new approach, and the simplicity of the present model notwithstanding, we propose to view this latter perspective-neutral quantum gravity theory as the structure containing and connecting all the different quantum reference system perspectives that one refers to when one speaks about `quantum general covariance' as in \cite{Giacomini:2017zju}. This will be further elaborated on in \cite{Hoehn:2018aqt, Hoehn:2018whn}, where it will also inspire a new perspective on the `wave function of the universe'.

\item[{\it Relational quantum mechanics and perspectives.}] In his seminal paper \cite{Rovelli:1995fv} on relational quantum mechanics, Rovelli suggested
{\it ``... to investigate the extent to which the noticed consistency between different observers' descriptions, which I believe characterizes quantum mechanics so marvellously, could be taken as the missing input for reconstructing the full formalism."} Whether or not a consistency among different observers' descriptions can be used in a reconstruction of quantum theory remains an open question. In fact, meanwhile, the formalism has been reconstructed without it, while still being compatible with relational quantum mechanics \cite{Hoehn:2014uua,Hoehn:2015zom} (see \cite{Hoehn:2016otu} for a summary).{\footnote{{There exist further operational reconstructions of quantum theory, see e.g.\ \cite{Hardy:2001jk,Dakic:2009bh,masanes2011derivation,chiribella2011informational, Barnum:2014fk, goyal2010information }, which however pursue a conceptually different route than that proposed in \cite{Rovelli:1995fv}.}}} However, in line with our perspective-neutral approach of sec.\ \ref{sec_meta}, this consistency among different observer perspectives seems to be rather a characterizing feature of physics in general.{\footnote{{See also \cite{Mueller:2017cdn,Mueller:2017jwa} for a striking recent approach to deriving an inter-observer consistency  from subjective individual perspectives using algorithmic information theory.}}}

\end{description}

\section*{Acknowledgments}
We are grateful to \v{C}aslav Brukner for numerous discussions on the topic and to Lucien Hardy for conversations over the interplay of quantum reference frames and indefinite causal structures. PH is indebted to Tim Koslowski for intensive conversations over perspective changes in quantum theory in the past years and, in particular, for suggesting the model of \cite{Vanrietvelde:2018dit}, which ultimately also inspired the simplified version of this manuscript. PH would also like to thank Sylvain Carrozza, Bianca Dittrich, Henrique (aka Heinrich) Gomes, Markus M\"uller, Dennis R\"atzel and Wolfgang Wieland for helpful discussions. The project leading to this publication has initially received funding from the European Union's Horizon 2020 research
and innovation programme under the Marie Sklodowska-Curie grant agreement No 657661 (awarded to PH). PH also acknowledges support through a Vienna Center for Quantum Science and Technology Fellowship. F.G. and E.C.R. acknowledge support from the John Templeton Foundation, Project 60609, ``Quantum Causal Structures'', from the research platform ``Testing Quantum and Gravity Interface with Single Photons'' (TURIS), and the Austrian Science Fund (FWF) through the project I-2526-N27 and the doctoral program ``Complex Quantum Systems'' (CoQuS) under Project W1210-N25. This publication was made possible through the support of a grant from the John Templeton Foundation. The opinions expressed in this publication are those of the authors and do not necessarily reflect the views of the John Templeton Foundation.

\newpage
\onecolumn

\appendix

\section{Lagrangian with translational invariance}\label{app_lag1d}

For simplicity, we shall take the $N$ particles to be of unit mass 
and the configuration manifold as $\cq=\mathbb{R}^N$. The Lagrangian on the tangent bundle $T\cq\simeq\mathbb{R}^{2N}$ reads
\ba
L=\f{1}{2}\,\sum_{i=1}^N\,\dot{q}_i^2\,\,\,-\,\,\,\underbrace{\f{1}{2N}\,\left(\sum_{i=1}^N\,\dot{q}_i\right)^2}_{E_{\rm kin}^{\rm cm}}\,\,\,-\,\,\,\,\,V\left(\{q_i-q_j\}_{i,j=1}^N\right)\,.\label{Lag1d}
\ea
We have subtracted the kinetic energy of the center of mass so that only the motion relative to the latter contributes to the energy. The potential is translation invariant. In consequence, this Lagrangian is singular and features a gauge symmetry: it is invariant under global translations
\ba
(q_i,\dot{q}_i)\,\,\,\mapsto\,\,\,(q_i+f(t),\dot{q}_i+\dot{f}(t)),\label{gauge1d}
\ea
where $f(t)$ is an arbitrary function of time that does not depend on particle $i$. In particular, the equations of motion are underdetermined and read
\ba
-\f{\p V}{\p q_i}=\ddot{q}_i-\f{1}{N}\,\sum_{j=1}^N\,\ddot{q}_j\,,
\ea
so that there are only $N-1$ independent equations as their sum implies
\ba
\sum_{i=1}^N\,\f{\p V}{\p q_i}=0\,,
\ea
which is automatically satisfied for a translation invariant potential.

The physical interpretation is clear: the localizations $q_i(t)$ and motions $\dot{q}_i(t)$ of the $N$ particles with respect to the Newtonian background space have no physical meaning, but are gauge dependent. Only the relative localization and motion of the particles is physically relevant, thereby providing a toy model for Mach's principle. Thanks to the symmetry, physics is here relational. 

This becomes especially explicit in the canonical formulation on which we shall henceforth focus. The Legendre transformation to the phase space $T^*\cq\simeq\mathbb{R}^{2N}$, in coordinates $(q_i,\dot{q}_i)\mapsto(q_i,p_i)$, where
\ba
p_i=\f{\p L}{\p \dot{q}_i}=\dot{q}_i-\f{1}{N}\,\sum_{j=1}^N\,\dot{q}_j\,,\label{leg}
\ea
fails to be surjective and evidently maps onto the $(2N-1)$-dimensional (primary) constraint surface defined by
(\ref{constraint}),
in line with the symmetry of the Lagrangian. 

\section{Switching internal perspectives as a gauge transformation}\label{app_clswitch}

The embedding map of the reduced phase space in $A$ perspective into the constraint surface reads 
\ba
\iota_{BC|A}:\cp_{BC|A}&\hookrightarrow&\cc\nn\\
(q_{i\neq A},p_{i\neq A})&\mapsto& \left(q_{i\neq A},p_{i\neq A},q_A=0,p_A=-\sum_{i\neq A}\,p_i\right)
\ea
and its image is precisely $\cc\cap\cg_{BC|A}$. Conversely, we can also define a projection
\ba
\pi_{BC|A}:\cc\cap\cg_{BC|A}&\rightarrow&\cp_{BC|A}\nn\\
\left(q_{i\neq A},p_{i\neq A},q_A=0,p_A=-\sum_{i\neq A}\,p_i\right)&\mapsto& (q_{i\neq A},p_{i\neq A})\,,
\ea
that drops all redundant information so that $\pi_{BC|A}\circ\iota_{BC|A}=\text{Id}_{\cp_{BC|A}}$. Clearly, the same structures can be constructed for $C$ perspective. 

Now what is the gauge transformation that takes us from $\cc\cap\cg_{BC|A}$ to $\cc\cap\cg_{AB|C}$, where $\cg_{AB|C}$ is defined by $q_C=0$?
Denote the flow on the constraint surface generated by (\ref{constraint}) by $\alpha_P^s$, where $s$ is the flow parameter. The gauge transformation of a phase space function $F$ corresponds to transporting the argument along the flow $\alpha^s_P\cdot F(x)=F(\alpha_P^s(x))$ with $x$ a point on the constraint surface. Explicitly, it reads
\begin{equation}\label{}
\alpha^s_P\cdot F(x)= \sum_{k=0}^\infty \frac{s^k}{k!} \{F,P\}_k(x)\,,
\end{equation}
where  $\{F,P\}_k = \{ \ldots \{\{F,P \}, P \}, \ldots, P \}$ is the $k$-nested Poisson bracket of $F$ with $P$.

Using (\ref{gaugeTransfo}), these gauge transformations are easy to evaluate for the canonical variables
\begin{equation}
\alpha^s_P\cdot q_i (x)= q_i(x) + s\,,\q\q\q\q\q
\alpha^s_P\cdot p_i (x) = p_i (x)\,.
\end{equation}
Hence, jumping from the reference frame of $A$ to the reference frame of, say, $C$ corresponds to the gauge transformation
\begin{equation}\label{}
\alpha_{A \rightarrow C} := \alpha^{-q_C(x)}_P\,,
\end{equation}
i.e.\ to flowing with `parameter distance' $s=-q_C(x)$ (where $q_C(x)$ is the actual value of the relative distance of $A$ and $C$ prior to the transformation), as it verifies $\alpha_{A \rightarrow C}\cdot q_C(x)=0$ and $\alpha_{A \rightarrow C}\cdot q_A(x)=-q_C(x)$.

It is clear that altogether this defines a map, depicted in the diagram of sec.\ \ref{sec_clswitch},
\ba
\cs_{A\to C}:=\pi_{AB|C}\circ\alpha_{A\to C}\circ\iota_{BC|A}:\cp_{BC|A}\rightarrow\cp_{AB|C}\,.
\ea
Taking into account the swap of non-redundant Dirac observable from $q_B-q_A$ to $q_B-q_C$ (and the inverse switch of redundant Dirac observable) through the $A,C$ label exchange, it reads in coordinates:
\ba
(q_B,p_B,q_C,p_C)\mapsto\left(q'_A=-q_C,p'_A=-p_B-p_C,q'_B=q_B-q_C,p'_B=p_B\right)\,.
\ea

\section{Physical inner product for Dirac quantization}\label{app_PIP}

The improper projector (\ref{projp}) $\delta(\hat{P})$ defines equivalence classes of states in $\mathcal{H}^{kin}$ that are mapped to the same solution $\ket{\phi}^{\rm phys}$. One can define an inner product between $\ket{\psi}^{\rm phys}$ and $\ket{\phi}^{\rm phys}$ 
by using any member $\ket{\psi}^{\rm kin}$, $\ket{\phi}^{\rm kin}$ of their respective equivalence classes:
\begin{equation}\label{PIP1}
(\psi^{\rm phys},\phi^{\rm phys})_{\rm phys} := {}^{\rm kin}\bra{\psi} \delta(\hat{P})\ket{\phi}^{\rm kin}\,,
\end{equation}
where $\langle\cdot|\cdot\rangle$ is the original inner product of $\ch^{\rm kin}$. Since $\delta(\hat{P})$ is symmetric in $\ch^{\rm kin}$, this construction is independent on which representative is chosen from each equivalence class. Through Cauchy completion (and other technical subtleties which we shall here ignore), the space of solutions to (\ref{hatp}) can thereby be turned in a proper Hilbert space $\ch^{\rm phys}$.

Using (\ref{redphysstate}), the physical inner product in momentum representation takes either of the following equivalent forms:
\ba
(\psi^{\rm phys},\phi^{\rm phys})_{\rm phys} &=& \int \mathrm{d} p_B \ \mathrm{d}p_C \ [\psi_{BC|A}(p_B,p_C)]^* \phi_{BC|A}(p_B,p_C)\nn\\
&=& \int \mathrm{d} p_A \ \mathrm{d}p_C \ [\psi_{AC|B}(p_A,p_C)]^* \phi_{AC|B}(p_A,p_C)\label{PIP2}\\
&=& \int \mathrm{d} p_A \ \mathrm{d}p_B \ [\psi_{AB|C}(p_A,p_B)]^* \phi_{AB|C}(p_A,p_B)\,,\nn
\ea
i.e., essentially just drops a redundant (and singular) momentum integration.

Next, we show that the `Page-Wootters like' projection (\ref{relativePsi}) is consistent with the inner products. More precisely, if the inner product on the transformed set $\ch^{\rm phys}_{A,BC}:=\hat{\mathcal{T}}_{A,BC}(\ch^{\rm phys})$ is defined, in analogy to (\ref{PIP1}), as
\ba\label{PIP3}
(\psi_{A,BC}|\phi_{A,BC})_{A,BC}&:=&{}^{\rm kin}\braket{\hat{\mathcal{T}}_{A,BC}\,\psi\,|\,\phi}_{A,BC}\nn\\
&=&{}^{\rm kin}\bra{\psi}\hat{\mathcal{T}}_{A,BC}^\dag\, \ket{\phi}_{A,BC}\\\
&=& \int \mathrm{d} p_B \ \mathrm{d}p_C \ [\psi_{BC|A}(p_B,p_C)]^* \phi_{BC|A}(p_B,p_C)\,,\nn
\ea
 then $\hat{\mathcal{T}}_{A,BC}$ indeed defines an isometry from $\ch^{\rm phys}$ to $\ch^{\rm phys}_{A,BC}$. The last line also coincides with the inner product on the reduced Hilbert space $\ch_{BC|A}$ of sec.\ \ref{reducedQuantization} and so $\hat{\mathcal{T}}_{A,BC}$, followed by the projection (\ref{relativePsi}), also defines an isometry from $\ch^{\rm phys}$ to $\ch_{BC|A}$.

Given the transformations (\ref{xConjugationB}--\ref{HDiracReduced}), it is also clear that 
\ba
(\psi^{\rm phys},\hat{O}\,\psi^{\rm phys})_{\rm phys} &\equiv& (\psi_{A,BC}| \Hat{\mathcal{T}}_{A,BC} \ \hat{O} \ (\Hat{\mathcal{T}}_{A,BC})^\dagger\,|\,\psi_{A,BC})_{A,BC}\nn\\
&=& \langle \psi|_{BC|A}\,\hat{O}_{BC|A}\,|\psi\rangle_{BC|A}\,,\label{pexp}
\ea
where $\hat{O}$ is a relevant Dirac observable containing $B$ and $C$ information and $\hat{O}_{BC|A}$ is the corresponding reduced observable on the reduced Hilbert space $\ch_{BC|A}$. Hence, expectation values of relevant Dirac observables on $\ch^{\rm phys}$ coincide with those of the correctly transformed observables in $\ch_{BC|A}$.

\section{Mathematical non-uniqueness of constraint trivialization}\label{app_nonunique}

{The trivialization amounts to transforming the constraint such that it acts only on the reference frame degrees of freedom and the latter become completely fixed and redundant. In the present model, given the linear structure of the constraint $\hat P$, this means fixing the momentum of the chosen reference frame, e.g.\ of $A$. Hence, the trivialization (\ref{TA}) is unique only up to the number to which we fix $A$'s momentum.} 
For example, if instead we chose
\ba
\hat{\mathcal{T}}'_{A,BC} = \exp{ \Big( i\, \Hat{q}_A (\Hat{p}_B + \Hat{p}_C+k) \Big) }\,,\label{altTA}
\ea
where $k\in\mathbb{R}$, we would have {
\ba
\hat{\mathcal{T}}'_{A,BC}\,\hat P\, \left(\hat{\mathcal{T}}'_{A,BC}\right)^\dag=\hat p_A-k\,\label{nochwas}
\ea 
and}
\begin{equation}
    \ket{\psi}_{A,BC} \,\, = \,\,\ket{p=k}_A \otimes \ket{\psi}_{BC|A}\,.
\end{equation}
Yet, also in this case, does one find
\ba
\Hat{\mathcal{T}}'_{A,BC} \ \hat{H}_{\rm tot} \ (\Hat{\mathcal{T}}'_{A,BC})^\dagger\, \ket{\psi}_{A,BC} = \ket{p=k}_A \otimes \Hat{H}_{BC|A}\,\ket{\psi}_{BC|A}
\,,
\ea
and, in fact, all of the relevant structures (\ref{PIP3}, \ref{xConjugationB}, \ref{seqn}, \ref{pexp}) are actually independent of the choice of $k$. The non-uniqueness of the transformation thereby has no physical consequences and only affects the irrelevant information in the $A$-slot. {Up to the irrelevant number $k$ the trivialization is unique.}

\section{Transformation between two quantum reference frames} \label{transformation}

Here we shall prove the claim of sec.\ \ref{sec_qswitch1d}. Writing an arbitrary state in $\ch_{BC|A}$ as in (\ref{redAstate}, \ref{relativePsi}),
\begin{equation}
\ket{\psi}_{BC|A} = \int \mathrm{d} p_B \ \mathrm{d}p_C \  \  \psi_{BC|A}(p_B,p_C) \ket{p_B}_B \ket{p_C}_C\,,
\end{equation}
one finds
 \ba
 \hat{\mathcal{S}}_{A\to C}\,\ket{\psi}_{BC|A} &=&\int\,\mathrm{d}p'_C\,\mathrm{d}p_B\,\mathrm{d}p_C\,\psi_{BC|A}(p_B,p_C)\nn\\
 &&{}_C\bra{p'_C}\,\exp{ \Big( i\, \Hat{q}_C (\Hat{p}_A + \Hat{p}_B) \Big) }\,\exp{ \Big( -i\, \Hat{q}_A (\Hat{p}_B + \Hat{p}_C) \Big) }\,\ket{p=0}_A\,\ket{p_B}_B\ket{p_C}_C\nn\\
&=&\int\,\mathrm{d}p_B\,\mathrm{d}p_C\,\psi_{BC|A}(p_B,p_C)\,\ket{-p_B-p_C}_A\ket{p_B}_B  \,.\label{app_SAC}
 \ea
Recalling from (\ref{redphysstate}) that
\begin{equation}
\psi_{AB|C}(p_A,p_B) = \psi_{BC|A}(p_B,-p_A-p_B)\label{app1}
\end{equation}
and using the change of variables $p_A=-p_B-p_C$, we obtain from (\ref{app_SAC})
\ba
 \hat{\mathcal{S}}_{A\to C}\,\ket{\psi}_{BC|A}&=&\int\,\mathrm{d}p_A\,\mathrm{d}p_B\,\psi_{AB|C}(p_A,p_B)\,\ket{p_A}_A\ket{p_B}_B\nn\\
 &=&\ket{\psi}_{AB|C}  \,.
 \ea
 
This transformation is equivalent to 
\ba
 \hat{\mathcal{S}}_{A\to C}=\hat{\mathcal{P}}_{CA} \,e^{{i}\, \Hat{q}_C \Hat{p}_B}\,,
\ea
where $\hat{\mathcal{P}}_{CA}$ is the parity-swap operator defined in \cite{Giacomini:2017zju} on position eigenstates as
\ba
\hat{\mathcal{P}}_{CA}\,\ket{x}_C=\ket{-x}_A\,.
\ea
Note the similarity to the action of the gauge transformation $\alpha_{A\to C}$ in Appendix \ref{app_clswitch}. 

Indeed, it can be checked that on momentum eigenstates this yields
\begin{equation}\label{}
    \begin{split}
    \ket{-p_B-p_C}_A \ket{p_B}_B &= \hat{\mathcal{P}}_{CA} \ket{p_B + p_C}_C \ket{p_B}_B \\
    &= \hat{\mathcal{P}}_{CA} \,e^{i\, \Hat{q}_C \Hat{p}_B} \ket{p_C}_C \ket{p_B}_B\,, \\
\end{split}
\end{equation}
so that, upon using again (\ref{app1}) and the variable redefinition,
\begin{equation}
\begin{split}
    \ket{\psi}_{AB|C} &= \hat{\mathcal{P}}_{CA} \,e^{i\, \Hat{q}_C \Hat{p}_B} \int \mathrm{d} p_B \ \mathrm{d}p_C \  \  \psi_{BC|A}(p_B,p_C) \ket{p_C}_C \ket{p_B}_B \\
    &= \hat{\mathcal{P}}_{CA} \,e^{i\, \Hat{q}_C \Hat{p}_B} \ket{\psi}_{BC|A}\,.
    \end{split}
\end{equation}


\begin{thebibliography}{10}

\bibitem{Giacomini:2017zju}
F.~Giacomini, E.~Castro-Ruiz, and {\v{C}}.~Brukner, ``Quantum mechanics and the
  covariance of physical laws in quantum reference frames,'' \href{http://dx.doi.org/10.1038/s41467-018-08155-0}{{\em Nature
  communications} {\bfseries 10} no.~1, (2019) 494},

\href{https://arxiv.org/abs/1712.07207}{{\ttfamily arXiv:1712.07207 [quant-ph]}}.

\bibitem{Vanrietvelde:2018dit}
A.~Vanrietvelde, P.~A. H\"ohn, and F.~Giacomini, ``{Switching quantum reference
  frames in the N-body problem and the absence of global relational
  perspectives},''
\href{http://arxiv.org/abs/1809.05093}{{\ttfamily arXiv:1809.05093
  [quant-ph]}}.

\bibitem{Hoehn:2018aqt}
P.~A. H\"ohn and A.~Vanrietvelde, ``{How to switch between relational quantum
  clocks},''
\href{http://arxiv.org/abs/1810.04153}{{\ttfamily arXiv:1810.04153 [gr-qc]}}.

\bibitem{Hoehn:2018whn}
P.~A. H\"ohn, ``{Switching Internal Times and a New Perspective on the Wave
  Function of the Universe},''
  \href{http://dx.doi.org/10.3390/universe5050116}{{\em Universe} {\bfseries 5}
  no.~5, (2019) 116},
\href{http://arxiv.org/abs/1811.00611}{{\ttfamily arXiv:1811.00611 [gr-qc]}}.

\bibitem{pqps}
P.~A. H\"ohn, ``Effective changes of quantum reference systems in quantum phase
  space,'' {\em to appear} (2020) .

\bibitem{Aharonov:1967zza}
Y.~Aharonov and L.~Susskind, ``{Charge Superselection Rule},''
\href{http://dx.doi.org/10.1103/PhysRev.155.1428}{{\em Phys. Rev.} {\bfseries
  155} (1967) 1428--1431}.

\bibitem{DeWitt:1967yk}
B.~S. DeWitt, ``{Quantum theory of gravity. I. The canonical theory},''
\href{http://dx.doi.org/10.1103/PhysRev.160.1113}{{\em Phys.Rev.} {\bfseries
  160} (1967) 1113--1148}.

\bibitem{Rovelli:2004tv}
C.~Rovelli, {\em {Quantum Gravity}}.
\href{https://doi.org/10.1017/CBO9780511755804}{Cambridge University Press,
2004.}
\newblock

\bibitem{Barbour295}
J.~Barbour and B.~Bertotti, ``Mach's principle and the structure of dynamical
  theories,'' \href{http://dx.doi.org/10.1098/rspa.1982.0102}{{\em Proceedings
  of the Royal Society of London A: Mathematical, Physical and Engineering
  Sciences} {\bfseries 382} no.~1783, (1982) 295--306}.

\bibitem{mercati2018shape}
F.~Mercati, {\em Shape Dynamics: Relativity and Relationalism}.
\href{https://doi.org/10.1093/oso/9780198789475.001.0001}{Oxford University Press, 2018}.

\bibitem{Rovelli:1990pi}
C.~Rovelli, ``{Quantum reference systems},''
\href{http://dx.doi.org/10.1088/0264-9381/8/2/012}{{\em Class.Quant.Grav.}
  {\bfseries 8} (1991) 317--332}.

\bibitem{Rovelli:1990ph}
C.~Rovelli, ``{What is observable in classical and quantum gravity?},''
\href{http://dx.doi.org/10.1088/0264-9381/8/2/011}{{\em Class.Quant.Grav.}
  {\bfseries 8} (1991) 297--316}.

\bibitem{Kuchar:1991qf}
K.~Kucha\v{r}, ``{Time and interpretations of quantum gravity},''
  \href{http://dx.doi.org/10.1142/S0218271811019347}{{\em
  Int.J.Mod.Phys.Proc.Suppl.} {\bfseries D20} (2011) 3--86}.
Originally published in the Proc. 4th Canadian Conf. on General Relativity and
  Relativistic Astrophysics, eds. G. Kunstatter, D. Vincent and J. Williams
  (World Scientific, Singapore, 1992).

\bibitem{Isham:1992ms}
C.~Isham, ``{Canonical quantum gravity and the problem of time},'' in {\em
  {Integrable Systems, Quantum Groups, and Quantum Field Theories}},
  pp.~157--287,
\href{http://dx.doi.org/10.1007/978-94-011-1980-1_6}{Kluwer Academic Publishers, 1993}, \href{http://arxiv.org/abs/gr-qc/9210011}{{\ttfamily arXiv:gr-qc/9210011
  [gr-qc]}}.

\bibitem{Brown:1994py}
J.~D. Brown and K.~V. Kucha\v{r}, ``{Dust as a standard of space and time in
  canonical quantum gravity},''
  \href{http://dx.doi.org/10.1103/PhysRevD.51.5600}{{\em Phys.Rev.} {\bfseries
  D51} (1995) 5600--5629},
\href{http://arxiv.org/abs/gr-qc/9409001}{{\ttfamily arXiv:gr-qc/9409001
  [gr-qc]}}.

\bibitem{Dittrich:2004cb}
B.~Dittrich, ``{Partial and complete observables for Hamiltonian constrained
  systems},'' \href{http://dx.doi.org/10.1007/s10714-007-0495-2}{{\em
  Gen.Rel.Grav.} {\bfseries 39} (2007) 1891--1927},
\href{http://arxiv.org/abs/gr-qc/0411013}{{\ttfamily arXiv:gr-qc/0411013
  [gr-qc]}}.

\bibitem{Dittrich:2005kc}
B.~Dittrich, ``{Partial and complete observables for canonical General
  Relativity},'' \href{http://dx.doi.org/10.1088/0264-9381/23/22/006}{{\em
  Class.Quant.Grav.} {\bfseries 23} (2006) 6155--6184},
\href{http://arxiv.org/abs/gr-qc/0507106}{{\ttfamily arXiv:gr-qc/0507106
  [gr-qc]}}.

\bibitem{Tambornino:2011vg}
J.~Tambornino, ``{Relational observables in gravity: A review},'' \href{https://doi.org/10.3842/SIGMA.2012.017}{{\em SIGMA}
  {\bfseries 8} (2012) 017},
\href{http://arxiv.org/abs/1109.0740}{{\ttfamily arXiv:1109.0740 [gr-qc]}}.

\bibitem{Thiemann:2007zz}
T.~Thiemann, \href{https://doi.org/10.1017/CBO9780511755682}{\em {Modern Canonical Quantum General Relativity}}.
\newblock Cambridge University Press,
2007.
\newblock

\bibitem{Dittrich:2016hvj}
B.~Dittrich, P.~A. H\"ohn, T.~A. Koslowski, and M.~I. Nelson, ``{Can chaos be
  observed in quantum gravity?},''
  \href{http://dx.doi.org/10.1016/j.physletb.2017.02.038}{{\em Phys. Lett.}
  {\bfseries B769} (2017) 554--560},
\href{http://arxiv.org/abs/1602.03237}{{\ttfamily arXiv:1602.03237 [gr-qc]}}.

\bibitem{Dittrich:2015vfa}
B.~Dittrich, P.~A. H\"ohn, T.~A. Koslowski, and M.~I. Nelson, ``{Chaos, Dirac
  observables and constraint quantization},''
\href{http://arxiv.org/abs/1508.01947}{{\ttfamily arXiv:1508.01947 [gr-qc]}}.

\bibitem{Bojowald:2010xp}
M.~Bojowald, P.~A. H\"ohn, and A.~Tsobanjan, ``{An Effective approach to the
  problem of time},''
  \href{http://dx.doi.org/10.1088/0264-9381/28/3/035006}{{\em Class. Quant.
  Grav.} {\bfseries 28} (2011) 035006},
\href{http://arxiv.org/abs/1009.5953}{{\ttfamily arXiv:1009.5953 [gr-qc]}}.

\bibitem{Bojowald:2010qw}
M.~Bojowald, P.~A. H\"ohn, and A.~Tsobanjan, ``{Effective approach to the
  problem of time: general features and examples},''
  \href{http://dx.doi.org/10.1103/PhysRevD.83.125023}{{\em Phys.Rev.}
  {\bfseries D83} (2011) 125023},
\href{http://arxiv.org/abs/1011.3040}{{\ttfamily arXiv:1011.3040 [gr-qc]}}.

\bibitem{Hohn:2011us}
P.~A. H\"ohn, E.~Kubalova, and A.~Tsobanjan, ``{Effective relational dynamics
  of a nonintegrable cosmological model},''
  \href{http://dx.doi.org/10.1103/PhysRevD.86.065014}{{\em Phys.Rev.}
  {\bfseries D86} (2012) 065014},
\href{http://arxiv.org/abs/1111.5193}{{\ttfamily arXiv:1111.5193 [gr-qc]}}.

\bibitem{PhysRev.158.1237}
Y.~Aharonov and L.~Susskind, ``Observability of the sign change of spinors
  under $2\ensuremath{\pi}$ rotations,''
  \href{http://dx.doi.org/10.1103/PhysRev.158.1237}{{\em Phys. Rev.} {\bfseries
  158} (Jun, 1967) 1237--1238}.

\bibitem{PhysRevD.30.368}
Y.~Aharonov and T.~Kaufherr, ``Quantum frames of reference,''
  \href{http://dx.doi.org/10.1103/PhysRevD.30.368}{{\em Phys. Rev. D}
  {\bfseries 30} (Jul, 1984) 368--385}.

\bibitem{Bartlett:2007zz}
S.~D. Bartlett, T.~Rudolph, and R.~W. Spekkens, ``{Reference frames,
  superselection rules, and quantum information},''
\href{http://dx.doi.org/10.1103/RevModPhys.79.555}{{\em Rev. Mod. Phys.}
  {\bfseries 79} (2007) 555--609}, \href{https://arxiv.org/abs/quant-ph/0610030}{{\ttfamily arXiv:quant-ph/0610030}}.

\bibitem{bartlett2009quantum}
S.~D. Bartlett, T.~Rudolph, R.~W. Spekkens, and P.~S. Turner, ``Quantum
  communication using a bounded-size quantum reference frame,'' \href{http://dx.doi.org/10.1088/1367-2630/11/6/063013}{{\em New
  Journal of Physics} {\bfseries 11} no.~6, (2009) 063013}, \href{https://arxiv.org/abs/0812.5040}{{\ttfamily arXiv:0812.5040 [quant-ph]}}.

\bibitem{gour2008resource}
G.~Gour and R.~W. Spekkens, ``The resource theory of quantum reference frames:
  manipulations and monotones,'' \href{http://dx.doi.org/ 	10.1088/1367-2630/10/3/033023}{{\em New Journal of Physics} {\bfseries 10}
  no.~3, (2008) 033023}, \href{https://arxiv.org/abs/0711.0043}{{\ttfamily  	arXiv:0711.0043 [quant-ph]}}.

\bibitem{Palmer:2013zza}
M.~C. Palmer, F.~Girelli, and S.~D. Bartlett, ``{Changing quantum reference
  frames},'' \href{http://dx.doi.org/10.1103/PhysRevA.89.052121}{{\em Phys.
  Rev.} {\bfseries A89} no.~5, (2014) 052121},
\href{http://arxiv.org/abs/1307.6597}{{\ttfamily arXiv:1307.6597 [quant-ph]}}.

\bibitem{bartlett2006degradation}
S.~D. Bartlett, T.~Rudolph, R.~W. Spekkens, and P.~S. Turner, ``Degradation of
  a quantum reference frame,'' \href{http://dx.doi.org/ 	10.1088/1367-2630/8/4/058}{{\em New Journal of Physics} {\bfseries 8}
  no.~4, (2006) 58}, \href{https://arxiv.org/abs/quant-ph/0602069}{{\ttfamily arXiv:1307.6597 [quant-ph]}}.

\bibitem{smith2016quantum}
A.~R. Smith, M.~Piani, and R.~B. Mann, ``Quantum reference frames associated
  with noncompact groups: The case of translations and boosts and the role of
  mass,'' \href{http://dx.doi.org/10.1103/PhysRevA.94.012333}{{\em Physical Review A} {\bfseries 94} no.~1, (2016) 012333}, \href{https://arxiv.org/abs/1602.07696}{{\ttfamily arXiv:1602.07696 [quant-ph]}}.

\bibitem{poulin2007dynamics}
D.~Poulin and J.~Yard, ``Dynamics of a quantum reference frame,'' \href{http://dx.doi.org/ 	10.1088/1367-2630/9/5/156}{{\em New
  Journal of Physics} {\bfseries 9} no.~5, (2007) 156}, \href{https://arxiv.org/abs/quant-ph/0612126}{{\ttfamily  	arXiv:quant-ph/0612126}}.

\bibitem{PhysRevLett.111.020504}
M.~Skotiniotis, B.~Toloui, I.~T. Durham, and B.~C. Sanders, ``Quantum frameness
  for $cpt$ symmetry,''
  \href{http://dx.doi.org/10.1103/PhysRevLett.111.020504}{{\em Phys. Rev.
  Lett.} {\bfseries 111} (Jul, 2013) 020504}, \href{https://arxiv.org/abs/1306.6114}{{\ttfamily arXiv:1306.6114 [quant-ph]}}.

\bibitem{loveridge2017relativity}
L.~Loveridge, P.~Busch, and T.~Miyadera, ``Relativity of quantum states and
  observables,'' \href{http://dx.doi.org/10.1209/0295-5075/117/40004}{{\em EPL (Europhysics Letters)} {\bfseries 117} no.~4, (2017)
  40004}, \href{https://arxiv.org/abs/1604.02836}{{\ttfamily arXiv:arXiv:1604.02836 [quant-ph]}}.

\bibitem{pienaar2016relational}
J.~Pienaar, ``A relational approach to quantum reference frames for spins,''
  \href{https://arxiv.org/abs/1601.07320}{{\ttfamily  	arXiv:1601.07320 [quant-ph]}} (2016) .

\bibitem{angelo2011physics}
R.~M. Angelo, N.~Brunner, S.~Popescu, A.~J. Short, and P.~Skrzypczyk, ``Physics
  within a quantum reference frame,'' \href{http://dx.doi.org/10.1088/1751-8113/44/14/145304}{{\em Journal of Physics A: Mathematical
  and Theoretical} {\bfseries 44} no.~14, (2011) 145304}, \href{https://arxiv.org/abs/1007.2292}{{\ttfamily  	arXiv:1007.2292 [quant-ph]}}.

\bibitem{Hoehn:2014vua}
P.~A. H\"ohn and M.~P. M\"uller, ``{An operational approach to spacetime
  symmetries: Lorentz transformations from quantum communication},''
  \href{http://dx.doi.org/10.1088/1367-2630/18/6/063026}{{\em New J. Phys.}
  {\bfseries 18} no.~6, (2016) 063026},
\href{http://arxiv.org/abs/1412.8462}{{\ttfamily arXiv:1412.8462 [quant-ph]}}.

\bibitem{guerin2018observer}
P.~A. Gu\'erin and {\v{C}}.~Brukner, ``Observer-dependent locality of quantum
  events,'' \href{http://dx.doi.org/10.1088/1367-2630/aae742}{{\em New J. Phys.}
  {\bfseries 20}, 103031 (2018)},
\href{https://arxiv.org/abs/1805.12429}{{\ttfamily arXiv:1805.12429 [quant-ph]}}.

\bibitem{oreshkov2012quantum}
O.~Oreshkov, F.~Costa, and {\v{C}}.~Brukner, ``Quantum correlations with no
  causal order,'' \href{http://dx.doi.org/10.1038/ncomms2076}{{\em Nature communications} {\bfseries 3} (2012) 1092}, \href{https://arxiv.org/abs/1105.4464}{{\ttfamily arXiv:1105.4464 [quant-ph]}}.

\bibitem{Hardy:2018kbp}
L.~Hardy, ``{The Construction Interpretation: Conceptual Roads to Quantum
  Gravity},''
\href{http://arxiv.org/abs/1807.10980}{{\ttfamily arXiv:1807.10980
  [quant-ph]}}.

\bibitem{Hoehn:2017gst}
P.~A. H\"ohn, ``{Reflections on the information paradigm in quantum and
  gravitational physics},''
  \href{http://dx.doi.org/10.1088/1742-6596/880/1/012014}{{\em J. Phys. Conf.
  Ser.} {\bfseries 880} no.~1, (2017) 012014},
\href{http://arxiv.org/abs/1706.06882}{{\ttfamily arXiv:1706.06882 [hep-th]}}.

\bibitem{Dirac}
P.~A. Dirac, {\em {Lectures on Quantum Mechanics}}.
\href{https://doi.org/10.1142/6093}{Yeshiva University Press, 1964}.

\bibitem{Henneaux:1992ig}
M.~Henneaux and C.~Teitelboim, {\em {Quantization of Gauge Systems}}.
\newblock Princeton University Press,
1992.
\newblock

\bibitem{Rovelli:2013fga}
C.~Rovelli, ``{Why Gauge?},''
  \href{http://dx.doi.org/10.1007/s10701-013-9768-7}{{\em Found. Phys.}
  {\bfseries 44} no.~1, (2014) 91--104},
\href{http://arxiv.org/abs/1308.5599}{{\ttfamily arXiv:1308.5599 [hep-th]}}.

\bibitem{Gomes:2018dxs}
H.~Gomes, F.~Hopfm\"uller, and A.~Riello, ``{A unified geometric framework for
  boundary charges and dressings: non-Abelian theory and matter},'' \href{http://dx.doi.org/10.1016/j.nuclphysb.2019.02.020}{{\em Nuclear Physics B} {\bfseries 941} (2019) 249-315},
\href{http://arxiv.org/abs/1808.02074}{{\ttfamily arXiv:1808.02074 [hep-th]}}.

\bibitem{guillemin1982geometric}
V.~Guillemin and S.~Sternberg, ``Geometric quantization and multiplicities of
  group representations,'' \href{https://doi.org/10.1007/BF01398934}{{\em Inventiones mathematicae} {\bfseries 67} no.~3,
  (1982) 515--538}.

\bibitem{tian1998analytic}
Y.~Tian and W.~Zhang, ``An analytic proof of the geometric quantization
  conjecture of guillemin-sternberg,'' \href{https://doi.org/10.1007/s002220050223}{{\em Inventiones mathematicae}
  {\bfseries 132} no.~2, (1998) 229--259}.

\bibitem{hochs2008guillemin}
P.~Hochs and N.~Landsman, ``The guillemin--sternberg conjecture for noncompact
  groups and spaces,'' \href{https://doi.org/10.1017/is008001002jkt022}{{\em Journal of K-theory} {\bfseries 1} no.~3, (2008)
  473--533}, \href{https://arxiv.org/abs/math-ph/0512022}{{\ttfamily  arXiv:math-ph/0512022}}.

\bibitem{gotay1986constraints}
M.~J. Gotay, ``Constraints, reduction, and quantization,'' \href{ https://doi.org/10.1063/1.527026}{{\em Journal of
  mathematical physics} {\bfseries 27} no.~8, (1986) 2051--2066}.

\bibitem{Ashtekar:1982wv}
A.~Ashtekar and G.~t. Horowitz, ``{On the canonical approach to quantum
  gravity},''
\href{http://dx.doi.org/10.1103/PhysRevD.26.3342}{{\em Phys. Rev.} {\bfseries
  D26} (1982) 3342--3353}.

\bibitem{kucha1986covariant}
K.~Kucha\v{r}, ``Covariant factor ordering of gauge systems,'' \href{http://dx.doi.org/10.1103/physrevd.34.3044}{{\em Physical
  Review D} {\bfseries 34} no.~10, (1986) 3044}.

\bibitem{Ashtekar:1991hf}
A.~Ashtekar, {\em {Lectures on Nonperturbative Canonical Gravity}}, \href{http://dx.doi.org/10.1142/1321}{No.~6 in Advances series in astrophysics and cosmology. World
  Scientific,
1991}.

\bibitem{Schleich:1990gd}
K.~Schleich, ``{Is reduced phase space quantization equivalent to Dirac
  quantization?},''
\href{http://dx.doi.org/10.1088/0264-9381/7/8/028}{{\em Class. Quant. Grav.}
  {\bfseries 7} (1990) 1529--1538}.

\bibitem{Kunstatter:1991ds}
G.~Kunstatter, ``{Dirac versus reduced quantization: A Geometrical approach},''
\href{http://dx.doi.org/10.1088/0264-9381/9/6/005}{{\em Class. Quant. Grav.}
  {\bfseries 9} (1992) 1469--1486}.

\bibitem{Hajicek:1990eu}
P.~Hajicek and K.~V. Kuchar, ``{Constraint quantization of parametrized
  relativistic gauge systems in curved space-times},''
\href{http://dx.doi.org/10.1103/PhysRevD.41.1091}{{\em Phys. Rev.} {\bfseries
  D41} (1990) 1091--1104}.

\bibitem{Romano:1989zb}
J.~D. Romano and R.~S. Tate, ``{Dirac Versus Reduced Space Quantization of
  Simple Constrained Systems},''
\href{http://dx.doi.org/10.1088/0264-9381/6/10/017}{{\em Class. Quant. Grav.}
  {\bfseries 6} (1989) 1487}.

\bibitem{Loll:1990rx}
R.~Loll, ``{Noncommutativity of constraining and quantizing: A U(1) gauge
  model},''
\href{http://dx.doi.org/10.1103/PhysRevD.41.3785}{{\em Phys. Rev.} {\bfseries
  D41} (1990) 3785--3791}.

\bibitem{plyushchay1996dirac}
M.~S. Plyushchay and A.~V. Razumov, ``Dirac versus reduced phase space
  quantization for systems admitting no gauge conditions,'' \href{http://dx.doi.org/10.1142/S0217751X96000663}{{\em International
  Journal of Modern Physics A} {\bfseries 11} no.~08, (1996) 1427--1462}, \href{https://arxiv.org/abs/hep-th/9306017}{{\ttfamily arXiv:hep-th/9306017}}.

\bibitem{Hoehn:2014uua}
P.~A. H\"ohn, ``{Toolbox for reconstructing quantum theory from rules on
  information acquisition},''
  \href{http://dx.doi.org/10.22331/q-2017-12-14-38}{{\em Quantum} {\bfseries 1}
  no.~38, (2017) },
\href{http://arxiv.org/abs/1412.8323}{{\ttfamily arXiv:1412.8323 [quant-ph]}}.

\bibitem{Rovelli:1995fv}
C.~Rovelli, ``{Relational quantum mechanics},''
  \href{http://dx.doi.org/10.1007/BF02302261}{{\em Int.J.Theor.Phys.}
  {\bfseries 35} (1996) 1637--1678},
\href{http://arxiv.org/abs/quant-ph/9609002}{{\ttfamily arXiv:quant-ph/9609002
  [quant-ph]}}.

\bibitem{rovelli2018space}
C.~Rovelli, ``Space is blue and birds fly through it,'' \href{http://dx.doi.org/10.1098/rsta.2017.0312}{{\em Phil. Trans. R.
  Soc. A} {\bfseries 376} no.~2123, (2018) 20170312},
\href{https://arxiv.org/abs/1712.02894}{{\ttfamily  	arXiv:1712.02894 [physics.hist-ph]}}.

\bibitem{Gomes:2010fh}
H.~Gomes, S.~Gryb, and T.~Koslowski, ``{Einstein gravity as a 3D conformally
  invariant theory},''
  \href{http://dx.doi.org/10.1088/0264-9381/28/4/045005}{{\em
  Class.Quant.Grav.} {\bfseries 28} (2011) 045005},
\href{http://arxiv.org/abs/1010.2481}{{\ttfamily arXiv:1010.2481 [gr-qc]}}.

\bibitem{Gomes:2011zi}
H.~Gomes and T.~Koslowski, ``{The Link between General Relativity and Shape
  Dynamics},'' \href{http://dx.doi.org/10.1088/0264-9381/29/7/075009}{{\em
  Class. Quant. Grav.} {\bfseries 29} (2012) 075009},
\href{http://arxiv.org/abs/1101.5974}{{\ttfamily arXiv:1101.5974 [gr-qc]}}.

\bibitem{Hoehn:2018rfe}
P.~A. H\"ohn, M.~P. M\"uller, C.~Pfeifer, and D.~R\"atzel, ``{A local quantum
  Mach principle and the metricity of spacetime},''
\href{http://arxiv.org/abs/1811.02555}{{\ttfamily arXiv:1811.02555 [gr-qc]}}.

\bibitem{Barbour:2014bga}
J.~Barbour, T.~Koslowski, and F.~Mercati, ``{Identification of a gravitational
  arrow of time},''
  \href{http://dx.doi.org/10.1103/PhysRevLett.113.181101}{{\em Phys. Rev.
  Lett.} {\bfseries 113} no.~18, (2014) 181101},
\href{http://arxiv.org/abs/1409.0917}{{\ttfamily arXiv:1409.0917 [gr-qc]}}.

\bibitem{Barbour:2015sba}
J.~Barbour, T.~Koslowski, and F.~Mercati, ``{Entropy and the Typicality of
  Universes},''
\href{http://arxiv.org/abs/1507.06498}{{\ttfamily arXiv:1507.06498 [gr-qc]}}.

\bibitem{Hajicek:1986ky}
P.~H{\'a}j{\'\i}\v{c}ek, ``{Origin of nonunitarity in quantum gravity},''
\href{http://dx.doi.org/10.1103/PhysRevD.34.1040}{{\em Phys.Rev.} {\bfseries
  D34} (1986) 1040}.

\bibitem{Kempf:2000qz}
A.~Kempf and J.~R. Klauder, ``{On the implementation of constraints through
  projection operators},''
  \href{http://dx.doi.org/10.1088/0305-4470/34/5/307}{{\em J. Phys.} {\bfseries
  A34} (2001) 1019--1036},
\href{http://arxiv.org/abs/quant-ph/0009072}{{\ttfamily arXiv:quant-ph/0009072
  [quant-ph]}}.

\bibitem{Marolf:1995cn}
D.~Marolf, ``{Refined algebraic quantization: Systems with a single
  constraint},''
\href{http://arxiv.org/abs/gr-qc/9508015}{{\ttfamily arXiv:gr-qc/9508015
  [gr-qc]}}.

\bibitem{Marolf:2000iq}
D.~Marolf, ``{Group averaging and refined algebraic quantization: Where are we
  now?},''
\href{http://arxiv.org/abs/gr-qc/0011112}{{\ttfamily arXiv:gr-qc/0011112
  [gr-qc]}}.

\bibitem{Faddeev:1967fc}
L.~D. Faddeev and V.~N. Popov, ``{Feynman Diagrams for the Yang-Mills Field},''
\href{http://dx.doi.org/10.1016/0370-2693(67)90067-6}{{\em Phys. Lett. B}
  {\bfseries 25} (1967) 29--30}.

\bibitem{Fradkin:1975cq}
E.~S. Fradkin and G.~A. Vilkovisky, ``{Quantization of relativistic systems
  with constraints},''
\href{http://dx.doi.org/10.1016/0370-2693(75)90448-7}{{\em Phys. Lett.}
  {\bfseries 55B} (1975) 224--226}.

\bibitem{Batalin:1977pb}
I.~A. Batalin and G.~A. Vilkovisky, ``{Relativistic S Matrix of Dynamical
  Systems with Boson and Fermion Constraints},''
\href{http://dx.doi.org/10.1016/0370-2693(77)90553-6}{{\em Phys. Lett.}
  {\bfseries 69B} (1977) 309--312}.

\bibitem{Fradkin:1977xi}
E.~S. Fradkin and T.~E. Fradkina, ``{Quantization of Relativistic Systems with
  Boson and Fermion First and Second Class Constraints},''
\href{http://dx.doi.org/10.1016/0370-2693(78)90135-1}{{\em Phys. Lett.}
  {\bfseries 72B} (1978) 343--348}.

\bibitem{Dittrich:2011ke}
B.~Dittrich and P.~A. H\"ohn, ``{Canonical simplicial gravity},''
  \href{http://dx.doi.org/10.1088/0264-9381/29/11/115009}{{\em
  Class.Quant.Grav.} {\bfseries 29} (2012) 115009},
\href{http://arxiv.org/abs/1108.1974}{{\ttfamily arXiv:1108.1974 [gr-qc]}}.

\bibitem{Dittrich:2013jaa}
B.~Dittrich and P.~A. H\"ohn, ``{Constraint analysis for variational discrete
  systems},'' \href{http://dx.doi.org/10.1063/1.4818895}{{\em J. Math. Phys.}
  {\bfseries 54} (2013) 093505},
\href{http://arxiv.org/abs/1303.4294}{{\ttfamily arXiv:1303.4294 [math-ph]}}.

\bibitem{Hoehn:2014aoa}
P.~A. H\"ohn, ``{Classification of constraints and degrees of freedom for
  quadratic discrete actions},''
  \href{http://dx.doi.org/10.1063/1.4900926}{{\em J.Math.Phys.} {\bfseries 55}
  (2014) 113506},
\href{http://arxiv.org/abs/1407.6641}{{\ttfamily arXiv:1407.6641 [math-ph]}}.

\bibitem{Hoehn:2014qxa}
P.~A. H\"ohn, ``{Canonical linearized Regge Calculus: counting lattice
  gravitons with Pachner moves},''
  \href{http://dx.doi.org/10.1103/PhysRevD.91.124034}{{\em Phys. Rev.}
  {\bfseries D91} no.~12, (2015) 124034},
\href{http://arxiv.org/abs/1411.5672}{{\ttfamily arXiv:1411.5672 [gr-qc]}}.

\bibitem{Hoehn:2014wwa}
P.~A. H\"ohn, ``{Quantization of systems with temporally varying discretization
  II: Local evolution moves},'' \href{http://dx.doi.org/10.1063/1.4898764}{{\em
  J.Math.Phys.} {\bfseries 55} (2014) 103507},
\href{http://arxiv.org/abs/1401.7731}{{\ttfamily arXiv:1401.7731 [gr-qc]}}.

\bibitem{Page:1983uc}
D.~N. Page and W.~K. Wootters, ``{Evolution without evolution: Dynamics
  described by stationary observables},''
\href{http://dx.doi.org/10.1103/PhysRevD.27.2885}{{\em Phys. Rev.} {\bfseries
  D27} (1983) 2885}.

\bibitem{giacomini2019relativistic}
F.~Giacomini, E.~Castro-Ruiz, and {\v{C}}.~Brukner, ``Relativistic quantum
  reference frames: the operational meaning of spin,'' \href{http://dx.doi.org/10.1103/PhysRevLett.123.090404}{{\em Physical review
  letters} {\bfseries 123} no.~9, (2019) 090404},
\href{https://arxiv.org/abs/1811.08228}{{\ttfamily  	arXiv:1811.08228 [quant-ph]}}.

\bibitem{wigner1995remarks}
E.~P. Wigner, ``Remarks on the mind-body question,'' in \href{https://doi.org/10.1007/978-3-642-78374-6_20}{{\em Philosophical
  reflections and syntheses}, pp.~247--260.
\newblock Springer, 1995}.

\bibitem{deutsch1985quantum}
D.~Deutsch, ``Quantum theory as a universal physical theory,'' \href{https://doi.org/10.1007/BF00670071}{{\em
  International Journal of Theoretical Physics} {\bfseries 24} no.~1, (1985)
  1--41}.

\bibitem{brukner2017quantum}
{\v{C}}.~Brukner, ``On the quantum measurement problem,'' in \href{https://doi.org/10.1007/978-3-319-38987-5_5}{{\em Quantum [Un]
  Speakables II}, pp.~95--117.
\newblock Springer, 2017}, \href{https://arxiv.org/abs/1507.05255}{{\ttfamily  arXiv:1507.05255 [quant-ph]}}.

\bibitem{frauchiger2016single}
D.~Frauchiger and R.~Renner, ``Quantum theory cannot consistently describe the use of itself,'' \href{https://doi.org/10.1038/s41467-018-05739-8}{{\em
  Nature Communications} {\bfseries 9} no.~3711, (2018)
 }, \href{https://arxiv.org/abs/1604.07422}{{\ttfamily   	arXiv:1604.07422 [quant-ph]}}.

\bibitem{Hoehn:2015zom}
P.~A. H\"ohn and C.~S.~P. Wever, ``{Quantum theory from questions},''
  \href{http://dx.doi.org/10.1103/PhysRevA.95.012102}{{\em Phys. Rev.}
  {\bfseries A95} no.~1, (2017) 012102},
\href{http://arxiv.org/abs/1511.01130}{{\ttfamily arXiv:1511.01130
  [quant-ph]}}.

\bibitem{Hoehn:2016otu}
P.~A. H\"ohn, ``{Quantum theory from rules on information acquisition},''
  \href{http://dx.doi.org/10.3390/e19030098}{{\em Entropy} {\bfseries 19}
  (2017) 98},
\href{http://arxiv.org/abs/1612.06849}{{\ttfamily arXiv:1612.06849
  [quant-ph]}}.

\bibitem{Hardy:2001jk}
L.~Hardy, ``{Quantum theory from five reasonable axioms},''
\href{http://arxiv.org/abs/quant-ph/0101012}{{\ttfamily arXiv:quant-ph/0101012
  [quant-ph]}}.

\bibitem{Dakic:2009bh}
B.~Dakic and C.~Brukner, ``Quantum theory and beyond: Is entanglement
  special?,'' \href{https://doi.org/10.1017/CBO9780511976971.011}{{\em Deep Beauty: Understanding the Quantum World through
  Mathematical Innovation, Ed. H. Halvorson (Cambridge University Press, 2011)
  365-392} (11, 2009)} , \href{http://arxiv.org/abs/0911.0695}{{\ttfamily
 arXiv:0911.0695 [quant-ph]}}.
 
\bibitem{masanes2011derivation}
L.~Masanes and M.~P. M{\"u}ller, ``A derivation of quantum theory from physical
  requirements,'' \href{https://doi.org/ 	10.1088/1367-2630/13/6/063001}{{\em New Journal of Physics} {\bfseries 13} no.~6, (2011)
  063001}, \href{https://arxiv.org/abs/1004.1483}{{\ttfamily
  	arXiv:1004.1483 [quant-ph]}}.

\bibitem{chiribella2011informational}
G.~Chiribella, G.~M. D'Ariano, and P.~Perinotti, ``Informational derivation of
  quantum theory,'' \href{https://doi.org/10.1103/PhysRevA.84.012311}{{\em Physical Review A} {\bfseries 84} no.~1, (2011)
  012311}, \href{https://arxiv.org/abs/1011.6451}{{\ttfamily
  	 	arXiv:1011.6451 [quant-ph]}}.

\bibitem{Barnum:2014fk}
H.~Barnum, M.~P. M\"uller, and C.~Ududec, ``Higher-order interference and
  single-system postulates characterizing quantum theory,'', \href{https://doi.org/10.1088/1367-2630/16/12/123029}{{\em New Journal of Physics} {\bfseries 16}, (2011)
  123029},
  \href{http://arxiv.org/abs/1403.4147}{{\ttfamily  	arXiv:1403.4147 [quant-ph]}}.

\bibitem{goyal2010information}
P.~Goyal, ``From information geometry to quantum theory,'' \href{https://doi.org/10.1088/1367-2630/12/2/023012}{{\em New Journal of
  Physics} {\bfseries 12} no.~2, (2010) 023012}, \href{https://arxiv.org/abs/0805.2770}{{\ttfamily arXiv:0805.2770 [quant-ph]}}.

\bibitem{Mueller:2017cdn}
M.~P. M\"uller, ``{Law without law: from observer states to physics via
  algorithmic information theory},''
\href{http://arxiv.org/abs/1712.01826}{{\ttfamily arXiv:1712.01826
  [quant-ph]}}.

\bibitem{Mueller:2017jwa}
M.~P. M\"uller, ``{Could the physical world be emergent instead of fundamental,
  and why should we ask? (short version)},''
\href{http://arxiv.org/abs/1712.01816}{{\ttfamily arXiv:1712.01816
  [quant-ph]}}.

\end{thebibliography}
\providecommand{\href}[2]{#2}\begingroup\raggedright\endgroup

\end{document}